\documentclass[iop]{emulateapj}

\begin{document}
\title{Kinks and Dents in Protoplanetary Disks: Rapid Infrared Variability as Evidence for Large Structural Perturbations}
\author{Flaherty, K.M. \altaffilmark{1}, Muzerolle, J. \altaffilmark{2}, Rieke, G. \altaffilmark{1}, Gutermuth, R. \altaffilmark{3}, Balog, Z.\altaffilmark{4}, Herbst, W. \altaffilmark{5}, Megeath, S.T. \altaffilmark{6}}
\email{kflaherty@as.arizona.edu}

\altaffiltext{1}{Steward Observatory, University of Arizona, Tucson, AZ 85721}
\altaffiltext{2}{Space Telescope Science Institute, 3700 San Martin Dr., Baltimore, MD, 21218}
\altaffiltext{3}{Department of Astronomy, University of Massachusetts, Amherst, MA 01003}
\altaffiltext{4}{Max-Planck Institut f\"{u}r Astronomie, K\"{o}nigstuhl 17, D-69117 Heidelberg, Germany}
\altaffiltext{5}{Department of Astronomy, Wesleyan University, Middletown, CT 06459}
\altaffiltext{6}{Department of Physics and Astronomy, University of Toledo, Toledo, OH}

\begin{abstract}
We report on synoptic observations at 3.6 and 4.5\micron\ of young stellar objects in IC 348 with 38 epochs covering 40 days. We find that among the detected cluster members, 338 at [3.6] and 269 at both [3.6] and [4.5], many are variable on daily to weekly timescales with typical fluctuations of $\sim0.1$ mag. The fraction of variables ranges from  20\%\ for the diskless pre-main sequence stars to 60\%\ for the stars still surrounded by infalling envelopes. We also find that stars in the exposed cluster core are less variable than the stars in the dense, slightly younger, south-western ridge. This trend persists even after accounting for the underlying correlation with infrared SED type, suggesting that the change in variable fraction is not simply a reflection of the change in relative fraction of class I vs. class II sources across the cloud, but instead reflects a change in variability with age. We also see a strong correlation between infrared variability and X-ray luminosity among the class II sources. The observed variability most likely reflects large changes in the structure of the inner wall located at the dust sublimation radius. We explore the possibility that these structural perturbations could be caused by a hot spot on the star heating dust above the sublimation temperature, causing it to evaporate rapidly, and increasing the inner radius for a portion of the disk. Under a number of simplifying assumptions we show that this model can reproduce the size and timescale of the 3.6 and 4.5\micron\ fluctuations. Regardless of its source, the infrared variability indicates that the inner disk is not a slowly evolving entity, but instead is a bubbling, warped, dented mass of gas and dust whose global size and shape fluctuate in a matter of days.
\end{abstract}

\section{Introduction}

%\begin{itemize}
%\item Eiroa et al. 2002: SImultaneous optical and near-infrared monitoring of 18 Herbig Ae/Be and T Tauri stars. Most, but not all, show correlated optical and near-infrared fluctuations.
%\item Bary et al. 2009: IRS spectra over 11 T Tauri stars in taurus over 1.5 years. DG Tau and XZ Tau show significant variations in their silicate feature on month and year-long timescales. Possibly consistent with self-absorption due to two-temperature disk components. 
%\item Abraham et al. 2009 (plus updates): Change in silicate feature shape following an outburst in Ex Lupi. Crystalline silicate grains show up after outburst.
%\end{itemize}

Variability has been a defining characteristic of young pre-main sequence stars for over 60 years \citep{joy45}. These fluctuations occur not only over a wide range of timescales, but over a wide range of wavelengths and have been used to deduce various properties of these systems. Daily to weekly optical fluctuations up to 1-2 magnitudes are common \citep{her94}. Such behavior has been used to infer the presence of hot and cold spots covering a small to moderate fraction of the star and is often used to study the angular momentum evolution of young stars \citep[e.g.][]{her02,reb06}. X-ray variability, often associated with flares from magnetic reconnection events, is also common among young stars \citep{wol05,ste07}. Select young stars have even been observed to be variable at wavelengths as long as the radio \citep{fei99,for08}.

Infrared wavelengths are certainly no exception when it comes to young stellar variability. \citet{car01} find that many stars toward the star-forming cloud Orion A are variable in the near-infrared on timescales of days to weeks. \citet{mor09,mor11} find that $\sim$70\%\ of young stars are variable at 3.6 and 4.5\micron\ and \citet{bar05} even find a significant variable fraction at 10\micron\ on similarly short timescales. These wavelengths are particularly interesting because they probe the terrestrial planet forming region within a few AU of the star \citep{dal06}. The variability has the potential to provide information about a region of the disk that is well below the resolution limit of current telescopes.

These surveys have demonstrated the ubiquity of infrared variability, while more targeted campaigns of a small handful of objects have probed the details of disk structure. They have been able to infer vertical perturbations in the dust at the very inner edge of the disk \citep{hut94,juh07,sit08,esp11}. These perturbations produce a distinctive 'seesaw' behavior in the spectral energy distribution (SED) in which the short wavelength flux ($\lambda<8\micron$) increases as the inner disk grows and the long wavelength flux decreases due to the additional shadowing of the outer disk. It is difficult to reproduce a SED that appears to pivot around a point by the rotation of a vertical perturbation, without also changing the nature of that perturbation \citep{fla10}, suggesting that these structures are constructed and destroyed on short timescales. Inhomogeneous disk structure has also been seen in resolved observations \citep[e.g.][]{arn12} and has been inferred from rapid occultation events by dust clouds above the disk \citep{bou07,mor11,sit08,ale10}. The 'seesaw' behavior appears to be common among disks with large radial gaps and holes \citep{esp11,fla11,fla12} but may not be common among less evolved objects \citep[][Megeath et al. accepted]{kos12}. Among binary systems, variability in the optically thin emission has been observed \citep{ske10,nag12} and can be connected to the streamers of dust flowing through the cleared out region of the disk surrounding the binary stars \citep{nag12}.

Throughout these targeted studies, infrared variability has been used to infer new aspects of the disk structure, but only for a limited number of objects. Here we present an infrared monitoring campaign of the young ($\sim3$Myr), nearby ($\sim$320pc) cluster IC348 in an attempt to characterize variability for many young stellar objects over a wide range of evolutionary stages. In section 2 we describe our Spitzer warm-mission monitoring campaign and in section 3 we discuss our technique for selecting, and our sensitivity to, variable stars. Section 4 discusses variability among the cluster members as well as how this variability depends on various system parameters. We conclude in section 5 with a discussion of potential causes for these fluctuations.

\section{Data}
We performed repeated IRAC 3.6 and 4.5 \micron\ observations (PID 60160) of IC 348 over the course of 40 days from 3 Oct to 7 Nov 2009 to study the infrared variability of young stellar objects within this cluster. The cadence of these observations was varied from once every four hours to once every two days to capture a wide range of possible timescales for the variability. We mapped the cluster with a 5 by 5 grid, with each grid point separated by 260" and the total field centered at 03:44:20 +32:03:01. Since the fields of view of the [3.6] and [4.5] arrays do not completely overlap, our final map covered by both bands is not a square and has a total area of roughly 0.4$^{\circ}$ by 0.25$^{\circ}$. We chose a field center that is slightly south of the center of the main cluster core to cover the YSOs in the more highly obscured south-western ridge. The incomplete overlap between the [3.6] and [4.5] fields of view, and our choice of field center, results in many cluster members only having [3.6] data with no corresponding [4.5] data. Images were taken in HDR mode with a frame time of 12 seconds and 3 cycles at each position (to allow for removal of cosmic rays). The depth allowed us to detect sources down to [3.6],[4.5]$\sim$16 in each epoch, similar to earlier Spitzer studies by \citet{lad06} and \citet{mue07}. The IRAC data reduction pipeline is described in \citet{gut09} and \citet{mor12} with updates appropriate for the warm Spitzer mission described in Gutermuth et al in prep. This monitoring was part of a larger multi-wavelength campaign that includes MIPS [24] photometry (PID 40372), IRAC 3-8\micron\ cold mission photometry (PID 50596) (Muzerolle et al. in prep), Herschel PACS photometry (Balog et al. in prep) as well as ground-based spectroscopy \citep{fla11,fla12}.

\section{Defining Variables}
\begin{figure*}
\center
\includegraphics[scale=.45]{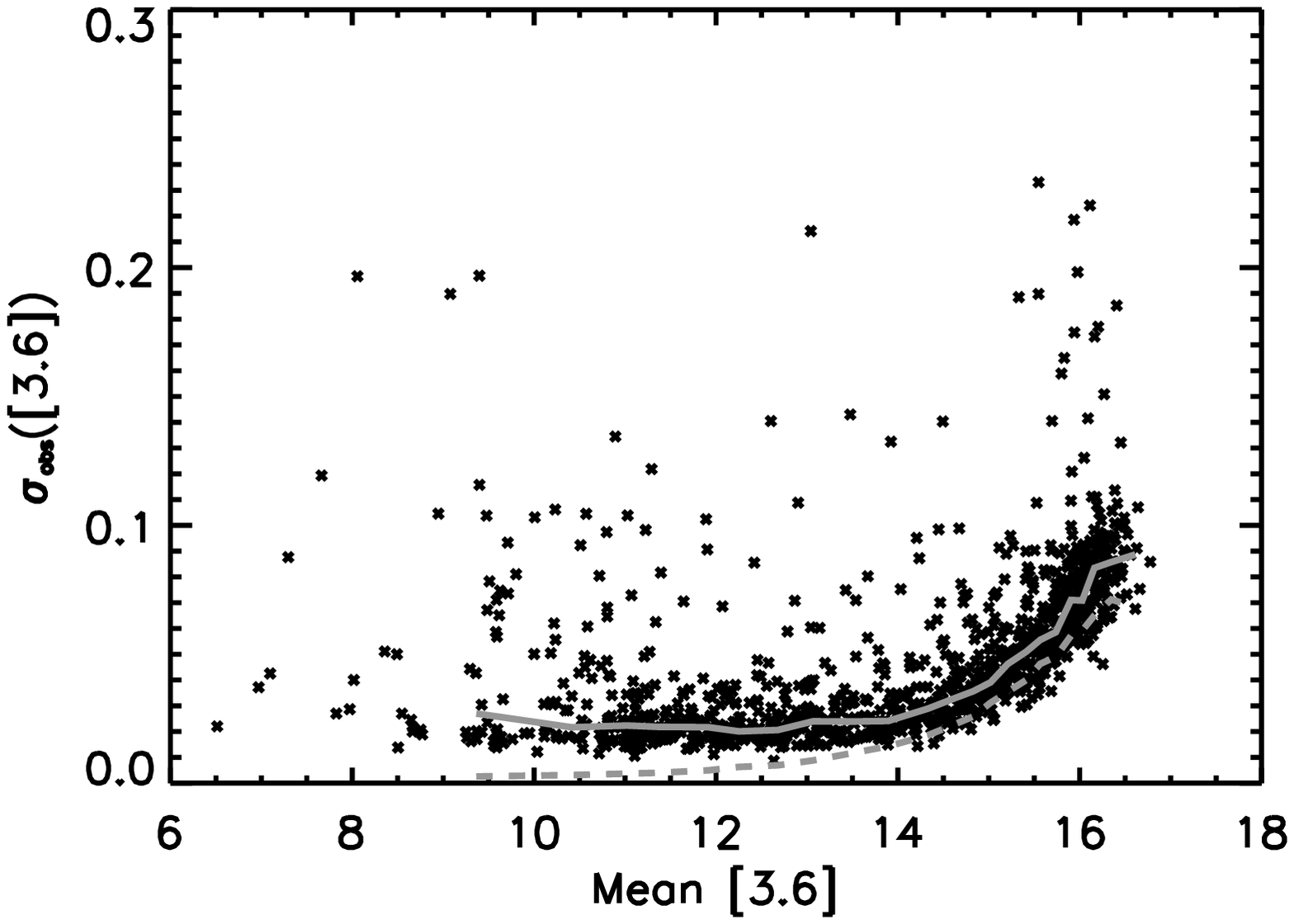}
\includegraphics[scale=.45]{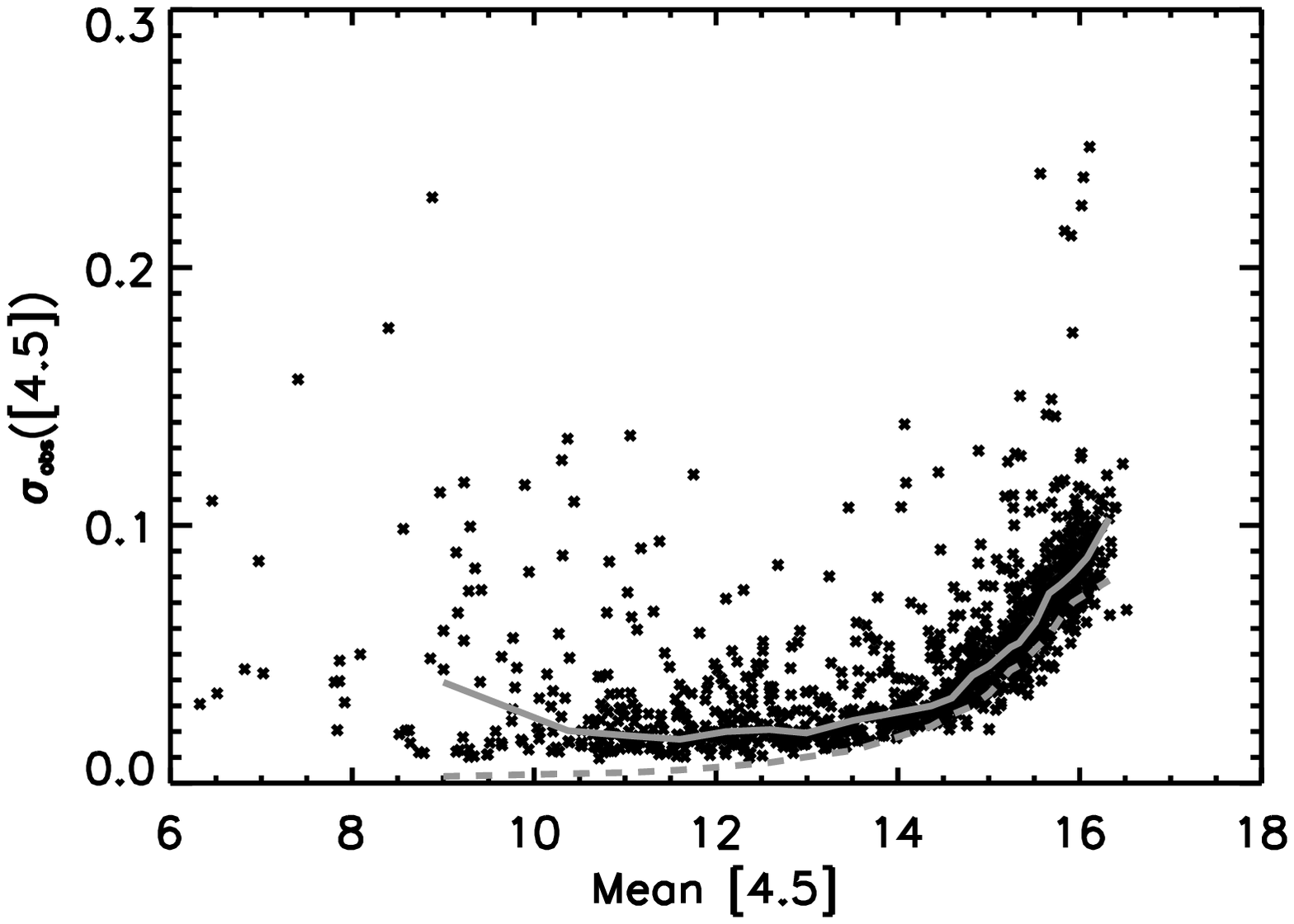}
\caption{Observed RMS versus magnitude for all sources, not just cluster members, detected in every epoch for [3.6] (left) and [4.5] (right). The photometric uncertainties (dashed line) systematically underestimate the true uncertainty as measured from the non-variable stars (solid line). We use the measured uncertainty of the non-variable stars as an estimate of the true uncertainty in the data. For clarity we have excluded the one source with $\sigma_{obs}$ greater than 0.3 from these plots. \label{rms}}
\end{figure*}

When defining a star as variable we strive for a method that maximizes sensitivity while still being applicable to many stars. We must first accurately characterize the uncertainty in the data. We start by calculating the rms, $\sigma_{obs}$, for every star that was detected in both [3.6] and [4.5] in all 38 epochs with a photometric uncertainty less than 0.1. The photometric uncertainty, as reported by the source extraction routines, is a combination of shot noise, uncertainty in background estimation and measured standard deviation in the sky annulus. The actual dispersion in the data may be larger than this due to detector effects that have not been fully corrected. The size of these systematic errors can be measured using the rms, $\sigma_{obs}$, of the non-variable sources. The $\sigma_{obs}$ is defined as:
\begin{equation}
\sigma_{obs}^2=\frac{n\Sigma^n_{i=1}w_i(m_i-\bar{m})^2}{(n-1)\Sigma^n_{i=1}w_i}
\end{equation}
where $n$ is the number of observations, $m_i$ is the magnitude on a given epoch, $\bar{m}$ is the average magnitude, $w_i=1/\sigma_i^2$ and $\sigma_i$ is the photometric noise. In Figure~\ref{rms} we show the $\sigma_{obs}$ in each band as a function of [3.6] and [4.5] magnitude. The main locus contains non-variable sources and traces the uncertainty in the data. The uncertainty is relatively flat for stars brighter than 13th magnitude, but increases sharply for fainter sources. We find that the $\sigma_{obs}$ is larger than the photometric uncertainties (shown as the dashed line) and the difference between $\sigma_{obs}$ and the photometric uncertainties increases from 13th to 10th magnitude. To more accurately estimate the uncertainties, we fit a line to the locus of non-variable sources (shown as a grey line in Figure~\ref{rms}). For each star in our sample, we use its average magnitude to interpolate along this line and derive the uncertainty that will replace its photometric uncertainty. Brighter than 10th magnitude there are not enough stars to create a well-defined locus and we assume the uncertainty is 0.017 and 0.014 mag for [3.6] and [4.5] respectively. A possible concern with this approach is that the nebulous emission associated with the molecular cloud creates a highly variable background that may lead to non-uniform uncertainties across the field. We examine the rms of those stars seen against the nebulosity, versus those seen off the nebulosity and find no difference in the location of the non-variable locus. This indicates that the nebulosity does not substantially effect our uncertainties and in turn does not change our sensitivity to variability. 

\begin{figure*}
\center
\includegraphics[scale=.45]{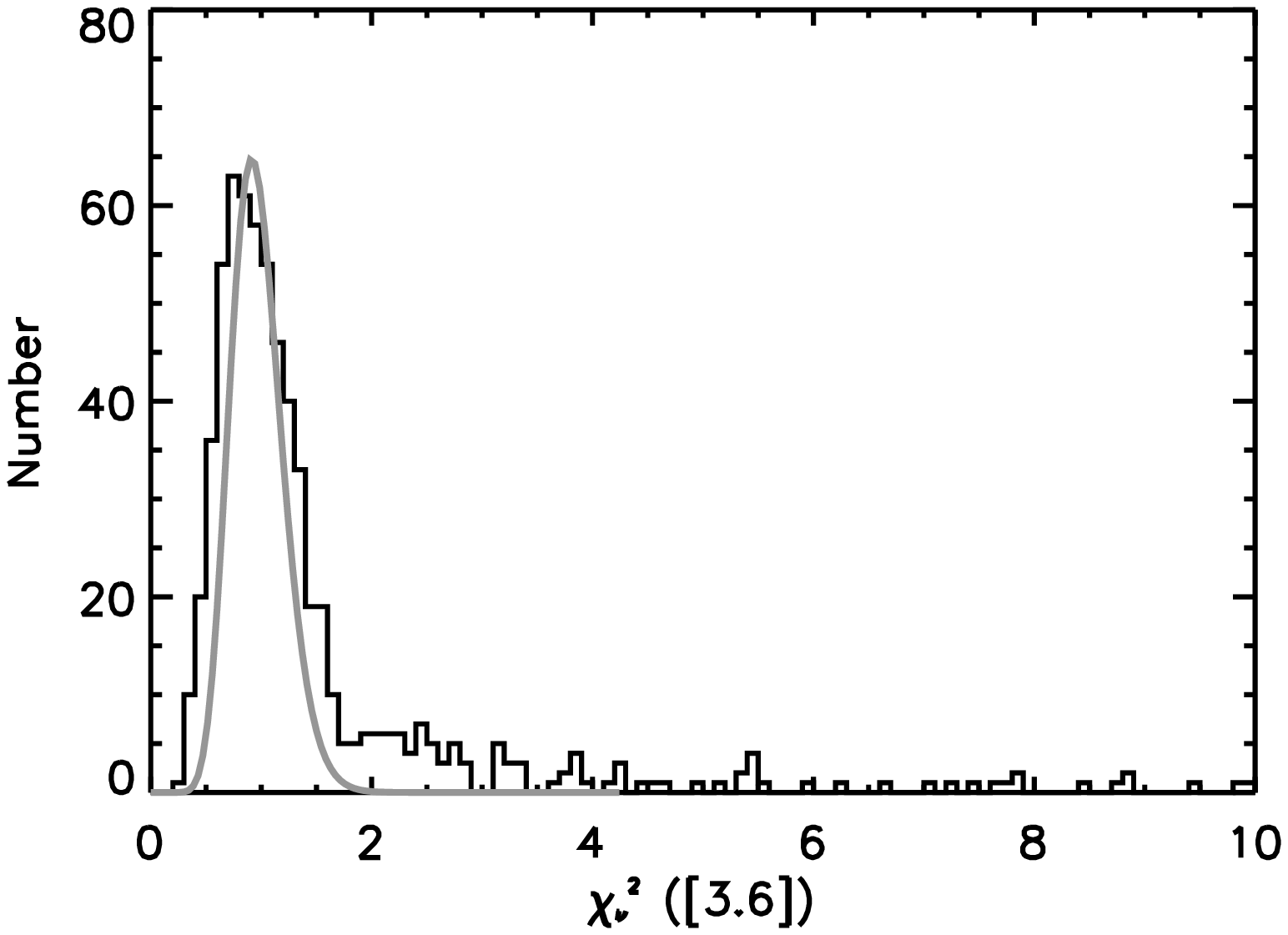}
\includegraphics[scale=.45]{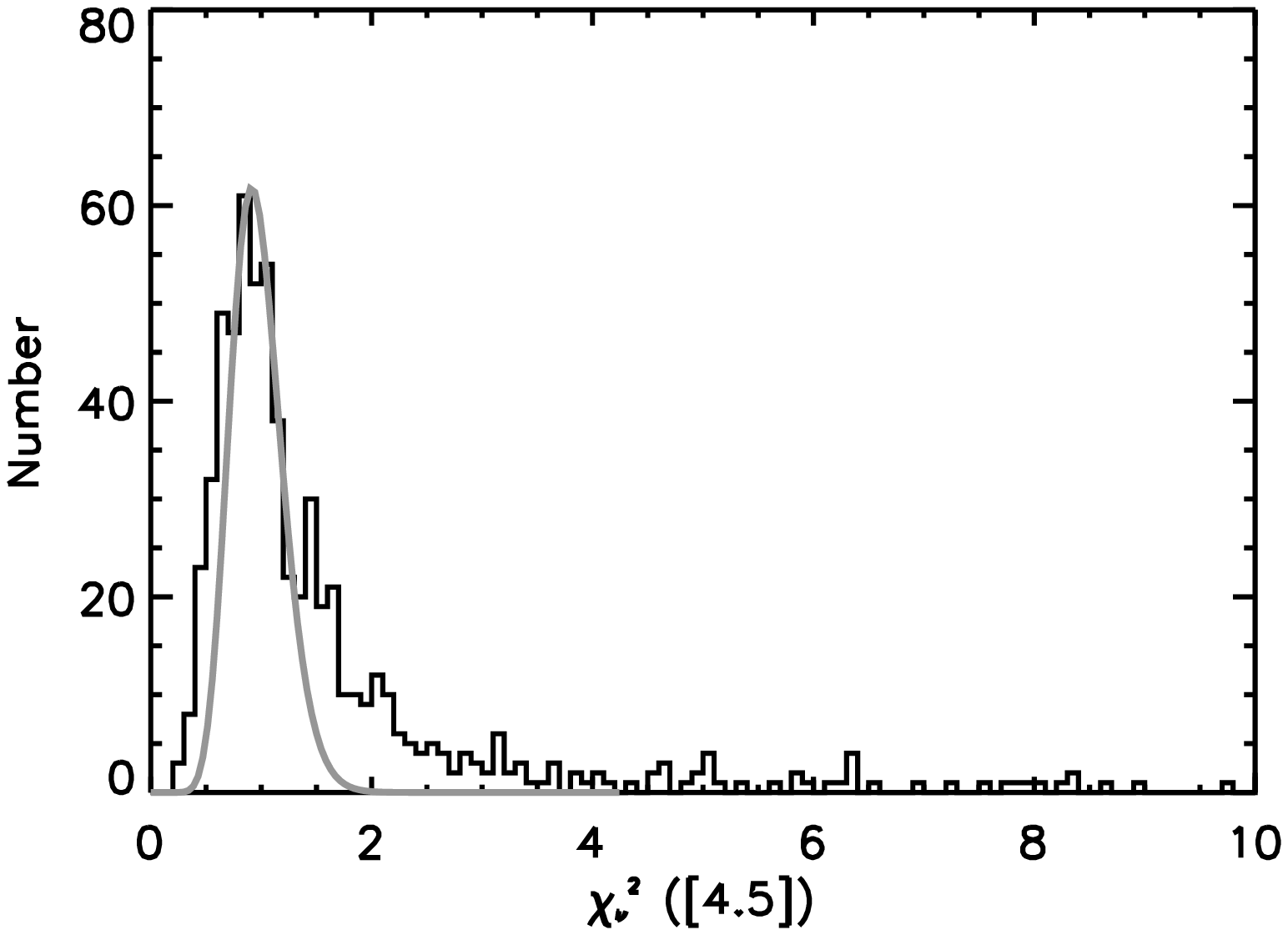}
\caption{Reduced $\chi^2$ distribution for [3.6] (left) and [4.5] (right). Solid grey line shows the expected distribution if there were no variable sources. The substantial population with $\chi^2_{\nu}>3$ are all variable in these bands. For clarity we have excluded the 40 and 29 points with $\chi^2_{\nu}$ ([3.6],[4.5] respectively) greater than 10. \label{chi}}
\end{figure*}

With an accurate estimate of the uncertainty in the data, we can determine if the spread in the flux of a particular star is larger than the uncertainties. This is quantified as the reduced chi-squared, $\chi^2_{\nu}$, which is defined in each band as:
\begin{equation}
\chi^2_{\nu}=\frac{1}{\nu}\Sigma^n_{i=1}\frac{(m_i-\bar{m})^2}{\sigma_{obs}^2}
\end{equation}
where $\nu=n-1$ is the number of degrees of freedom. Figure~\ref{chi} shows the distribution of $\chi^2_{\nu}$ in each band along with the expected chi-squared distribution given the number of degrees of freedom. The non-variable sources have $\chi_{\nu}^2$ centered at 1 with a spread that matches the expected distribution. The long tail toward high $\chi^2_{\nu}$ arises from the variable stars and any source with $\chi^2_{\nu}>3$ was taken to be variable. This boundary is confirmed by visual inspection of light curves that fit this criteria.

The use of the $\chi^2$ could be biased by a single data point in one band with anomalously high or low flux. In fact we find 13 cluster members (LRLL 39, 110, 154, 186, 286, 298, 329, 353, 478, 1477, 1707, 1719, 1923) for which the  $\chi_{\nu}^2$ changes by more than a factor of two in either [3.6] or [4.5] due to a discrepant data point resulting in a misclassification as a variable star. For these sources, as well as the rest of the cluster members, Table~\ref{tab_cluster} lists the membership information (RA, Dec, T$_{eff}$, L$_*$, etc.) while Tables~\ref{ch1_phot},\ref{ch2_phot} list the photometry. All show either one or two data points, in only one band, that are either much brighter or fainter than the rest of the light curve. Of these six are class III sources, three are evolved disks and four are class II sources (the definition of class III/evolved/class II is discussed below). The presence of this small number of dubious variables will not strongly affect our results, and we do not explicitly exclude them from our sample. 

\begin{figure}
\center
\includegraphics[scale=.5]{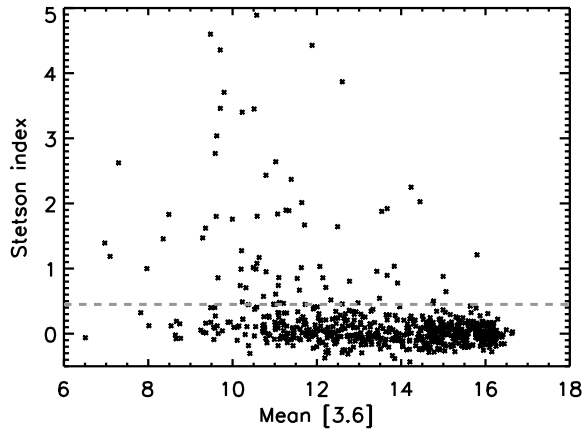}
\caption{Stetson index as a function of [3.6] for all sources detected in every epoch. The dashed line is our adopted boundary for splitting non-variable and variable sources. For clarity we have excluded the nine sources with a Stetson index greater than 5. \label{stetson}}
\end{figure}

Another statistical tool for selecting variables is the Stetson index \citep{ste96}, which takes advantage of both bands to look for correlated variability, and has been successfully used in the past to select variable young stellar objects \citep{car01}. The Stetson index is defined as:
\begin{equation}
S=\frac{\Sigma_{i=1}^n g_i sgn(P_i)\sqrt{|P_i|}}{\Sigma^n_{i=1}g_i}
\end{equation}
where $n$ is the number of epochs for which both [3.6] and [4.5] exist, $g_i$ is the weight given to each normalized residual (assumed to be 1 for our two band data), and $P_i=\delta_{1(i)}\delta_{2(i)}$ is the product of the normalized residuals. The normalized residual in each band is defined as:
\begin{equation}
\delta_i=\sqrt{\frac{n}{n-1}}\frac{m_i-\bar{m}}{\sigma_{obs}}
\end{equation}
Since many of the variable stars show correlated fluctuations in both bands (see discussion below) this index is particularly effective at picking out weak fluctuations that would not have been flagged based on the $\chi^2_{\nu}$. Figure~\ref{stetson} illustrates the distribution of Stetson indices, which shows a clear locus for the non-variable sources and a long tail from the variable stars. Stars with $S>0.45$ are taken to be variable. This boundary is defined based on visual inspection of light curves, and confirmed with the $\chi^2_{\nu}$ value, which independently indicates that many of these stars are variable. 

To search for periodic variables we detrend the light curve, followed by analysis with SigSpec \citep{ree07}. We first remove slow trends in the data by subtracting a second-order polynomial fit from the light curve. This is done to ensure that the mean of the light curve is constant over the visibility window, which is one of the required assumptions of many forms of time-series analysis \citep{bro02}. After subtracting the polynomial fit, we look for significant periods using the routine SigSpec \citep{ree07}. Unlike a Lomb-Scargle Periodogram, SigSpec takes into account both the amplitude and phase of the fourier transform of time-series data. It analyzes the significance of the amplitude and phase and derives a spectral significance, which corresponds to the probability that the signal is due to random noise. When searching for periods, we consider a significance threshold of 5 (a 1 in 10$^5$ chance that the signal is due to random noise) for a signal in one band or a threshold of 4 if the signal is present in both bands. Based on simulated light curves we find that we cannot accurately recover periods longer than 25 days (see discussion below) and we ignore any star that has a significant periodic signal longer than this limit. 

\subsection{Sensitivity}
\begin{figure*}
\center
\includegraphics[scale=.32]{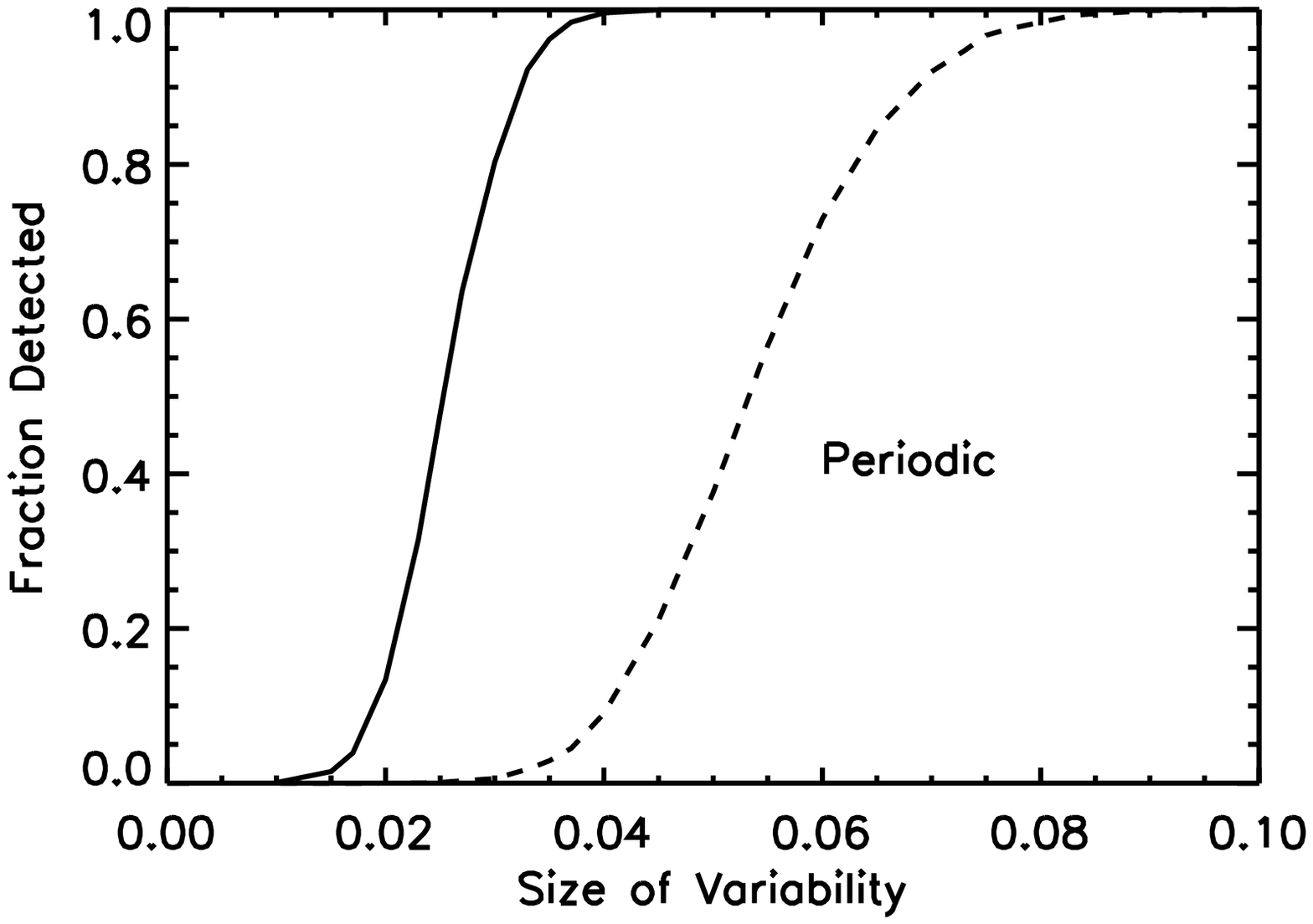}
\includegraphics[scale=.32]{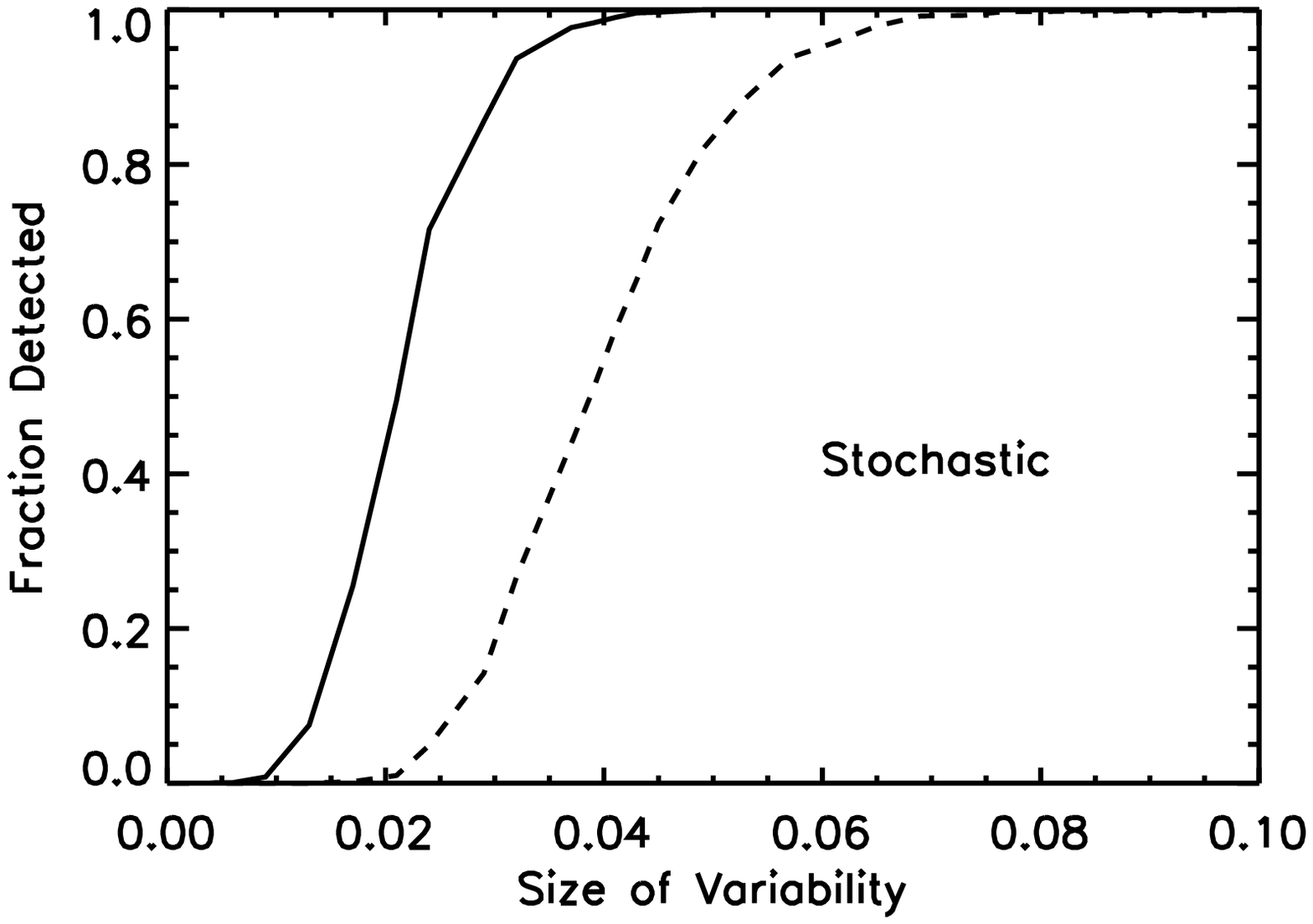}
\includegraphics[scale=.32]{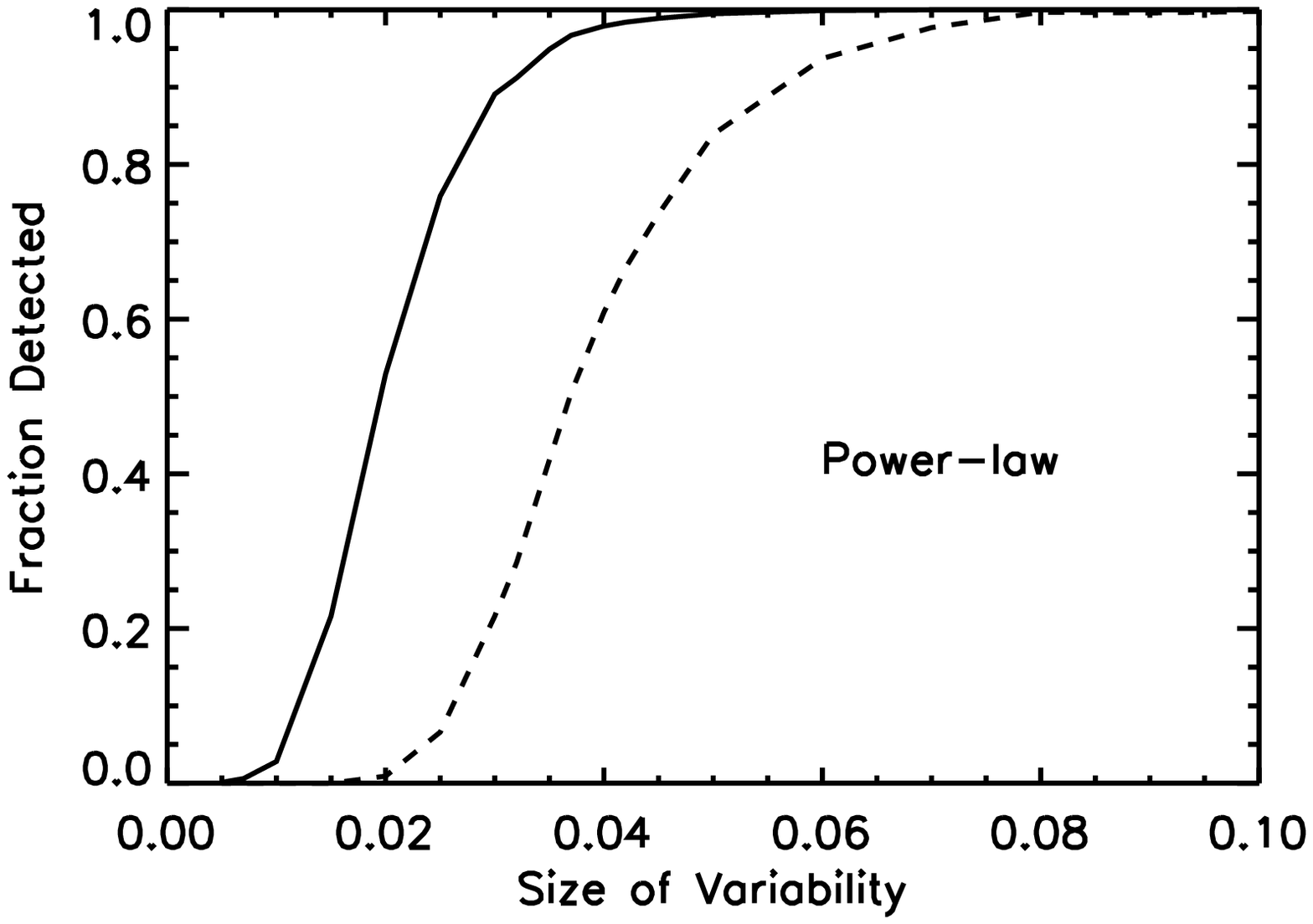}
\caption{Our sensitivity to variables as a function of the size of the variations. We simulate 10,000 13th magnitude stars with a period of 15 days (left panel), 5,000 stars with stochastic fluctuations with a timescale of 15 days (middle panel) and 5,000 stars with fluctuations with no characteristic timescale (right panel) with a given fluctuation size and measure the fraction of these simulated stars that are recovered as being variable using the same criteria as applied to the cluster sample. The solid line shows our sensitivity when both [3.6] and [4.5] data are included while the dashed line shows our sensitivity when only using the [3.6] data. Including the second band of data increases our sensitivity by about a factor of 2.\label{sens}}
\end{figure*}

\begin{figure}
\center
\includegraphics[scale=.45]{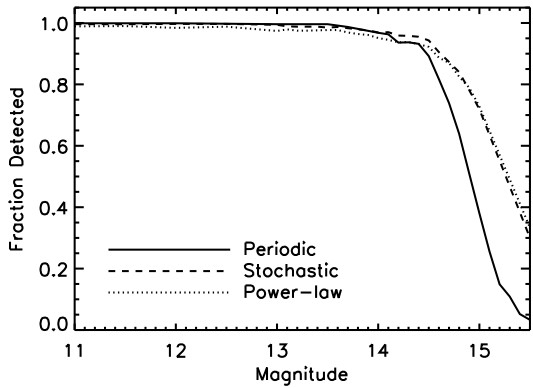}
\caption{Sensitivity as a function of magnitude. Simulated stars are either given periodic or stochastic fluctuations of 0.04 mag, and the fraction recovered is measured as a function of magnitude.Different lines refer to different models of the underlying light curve (see text for details). Our sensitivity severely drops off at about 14th mag. \label{mag_sens}}
\end{figure}

Given our wide range of fluxes, it is important to understand our sensitivity to fluctuations of different sizes as a function of magnitude. We can estimate our sensitivity by simulating light curves with a given fluctuation, sampling these light curves at our cadence, adding noise, and then calculating the $\chi_{\nu}^2$ and Stetson index to determine if they show variations. For the periodic stars we start with a sine curve with a period of 15 days with [3.6],[4.5]=13, fainter than the majority of cluster members, and the same fluctuation in each band. We create 10000 stars, add noise to each one, and measure how many are detected as periodic as a function of the size of the fluctuation. In figure~\ref{sens} we show that we can detect all of the stars with large fluctuations, but this sensitivity severely drops off at small fluctuations. Including data from both bands increases our sensitivity by a factor of 2, which makes up for the fewer sources that are detected in both bands. We estimate that we are able to detect $>$99\%\ of periodic variables with fluctuations larger than 0.035 mag. Beyond [3.6],[4.5]=14.5 our sensitivity severely decreases (Figure~\ref{mag_sens}). 

\begin{figure}
\center
\includegraphics[scale=.45]{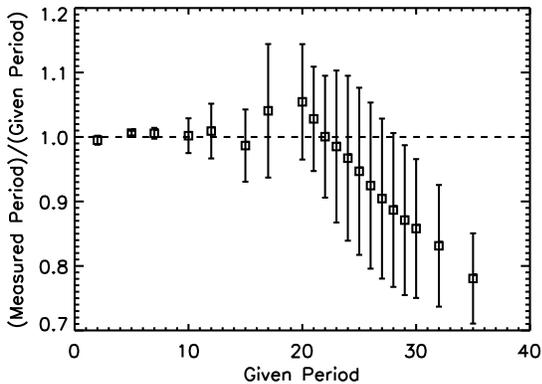}
\caption{Comparison of the period derived from the periodogram versus the given period for fake light curves. Error bars show the dispersion in the derived period. Above 25 days the measured period underestimates the actual period of the light curve and we do not mark stars as periodic if we derive a period longer than 25 days.\label{period_sens}}
\end{figure}

We are not uniformly sensitive to periodicity on all timescales because of our limited temporal coverage. Rapid fluctuations will not be detected as periodic because of our cadence and our finite observing window prevents us from deriving long periods. We estimate our sensitivity by calculating the period for our simulated stars and comparing this derived period to the model period. We find that the stars are not marked as periodic if their period is less than 2 days and we underestimate the period when it is longer than 25 days (Figure~\ref{period_sens}). For this reason we only mark a star as periodic if its period lies between 2 and 25 days.

\begin{figure*}
\center
\includegraphics[scale=.32]{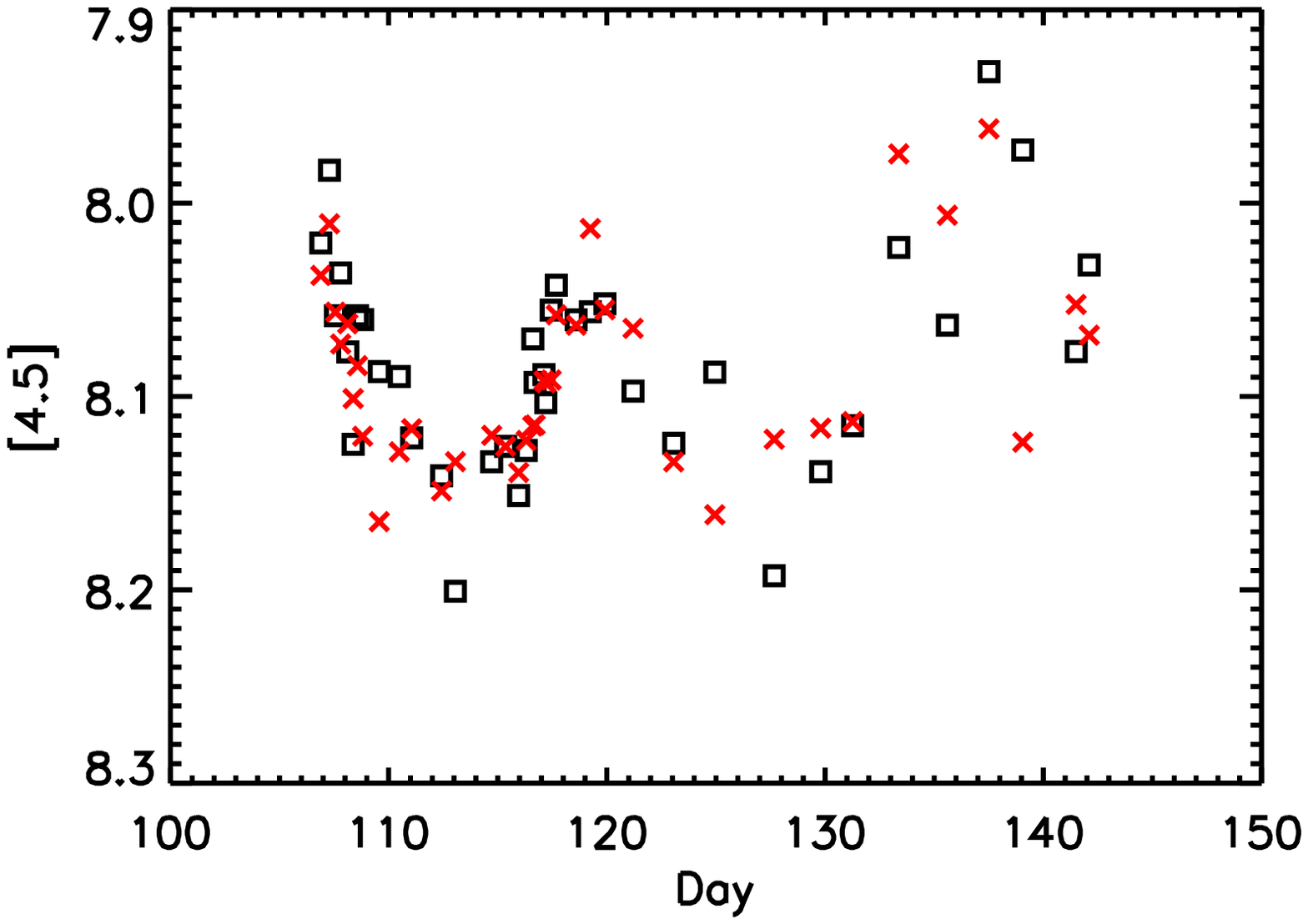}
\includegraphics[scale=.32]{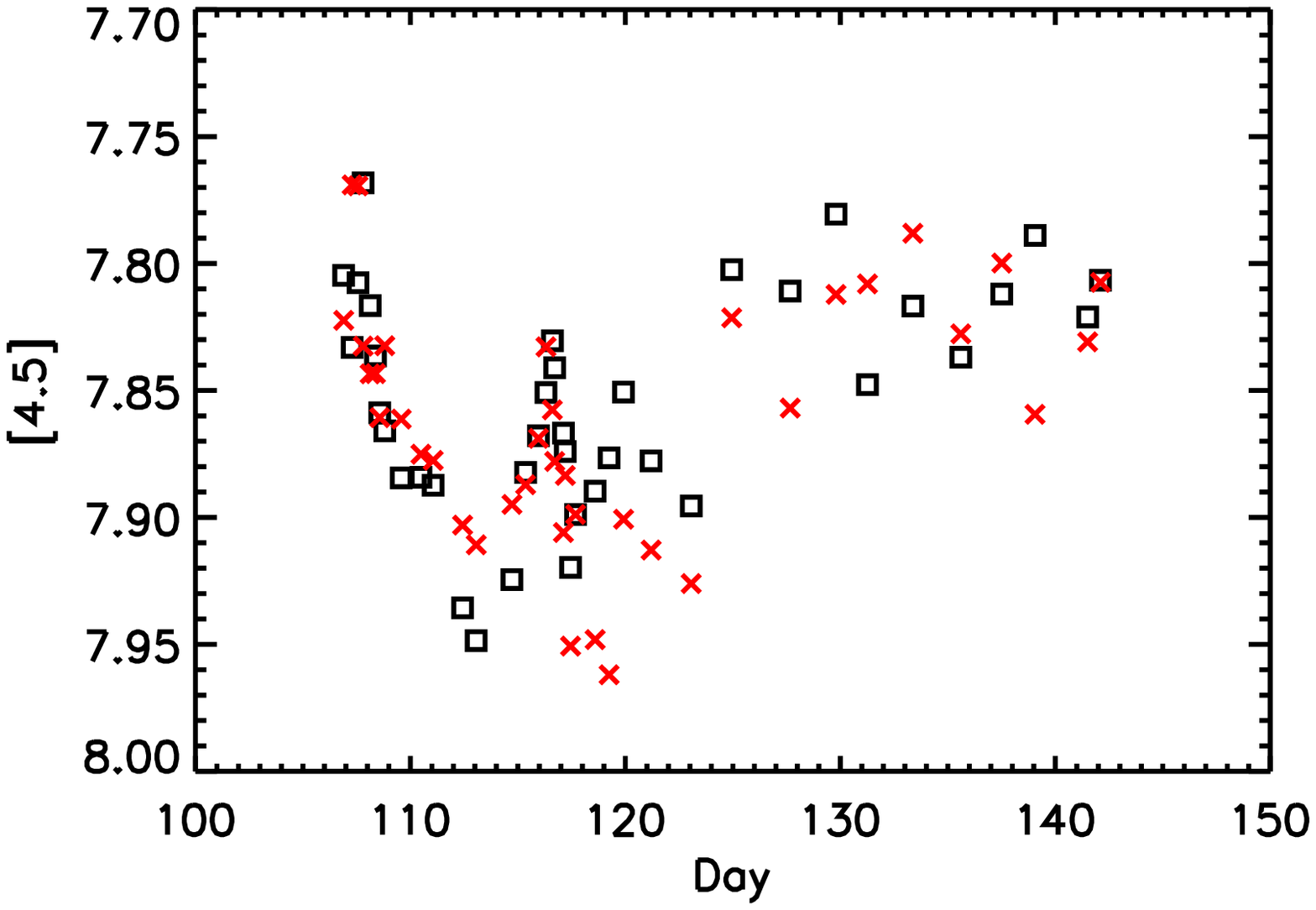}
\includegraphics[scale=.32]{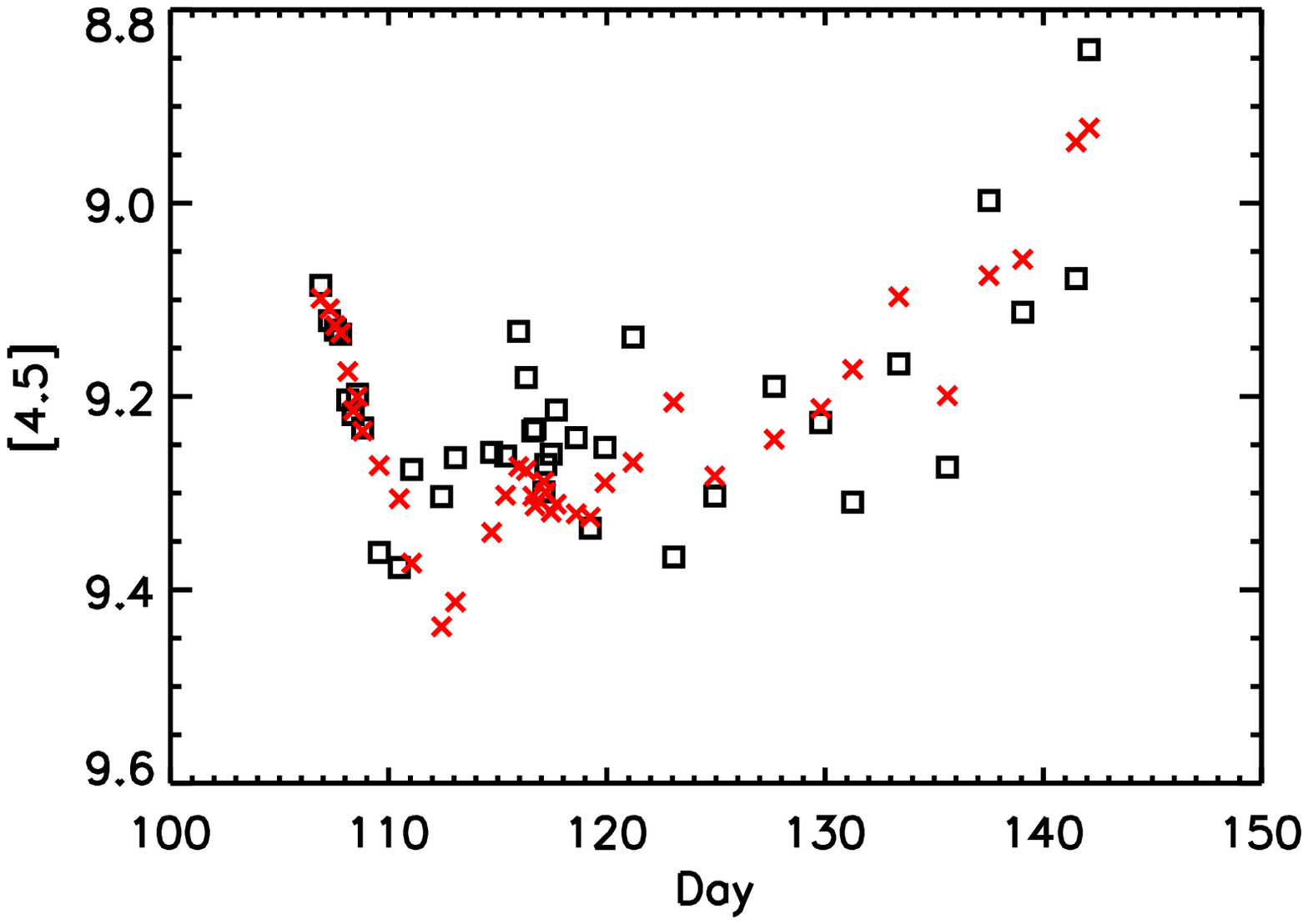}
\caption{[4.5] light curves of LRLL 15, 26 and 31 (red crosses, left to right) along with stochastic models (black squares) with variability timescale of 20 days. The stochastic model of \citet{kel09} is able to accurately reproduce the observed light curves, suggesting that it may be an accurate description of the underlying process that drives the variability, and can be useful for understanding the sensitivity of our survey \label{demonstrate_model}}
\end{figure*}

Estimating the sensitivity to aperiodic variations is more complicated. For periodic fluctuations we can use a sine curve as the basic model for the light curve, but no comparable model has been shown to model the light curves of aperiodic fluctuations accurately in young stellar objects. One potential model is the damped random walk model from \citet{kel09} that has been used to successfully model quasar variability. This continuous first order autoregressive process consists of a white noise perturbation that is exponentially damped at a characteristic timescale. This process is described by the following stochastic differential equation:
\begin{equation}
dX(t)=-\frac{1}{\tau}X(t)dt+\sigma\sqrt{dt}\epsilon(t)+bdt
\end{equation}
where $\tau$ is the relaxation time of the process $X(t)$ and $\epsilon(t)$ is a white noise process with zero mean and variance equal to 1. While this model has been connected to accretion disk physics, which may not be applicable here given the differences in accretion disks surrounding super-massive black holes and pre-main sequence stars, it can be used independent of a physical interpretation. We employ it simply as a statistical model and do not attempt to interpret the various terms as directly relating to disk physics.  In Figure~\ref{demonstrate_model} we show observed light curves for three cluster members along with simulated light curves using the model of \citet{kel09}. We are able to generate model light curves that accurately match the observations, allowing us to use this model to test our sensitivities. While this model has been updated to include fluctuations on a range of timescales \citep{kel11}, for simplicity we restrict ourselves to the single-timescale model. We construct light curves with $\tau=15$ days and then sample those light curves with our cadence. Figure~\ref{sens} shows our sensitivity to varying fluctuations for a star with [3.6],[4.5]=13 when using just the [3.6] data and when including both data sets. As with the periodic fluctuations we find that including both bands increases our sensitivity by a factor of 2. We estimate that we are able to detect $>$99\%\ of the variables with fluctuations larger than 0.04, similar to the periodic variables, although this conclusion is highly model dependent. Beyond [3.6],[4.5]=14.5 our sensitivity severely decreases (Figure~\ref{mag_sens}).

\begin{figure*}
\center
\includegraphics[scale=.4]{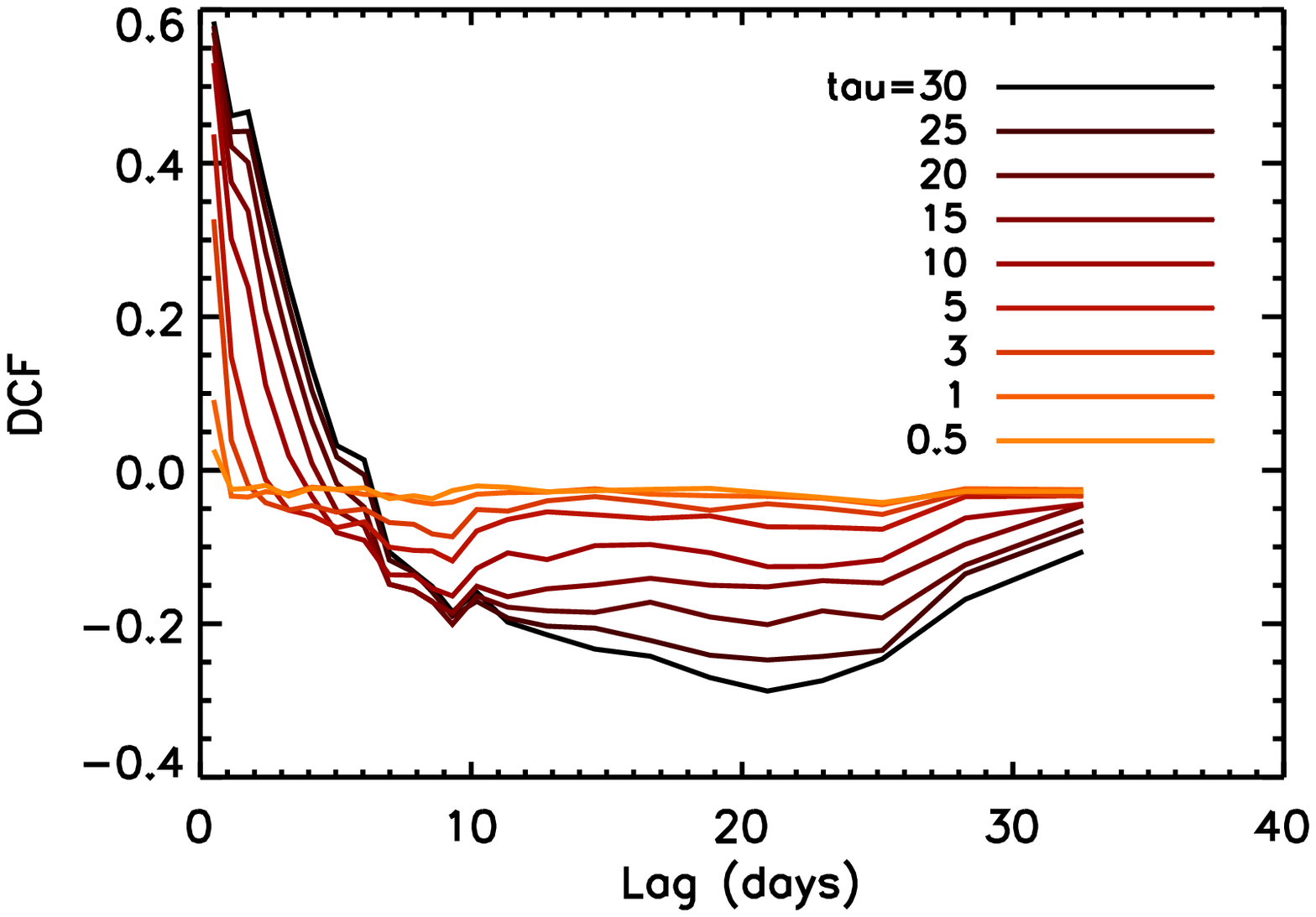}
\includegraphics[scale=.4]{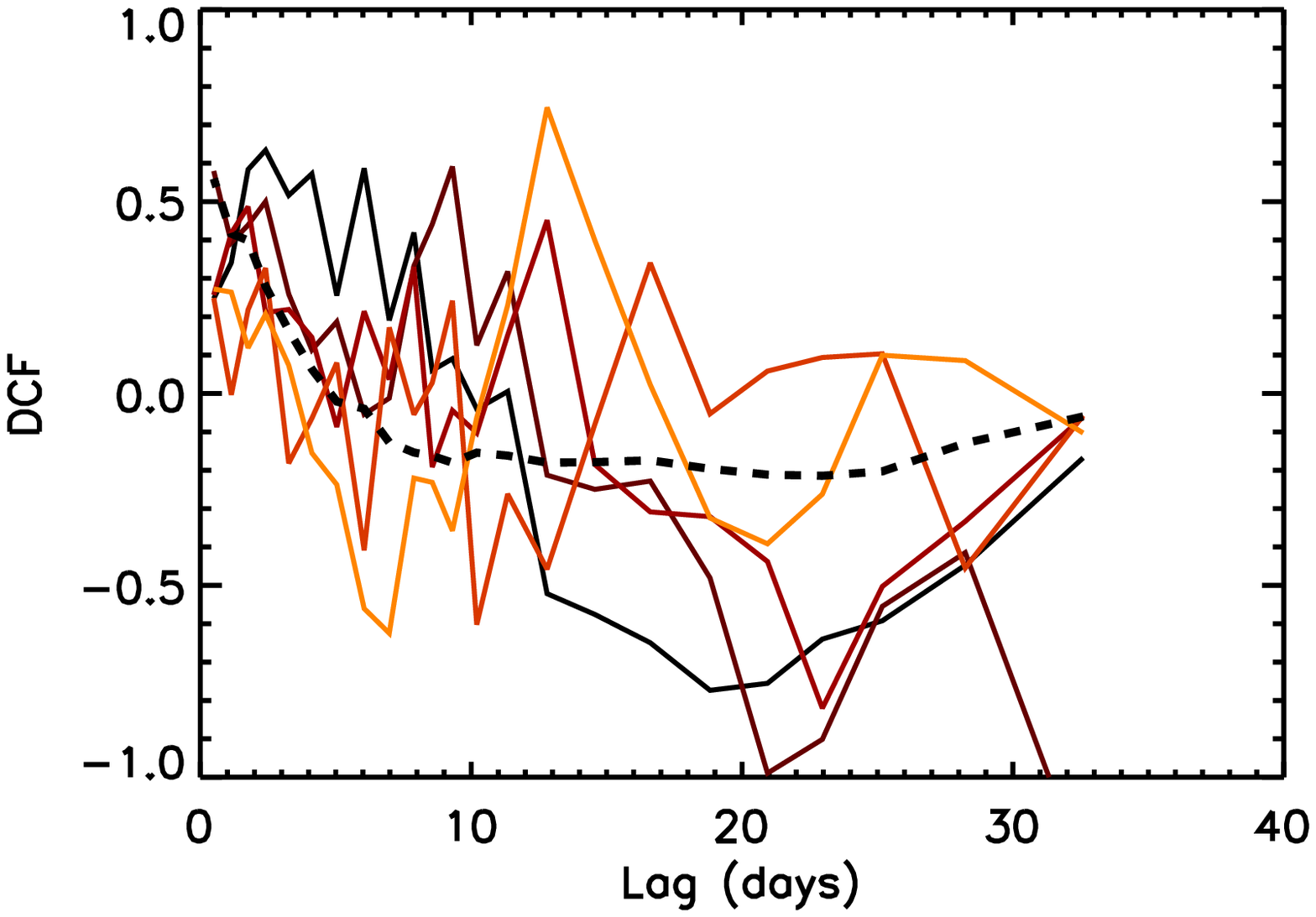}
\caption{On the left is the median discrete correlation function (DCF) for 5000 stars with aperiodic variability at different timescales. Light curves with timescales between 2 and 25 days have appreciable differences in the DCF. On the right we show the DCF for five light curves with a 20 day timescale (colored solid lines) along with the median DCF of 5000 stars with this timescale. The large variability in the shape of the DCF between different stars demonstrates the difficulty in deriving the variability timescale for a single star, despite the change in average shape of the DCF with timescale seen in the left-hand panel. \label{timescale_dcf}}
\end{figure*}

\begin{figure}
\center
\includegraphics[scale=.45]{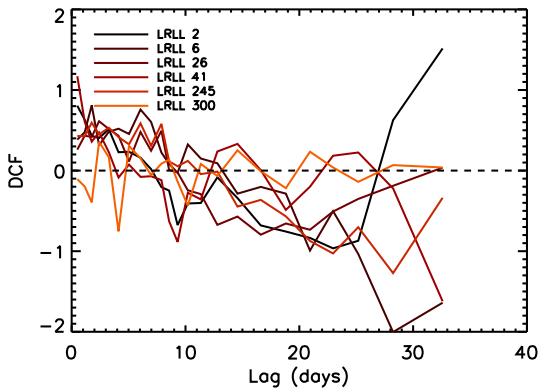}
\caption{Discrete correlation function for sample young stellar objects. While the shape of the DCF depends on the variability timescale, the limited sampling of our data makes the observed DCF too noisy to draw strong conclusions about the timescale. \label{dcf_examples}}
\end{figure}

This model, with its characteristic timescale, allows us to test our sensitivity to fluctuations on different timescales. Our cadence and total time coverage will strongly bias our sensitivity, and we can determine this sensitivity by 'observing' fake light curves at the same sampling rate as the actual observations.  To evaluate the fluctuations statistically as a function of timescale for the observed light curves, we choose to use the discrete correlation function, which is similar to the auto-correlation function but does not assume that the sampling is evenly spaced in time \citep{ede88}. To calculate the discrete correlation function we first calculate the unbinned discrete correlations, defined as:
\begin{equation}
UDCF_{ij}=\frac{(m_i-\bar{m})(m_j-\bar{m})}{(\sigma^2-e^2)}
\end{equation}
where $m_i,m_j$ are the magnitudes measured on epochs $i,j$, $\bar{m}$ is the average magnitude, $\sigma$ is the standard deviation of the magnitudes, and $e$ is the measurement error. By binning over M pairs we create the discrete correlation function, DCF($\tau$):
\begin{equation}
DCF(\tau)=\frac{1}{M}UDCF_{ij}
\end{equation}
We create simulated light curves as before, but instead of simply measuring the size of the fluctuations in these light curves we now calculate the DCF for each curve. To examine the general shape of the DCF for a particular timescale we average the DCF derived from each of our 5000 simulated light curves. We show the results for different timescales in figure~\ref{timescale_dcf}. The DCF for light curves with timescales below $\sim$2 days are completely flat. Above 25 days the shape of the DCF does not change, indicating that we are not able to constrain timescales longer than 25 days. Also in figure~\ref{timescale_dcf} we show the DCF for individual simulated light curves with a timescale of 20 days. The noisiness in each individual DCF indicates that we will be unable to use the observed DCF for the IC 348 cluster members to constrain the timescale accurately. In figure~\ref{dcf_examples} we show the DCF for a sub-sample of cluster members. These data also show the noisiness that prevents us from accurately deriving a timescale for the variability. In general the DCF are not completely flat, indicating that the light curves vary on timescales longer than 2 days.  Increasing the number of epochs, as well as the time baseline, would improve our ability to constrain the timescale of aperiodic fluctuations.

We can begin to probe the fluctuations on longer timescales by comparing our fall 2009 warm-mission [3.6] and [4.5] photometry with our March 2009 IRAC cold-mission data (Muzerolle et al. in prep)  as well as the GTO \citep[][PID 6]{lad06} and C2D \citep[][PID 178]{jor06} data taken in 2004. We find that 96\%(92\%) of the cluster members have a mean [3.6] ([4.5]) magnitude from March 2009 between the min and max magnitudes of the fall 2009 data. Only eight stars (LRLL 31, 35, 99B, 245, 382, 1683, 9099), 3\%\ of the cluster,  have both March 2009 [3.6] and [4.5] beyond the min and max of what is seen in our fall 2009 data. Roughly 6\%\ of the cluster members have both [3.6] and [4.5] in the GTO or C2D data outside of the range seen in our fall 2009 data. This suggests that large year-long fluctuations are uncommon among the cluster members, although more detailed analysis is needed.

Using the damped random walk model to estimate our sensitivity to fluctuations of different timescales and sizes is effective, but gives us a result that is highly model dependent. We can instead simulate the stochastic variables using a different model that has no characteristic time scale. The stochastic model described above has a power-spectrum density (PSD) that is a power law over many frequencies, but has a break at a characteristic frequency \citep{kel09}. We can also model the PSD with a power law ($\omega^{-2}$) with no break \citep{tim95}. This model also produces light curves that reasonably match our data and can be used to estimate the sensitivity. Light curve simulations are performed as before and the results are shown in Figure~\ref{sens}. We estimate that we are able to detect $>$99\%\ of the variables with fluctuations larger than 0.045, similar to the previous model. Our sensitivity also drops off beyond [3.6],[4.5]$=$14.5, as was seen with the stochastic light curve model.

Overall, we find that we are sensitive to almost all fluctuations larger than 0.04 mag, with decreasing sensitivity beyond [3.6],[4.5]=14. We can detect periodic fluctuations if the period is between 2 and 25 days, with diminished sensitivity to periods below 2 days and a systematic underestimation of the period when it is above 25 days. We have very little sensitivity to the exact timescale of aperiodic fluctuations, although there appear to be few hourly fluctuations or large year to year fluctuations.

\section{Cluster Variability}
\begin{figure*}
\center
\includegraphics[scale=.4]{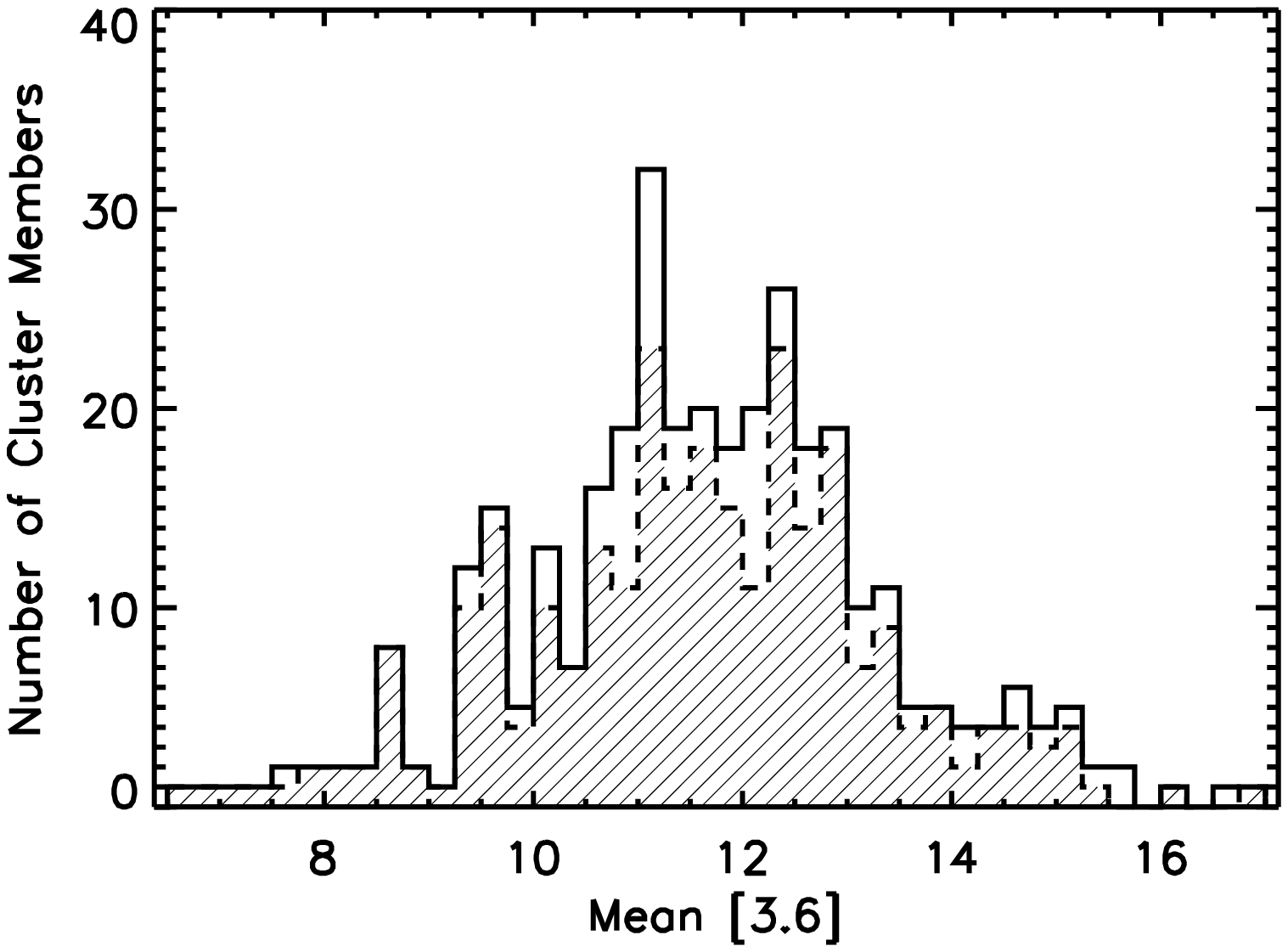}
\includegraphics[scale=.4]{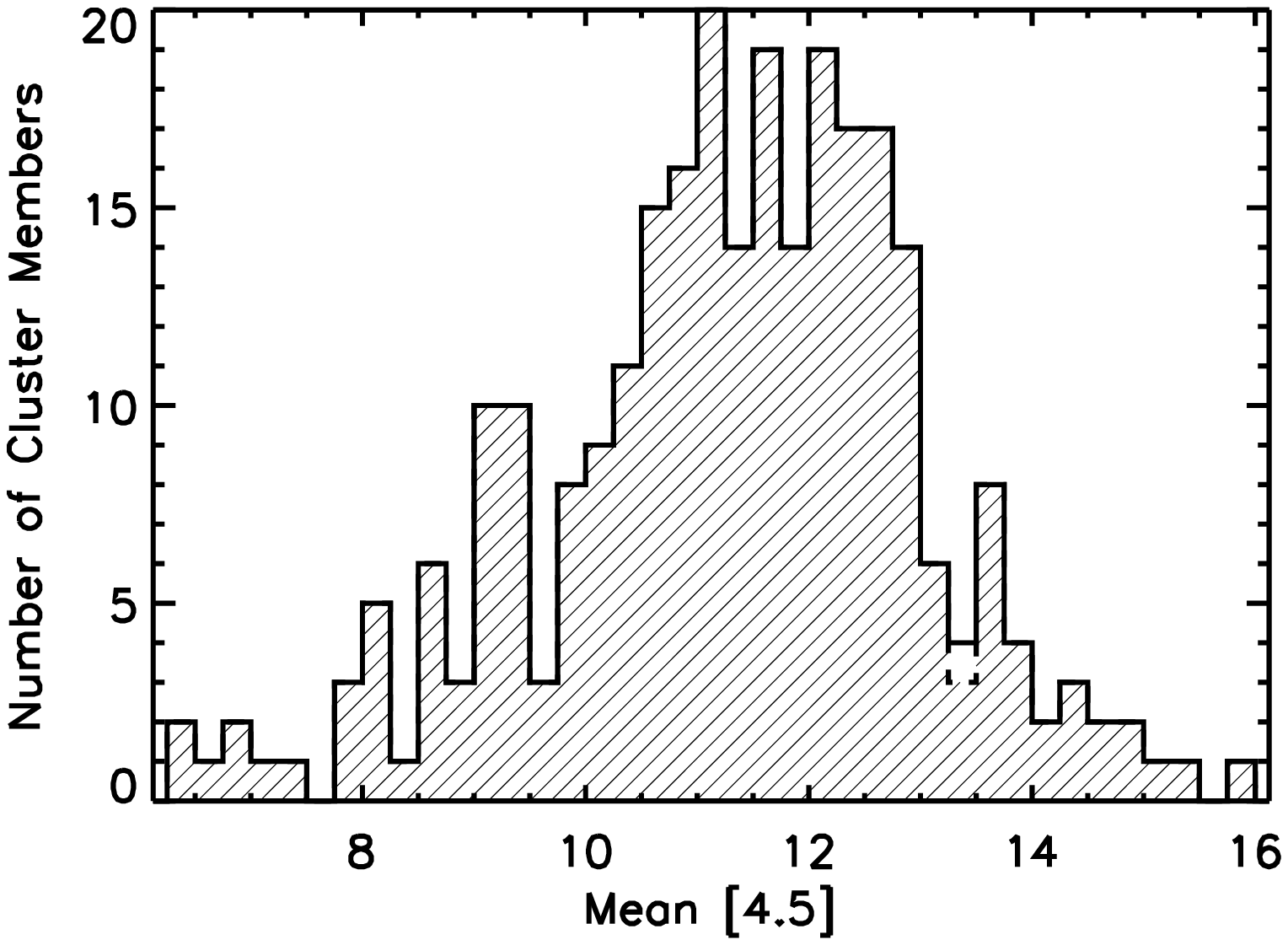}
\caption{Distribution of the [3.6] and [4.5] magnitudes for the cluster members detected on more than one epoch. The solid, unfilled histogram shows the stars detected in at least that band, while the dashed, filled histogram shows the stars detected in both bands. There are a substantial number of cluster members detected at only [3.6], but only one star was detected at [4.5] and not [3.6]. \label{mag_dist}}
\end{figure*}

With an understanding of the sensitivities and limitations of our data, we examine the IC 348 cluster \citep{luh03}. The cluster membership listed has been compiled from \citet{luh03,lad06,mue07,muz06,luh05} with the membership information (LRLL \#, RA, Dec, T$_{eff}$, etc.) listed in Table~\ref{tab_cluster} and the [3.6] and [4.5] photometry listed in Tables~\ref{ch1_phot} and ~\ref{ch2_phot} respectively. We detect 338 of 377 cluster members at [3.6] and 269 in both bands. The distribution of fluxes for the detected cluster members is shown in Figure~\ref{mag_dist}. The majority of the cluster members that are not detected at [3.6] are either outside the field of view (24 sources), are in a very early evolutionary state or have very low intrinsic luminosity that make them faint at these wavelengths (6 members), are close to another star, which confuses their photometry (3 members) or have very larger photometric uncertainties ($>0.1$mag, 2 sources). Due to the placement of the field, and the incomplete overlap between the [3.6] and [4.5] fields of view, there are a significant number of cluster members detected at [3.6] but not [4.5]. Only one source, located at the very edge of the field of view, is detected at [4.5], but not [3.6], most likely due to imperfect overlap between the two fields of view.  Most of the cluster members are brighter than 14th mag in both bands, which is roughly the magnitude at which the uncertainties start to increase dramatically. Of the cluster members detected at [3.6], 84 are variable and 254 are not variable, based only on the [3.6] data. For the cluster members detected in both bands, 125 are variable and 144 are not variable. In Table~\ref{tab_cluster} we mark the cluster members that we consider variable and non-variable, as well as whether or not this conclusion was derived from one band versus using both bands. The large difference in variability fraction reflects the large increase in sensitivity when both bands are examined together. For the rest of our analysis we consider only the 269 stars that have been detected in both bands because the increase in sensitivity makes up for the decrease in the sample size. When discussing variability trends, we will often distinguish between periodic variables and irregular variables. Keep in mind that when we say 'irregular variables' we really mean to say 'stars that do not have fluctuations larger than 0.035 mag on a period between 2 and 25 days.' Some of the 'irregular variables' may in fact be periodic, but at a level or on a timescale that we are not sensitive to. 

\subsection{What influences infrared variability?}

One distinct advantage of studying the IC 348 cluster is the wealth of ancillary data that is available in the literature. Its compactness and proximity make it an ideal target for optical spectroscopy \citep{luh03}, infrared photometry \citep{lad06,mue07}, X-ray imaging \citep{pre02,pre04} and optical photometric monitoring \citep{coh04,cie06,nor06}. These observations have led to a well-defined membership sample that is complete down to brown dwarf masses, along with measures of the H$\alpha$ emission line strength, infrared excess and X-ray luminosity for hundreds of members. Determining which of these system parameters influence the presence and strength of variability can constrain the physical model of the variability.

The parameters we examine, listed in Table~\ref{tab_cluster}, are a combination of values from the literature and those derived from our data. Stellar parameters, such as effective temperature (T$_{eff}$), stellar luminosity (L$_*$), H$\alpha$ equivalent width (EW), extinction (A$_V$), and X-ray luminosity (L$_X$) are all taken from the literature \citep{luh03,pre02}. To characterize the shape of the excess we use the slope, $\alpha$, of the infrared spectral energy distribution (SED), defined as $\lambda F_{\lambda}\propto \lambda^{\alpha}$ where $\alpha$ is measured over the 3-8\micron\ IRAC bands. The IRAC photometry was measured from a stack of the five consecutive days of monitoring taken in March 2009 (Muzerolle et al. in prep). We favor the added sensitivity from this deep stack in exchange for the possibility that the shape of the SED changes over time, which is occasionally but not frequently seen (Muzerolle et al. in prep). Infrared fluxes have been dereddened according to the extinction law of \citet{fla07} for those stars with measured extinction. The value of $\alpha_{IRAC}$ ranges from $\alpha=-3$ for purely photospheric emission (Class III objects), $-2.56<\alpha<-1.8$ for evolved disks, and $\alpha>-1.8$ for full disks \citep{lad06}.  Full disks can be further divided into class II sources ($-1.8<\alpha<0$) and class I sources ($\alpha>0$). Unless otherwise stated the variable $\alpha_{IRAC}$ refers to the slope of the dereddened, rather than observed, SED.

We can also characterize the excess based on the difference between the observed and the photospheric flux at 3.6\micron. A photosphere template is normalized to the dereddened J band flux and subtracted from the dereddened [3.6] flux (F(3.6)$_{obs}$/F(J)$_{obs}$-F(3.6)$_{phot}$/F(J)$_{phot}$) to produce a stellar flux normalized 3.6\micron\ excess. For the photosphere template we use a median of the flux from stars within IC 348 with no disk ($\alpha_{IRAC}<-2.56$) and an effective temperature within 5\%\ of the target star. We prefer using the cluster members to define the photosphere because the \citet{ken95} observed colors, as well as Kurucz model atmospheres, tend to overestimate the photospheric flux beyond K-band and thus underestimate the 3.6\micron\ excess. For the hottest stars, where there are few stars with similar effective temperatures, we use the \citet{ken95} observed colors. Both $\alpha_{IRAC}$ and the [3.6] excess trace the strength of the infrared emission, but do so in slightly different ways. The [3.6] excess, despite the larger uncertainties ($\sim0.01$), more directly traces the same part of the disk as the [3.6] and [4.5] photometry. A system with a large excess at 8$\micron$, but no excess at 3.6$\micron$ would have a large $\alpha_{IRAC}$ but the lack of dust capable of emitting in the [3.6] band may lead to less variability. 

The last parameter that we add is the position of the young stellar object within the cluster. IC 348 is composed of an exposed cluster core in the north-east and a highly obscured region in the south-west where most of the younger class I sources, along with all of the molecular outflows and sub-mm cores, reside \citep{her08}. The presence of these less-evolved sources indicates that the south-western ridge is a region of very recent star-formation. We characterize the position as the distance from HH 211, a prominent Herbig-Haro object in this region. 

\subsubsection{Irregular Variables}
During most of our analysis we only consider the irregular variables. The periodic variables represent a small subsample of the entire survey (121 irregular variables versus only 13 periodic variables, Table~\ref{tab_cluster}), and possibly have a different physical origin to their variability, which may bias our results. We include them in the figures, but consider the plausible causes of their variability separately in section 4.1.3

We start by considering the shape of the dereddened infrared SED, $\alpha_{IRAC}$ and how it correlates with the fraction of variables (f$_{var}$) and the size of the fluctuations in variable stars for the 239 cluster members with measurable variability and $\alpha_{IRAC}$.  We use linear regression to determine if there is a statistically significant correlation between $\alpha_{IRAC}$ and infrared variability. If the slope of the linear fit is significantly different from zero, then the two variables are correlated. This does require that we bin the stars in order to calculate f$_{var}$ and convert f$_{var}$ into a logit (=ln(f$_{var}$/(1-f$_{var}$))) in order to make it amenable to linear regression \citep{hos00}. The 239 stars with greater than two epochs of both [3.6] and [4.5] photometry, needed for measuring variability, and measurable infrared SED slopes, are split into four bins, each with an equal number of sources. The parameter f$_{var}$ is defined as the fraction of stars that exceed our variability criteria ($\chi^2>3$ in either [3.6] or [4.5] or S$>$0.45) and the size of the bin in parameter space is varied in order to ensure that each bin has an equal number of cluster members. To quantify the size of the variability we sum the $\chi^2_{\nu}$ in each band, which allows us to utilize information in both infrared bands. Again we perform linear regression to evaluate the strength of the trend statistically. 

\begin{figure*}
\center
\includegraphics[scale=.32]{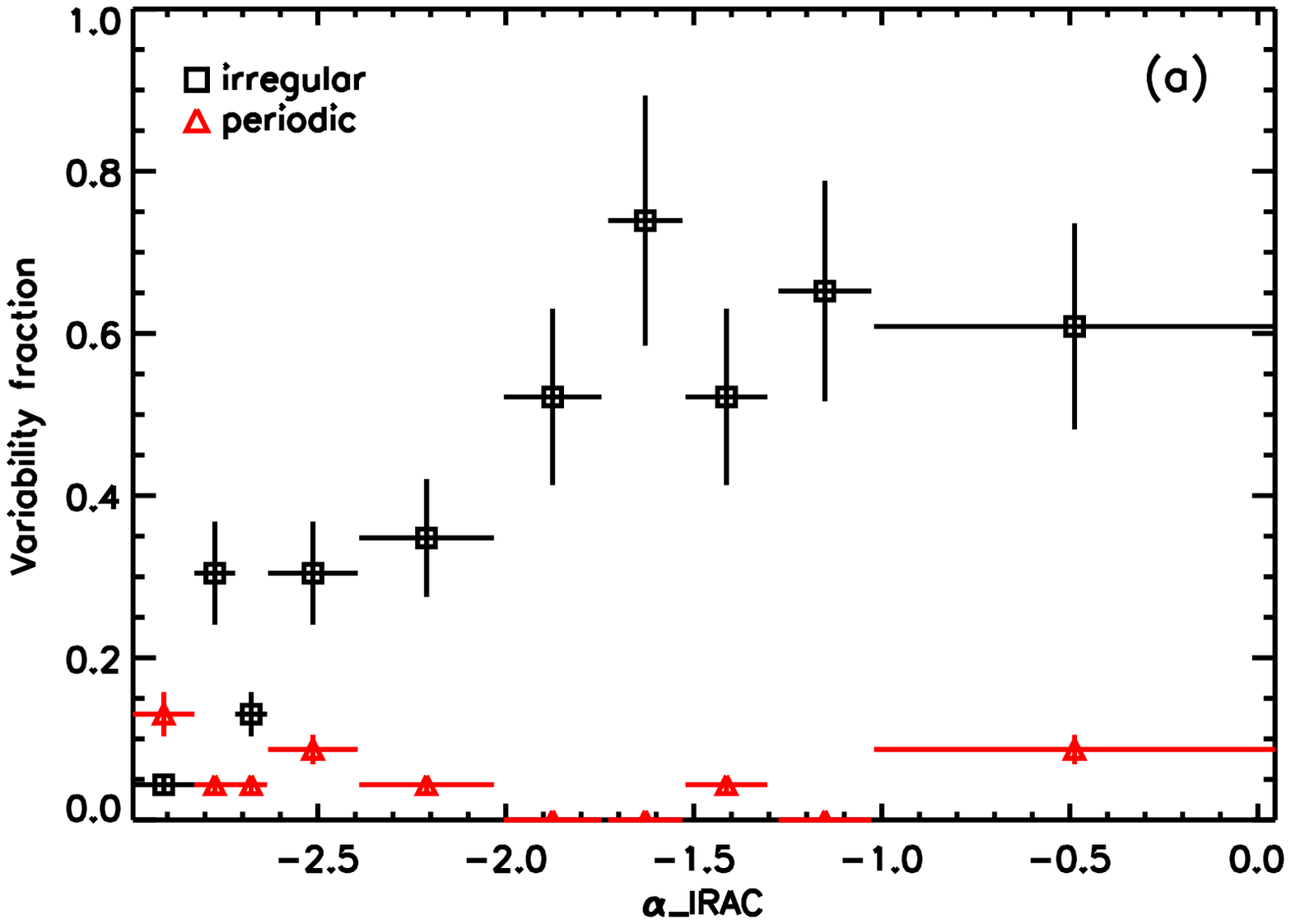}
\includegraphics[scale=.32]{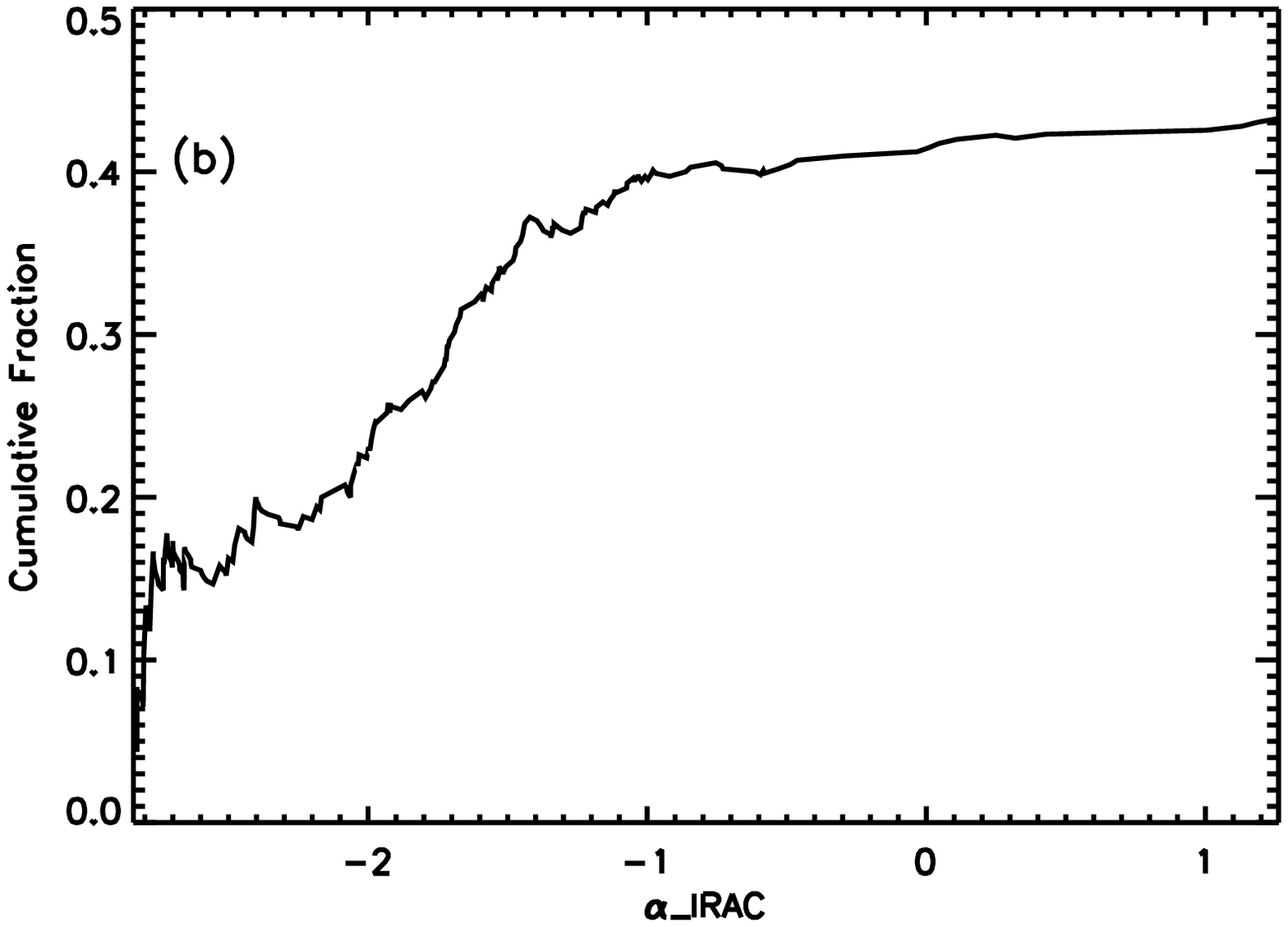}
\includegraphics[scale=.32]{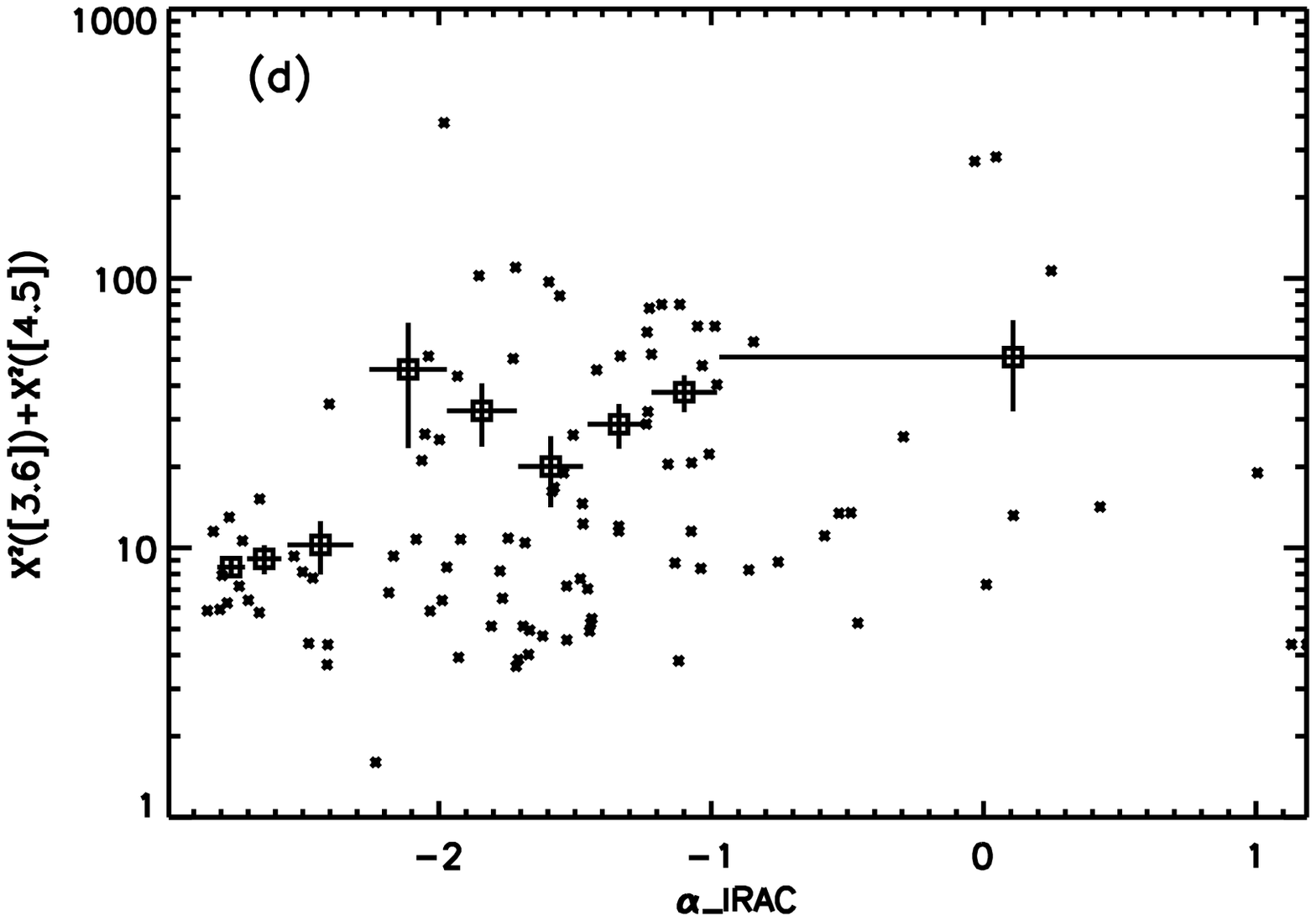}
\includegraphics[scale=.32]{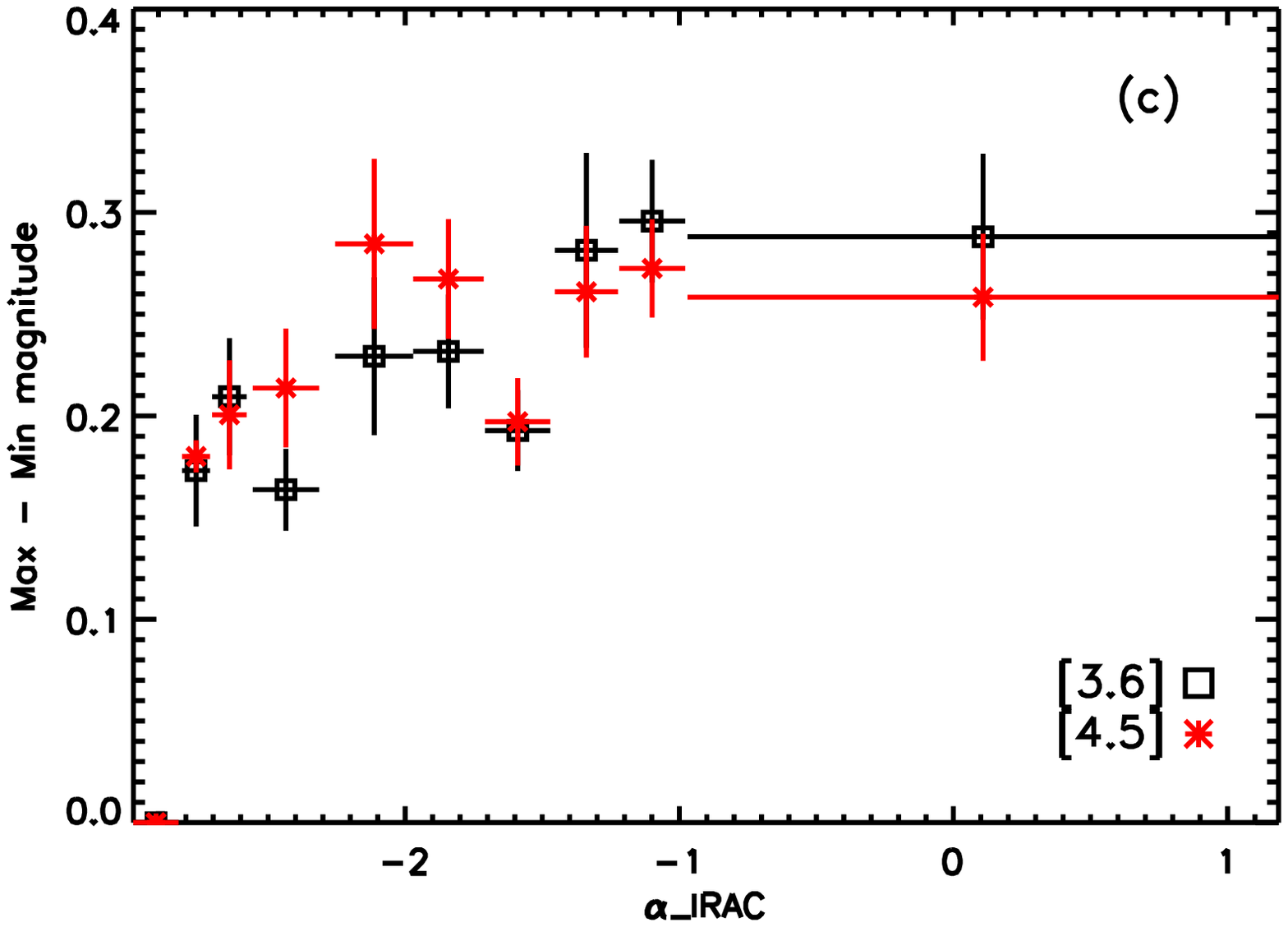}
\caption{Variability as a function of the shape of the infrared SED (239 members). (a) Fraction of cluster members that have irregular fluctuations (black squares) or periodic variability (red triangles). Each bin contains an equal number of cluster members and the horizontal error bars represent the size of the bin while the vertical error bars represent the poisson noise on f$_{var}$. (b) Cumulative fraction of irregular variables (c) Size of the fluctuations, measured using the sum of the $\chi^2_{\nu}$ in each band, for the irregular variables (black crosses, 120 members). Squares show the mean fluctuation in bins with equal numbers of cluster members, but not necessarily an equal number of variable stars. (d) Mean of the peak to peak fluctuations in magnitude of the irregular variables at [3.6] (black squares) and [4.5] (red stars). Horizontal error bars designate the bin size and vertical error bars are the error on the mean. The bin with the smallest $\alpha_{IRAC}$ does not have enough variable stars to define a mean, and is assigned a value of zero. \label{var_airac}}
\end{figure*}

We find a strong correlation between f$_{var}$ and $\alpha_{IRAC}$ (Figure~\ref{var_airac}a) with the variable fraction increasing from 20\%\ for those stars without disks to 60\%\ for stars with full disks/envelopes. We also show the cumulative fraction, although it isn't used to look for trends statistically, to visualize the correlation without the need for binning (Fig~\ref{var_airac}b). We also find a significant trend between the size of the variability and the SED slope (Figure~\ref{var_airac}c). The $\chi^2_{\nu}$, while useful statistically, is hard to interpret in terms of actual changes in the [3.6] and [4.5] magnitude. To more intuitively understand the size of the fluctuations in Figure~\ref{var_airac}d we calculate the max-min magnitude for each irregular variable and show the median of these values in each bin. These bins are the same as used when defining the variable fraction, with an equal number of cluster members, but not an equal number of irregular variables, in each bin. The typical size of the fluctuations increases from $\sim$0.15 mag to 0.3 mag from diskless to class I sources. 

This trend is consistent with previous studies of infrared variability \citep{mor09,mor11}. Splitting the cluster members into infrared classes using the dereddened 3-8\micron\ SED slope, we find that 79\%\ of class Is ($\alpha_{IRAC}>-0.5$), 61\%\ of class IIs ($-1.8<\alpha_{IRAC}<-0.5$), 44\%\ of evolved disks ($-2.56<\alpha_{IRAC}<-1.8$) and 20\%\ of class IIIs ($\alpha_{IRAC}<-2.56$) are variable (24,101,64,75 total stars in each group). The YSOVAR survey of Orion found variable fractions of 84\%, 70\% and 45\% for class I, II and III respectively \citep{mor11}. These variable fractions show the same trend with evolutionary stage as in IC 348, although the fractions are higher in each SED class. Orion is younger than IC 348, and it is possible that as stars evolve they become less variable, a possibility discussed in more detail below. In IC1396A the variable fractions for class I and class II sources are 75\% and 58\% respectively, similar to our results \citep{mor09}. 

\begin{figure*}
\center
\includegraphics[scale=.32]{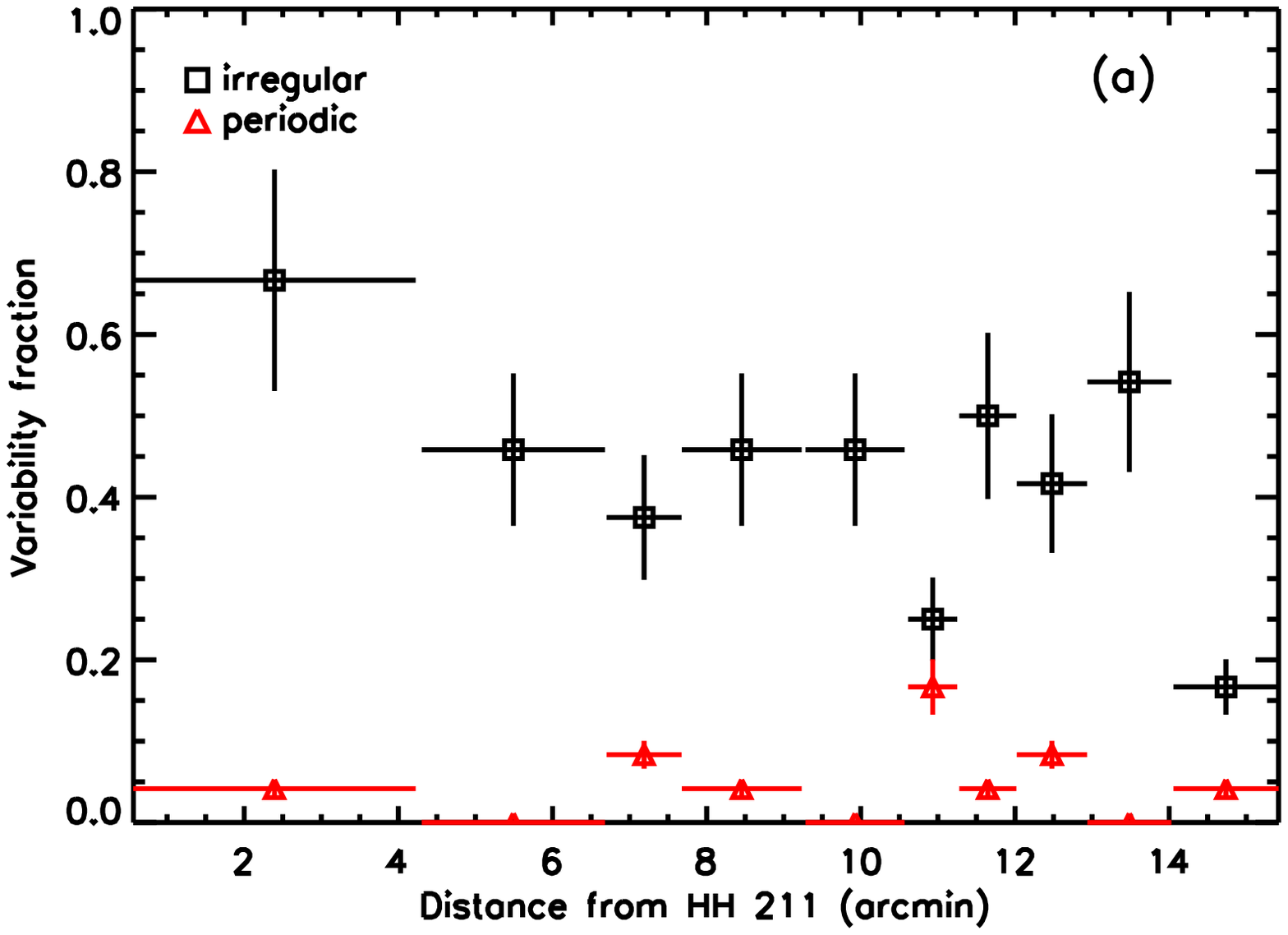}
\includegraphics[scale=.32]{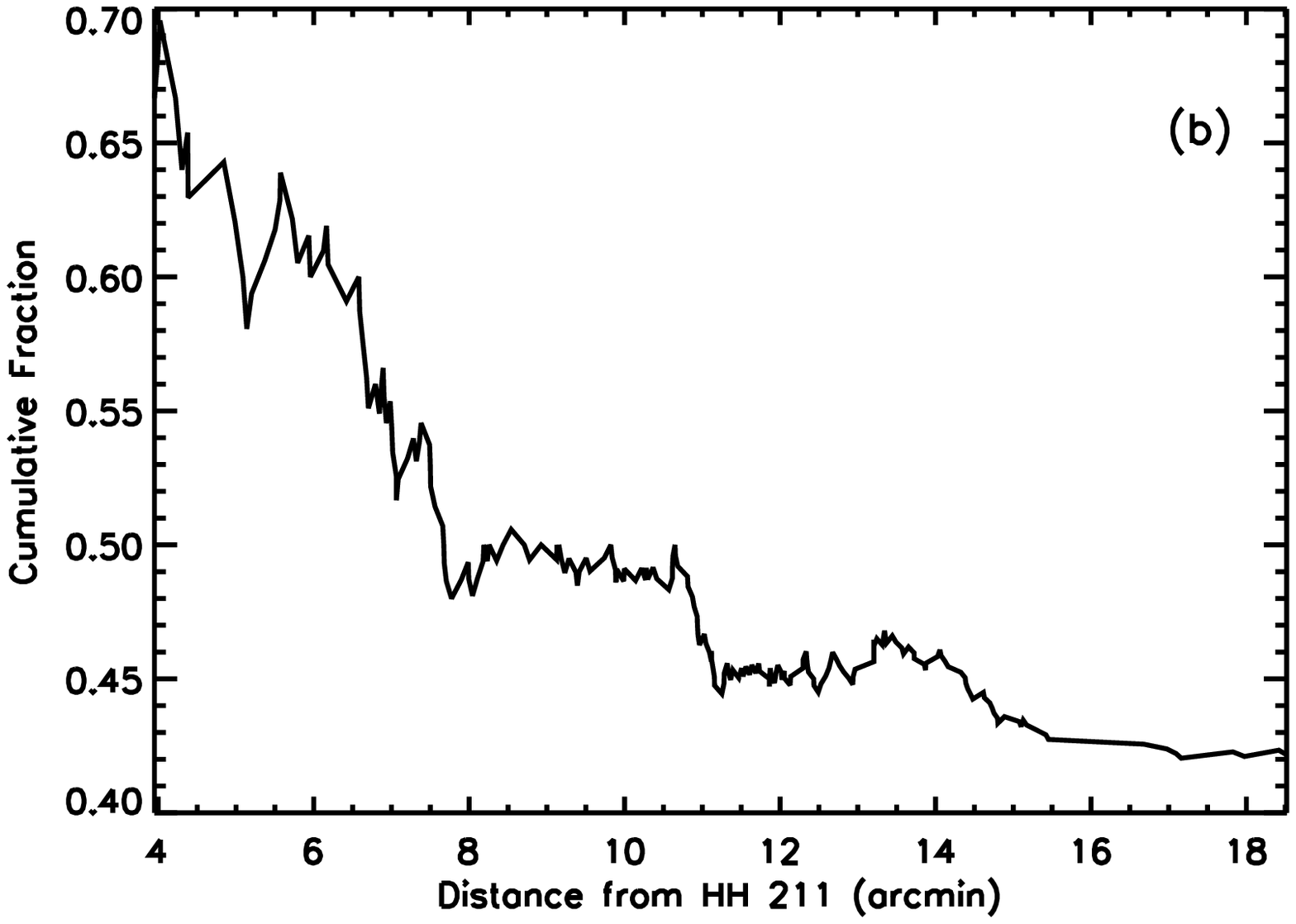}
\includegraphics[scale=.32]{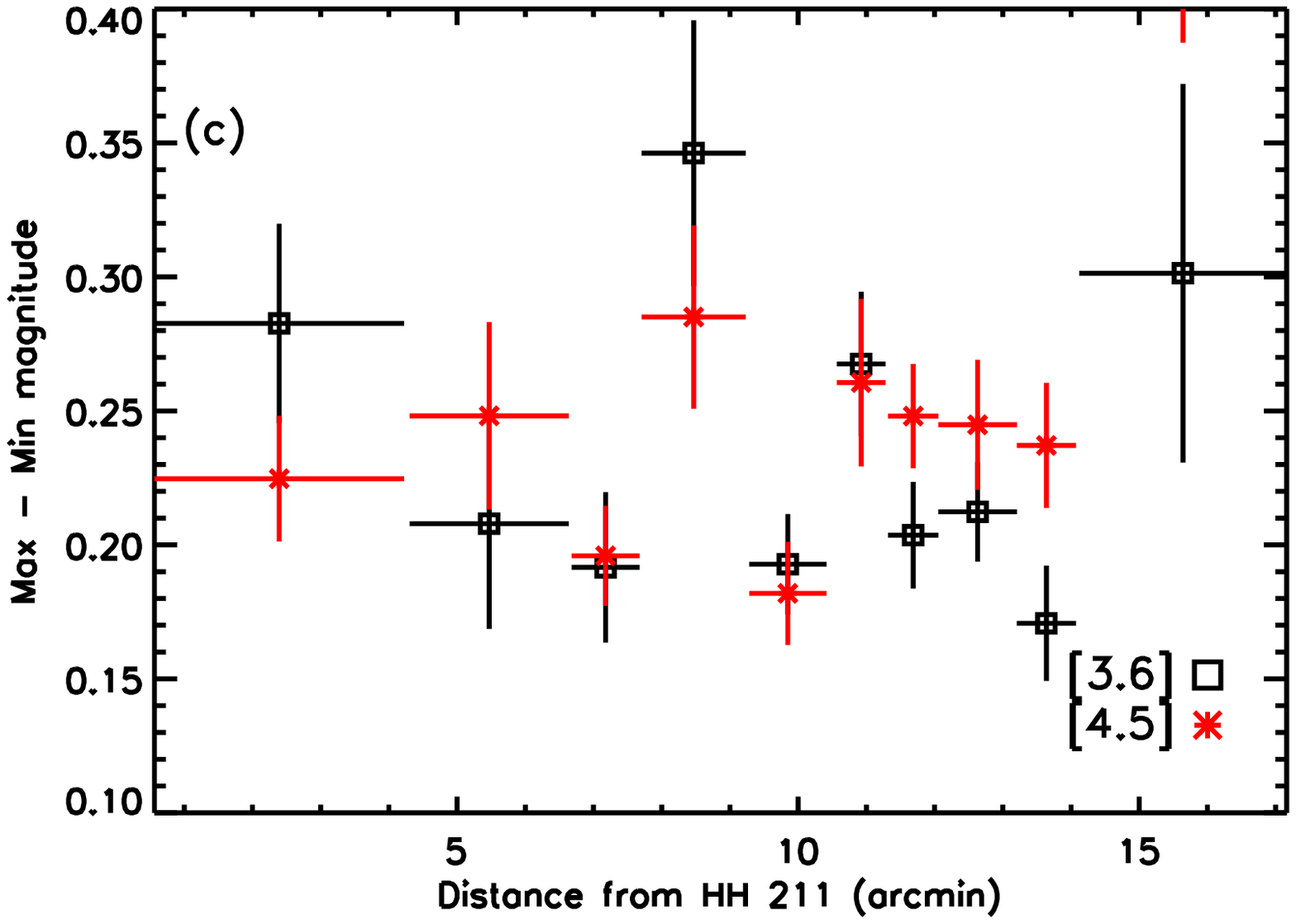}
\includegraphics[scale=.32]{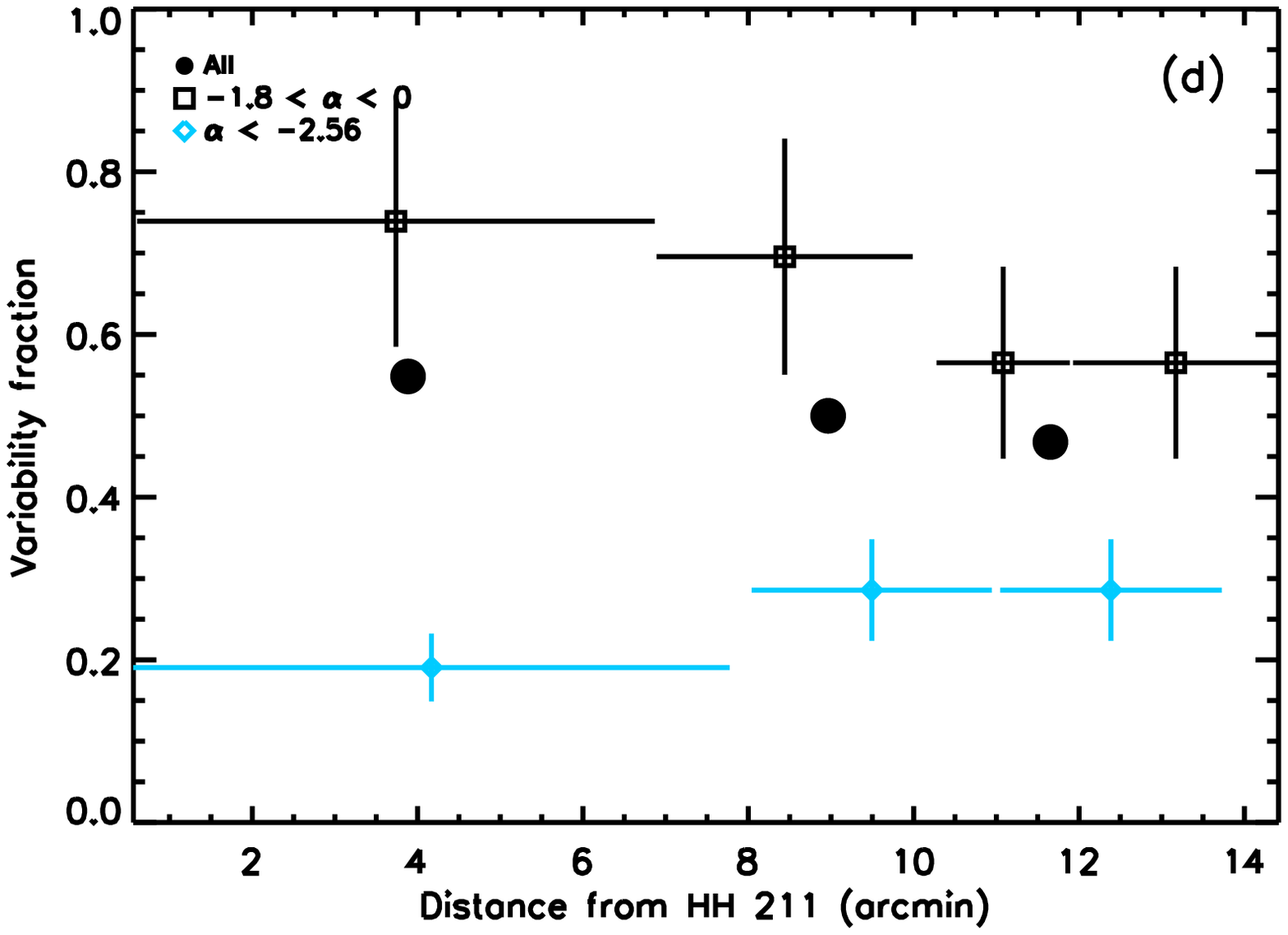}
\includegraphics[scale=.32]{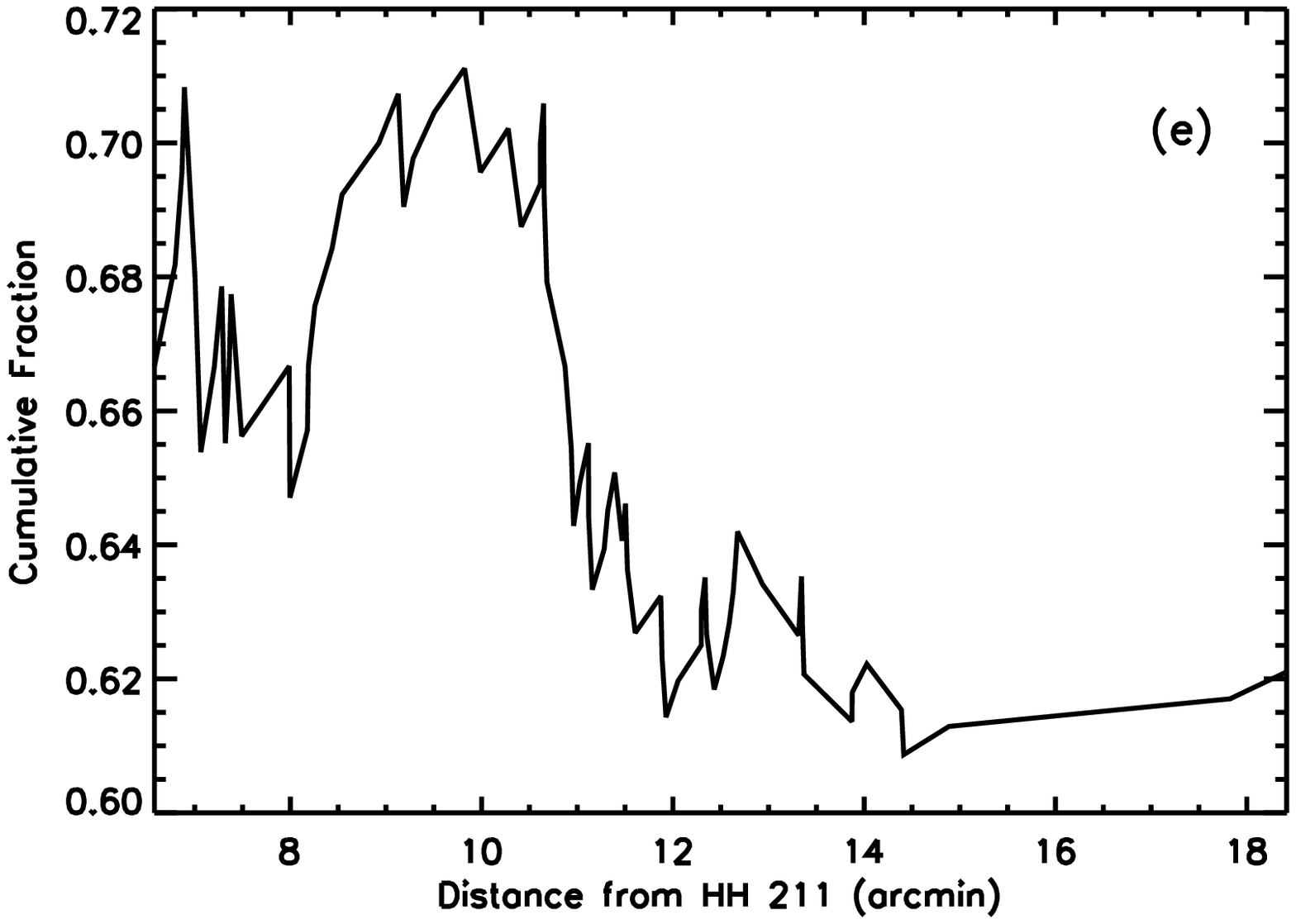}
\caption{(a) Variable fraction as a function of position (249 members). (b) Cumulative fraction. (c) Mean peak to peak fluctuations (d) Fraction of irregular variables split into class II sources (black squares, 95 members) and class III sources (blue diamonds, 84 members). (e) Cumulative fraction of irregular variables among the class II sources. There is a significant trend between f$_{var}$ and position even when looking at only the class II sources. \label{var_pos}}
\end{figure*}

\begin{figure}
\center
\includegraphics[scale=.5]{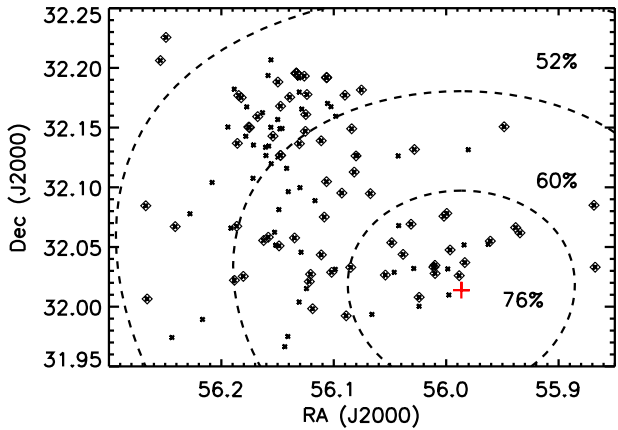}
\caption{Map of cluster members with $\alpha_{IRAC}>-1.8$ (ie. stars with disks, 95 members) marked as crosses, with variable stars marked with diamonds. The position of HH 211 is marked with a red plus sign surrounded by circles at a distance of 5, 10 and 15 arcminutes. The fraction of variable stars with disks is noted at each radius. There is a significant decrease in disk fraction with radius even at a fixed SED shape, indicating that variability depends strongly on position within the cluster.\label{map_fvar}}
\end{figure}

\begin{figure*}
\center
\includegraphics[scale=.32]{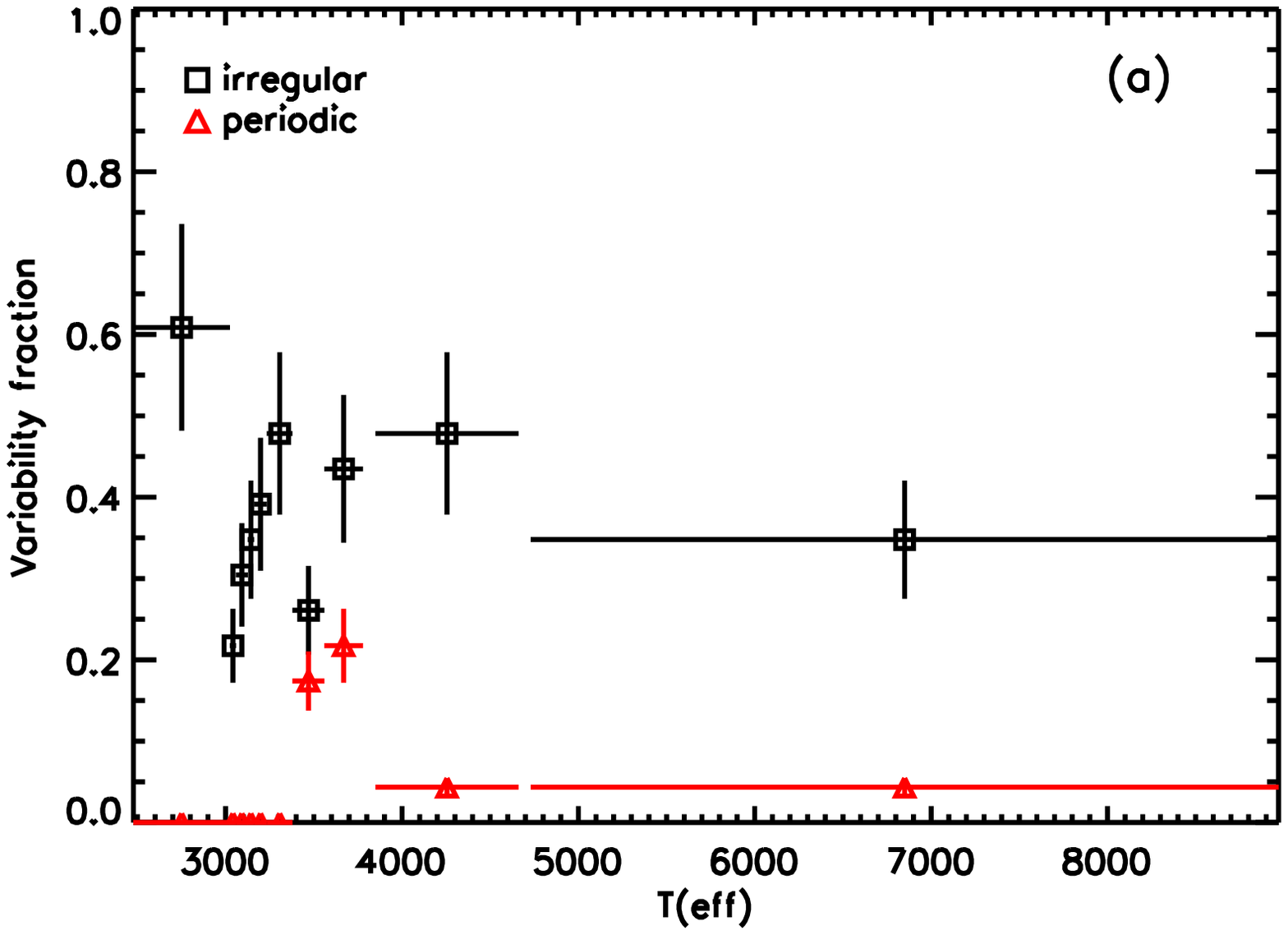}
\includegraphics[scale=.32]{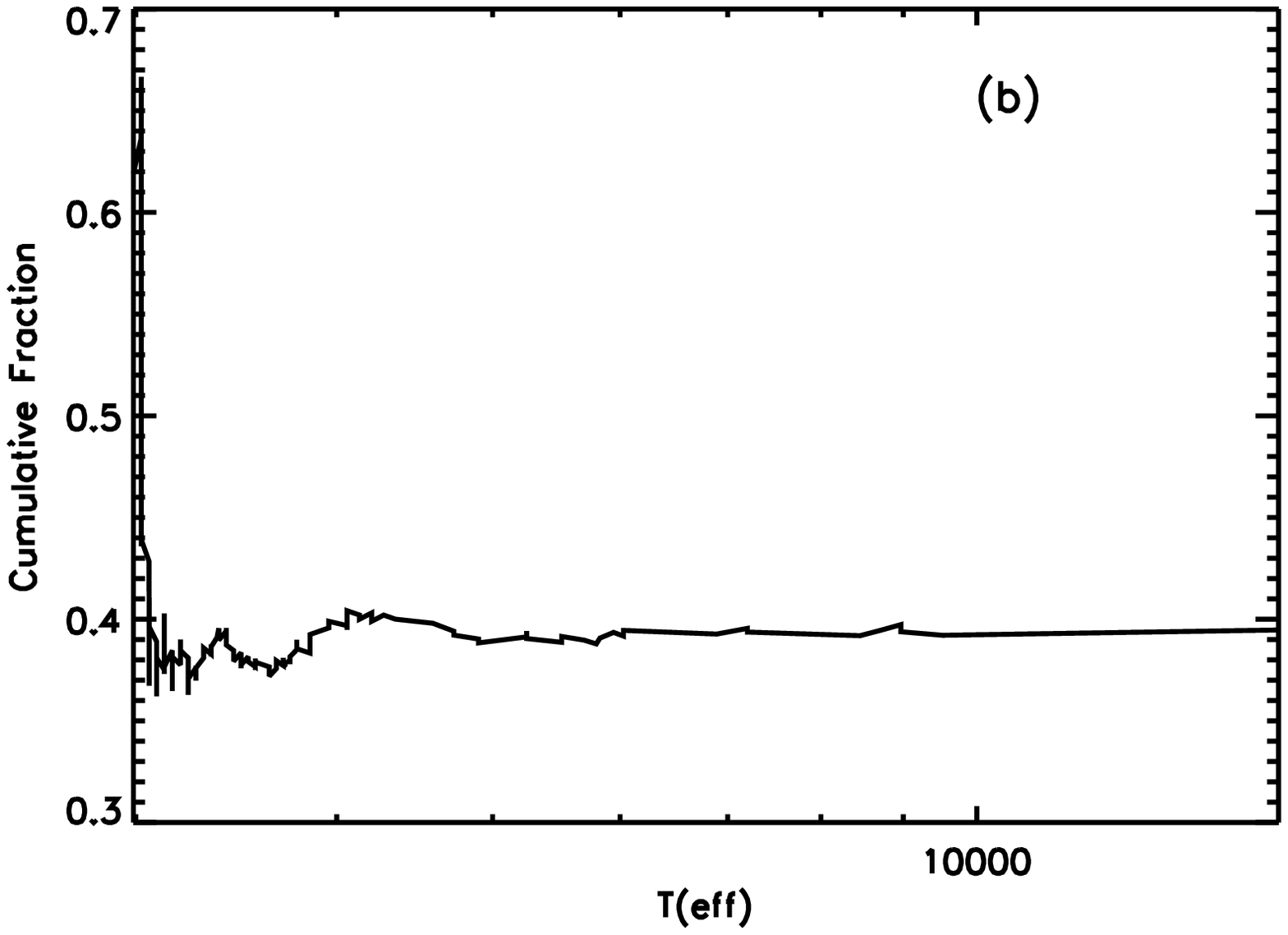}
\includegraphics[scale=.32]{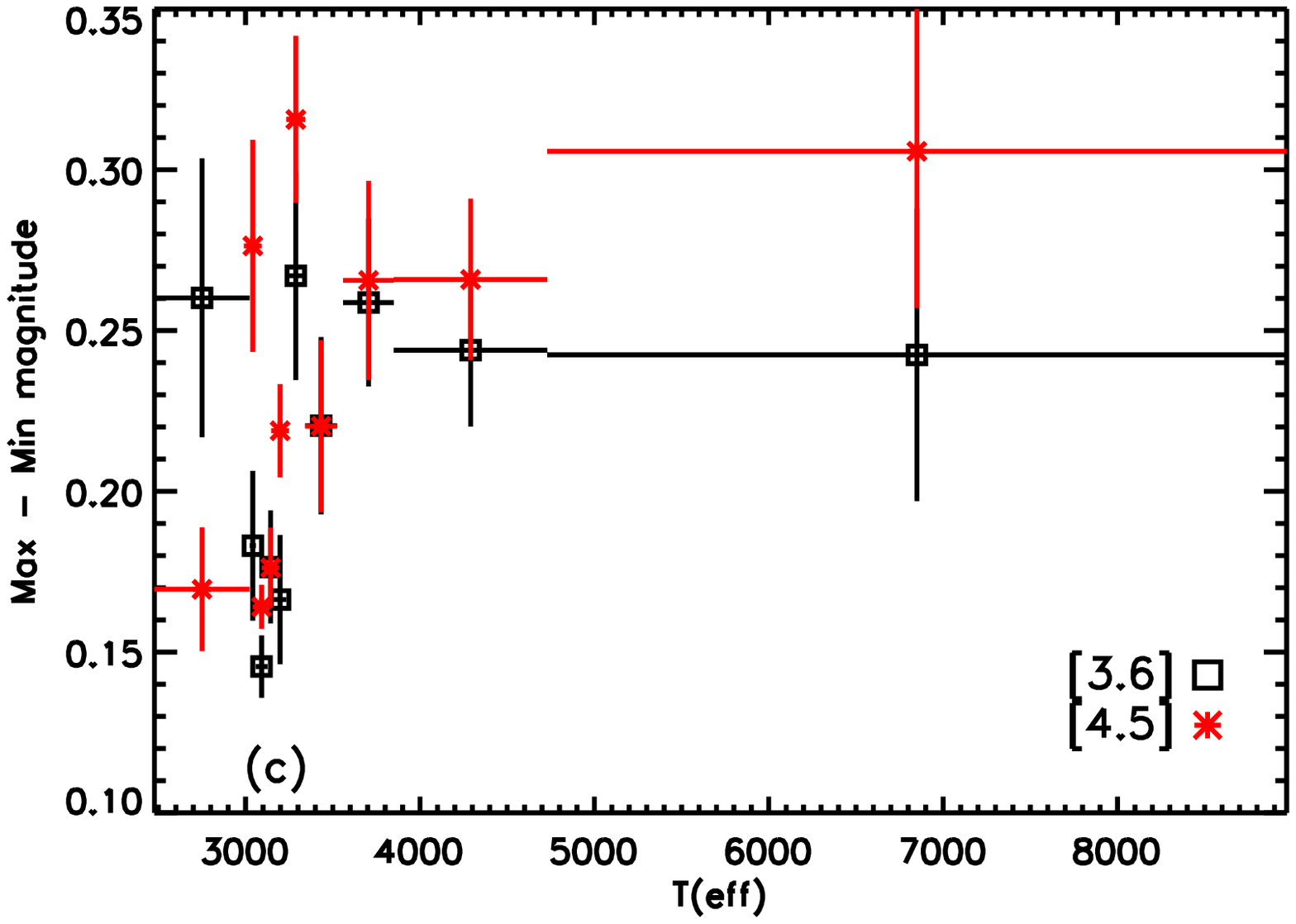}
\includegraphics[scale=.32]{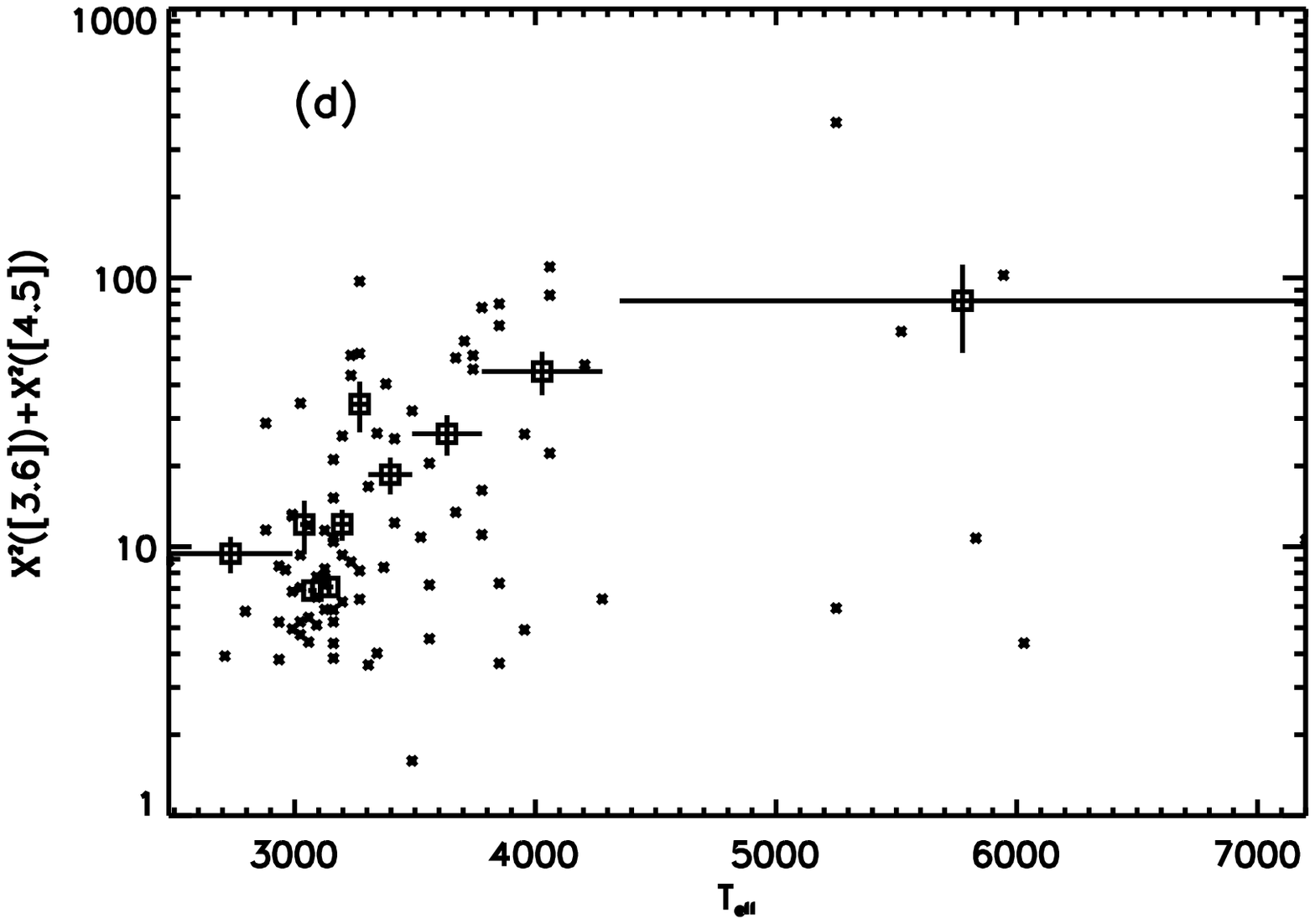}
\includegraphics[scale=.32]{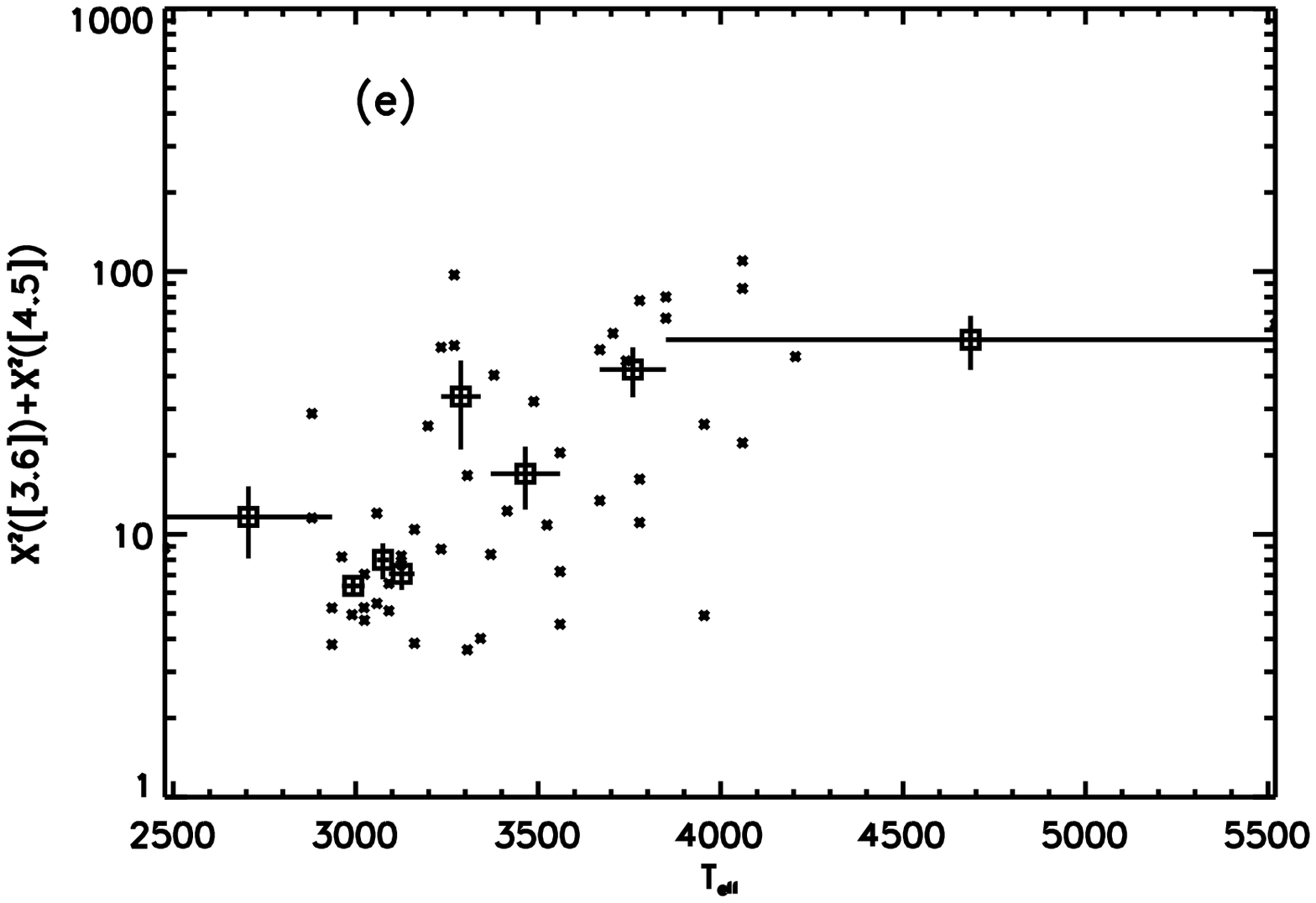}
\caption{(a) Variable fraction as a function of T$_{eff}$ (235 members). (b) Cumulative fraction. (c) Mean peak to peak fluctuations for irregular variables . (d) Size of the fluctuations for all irregular variables (86 members). (e) Size of the fluctuations for only the class II sources (86 members). There is a significant trend between$\chi^2_{\nu}$ and T$_{eff}$ even when looking at only the class II sources, but this is most likely an observational bias. \label{var_teff}}
\end{figure*}

\begin{figure*}
\center
\includegraphics[scale=.45]{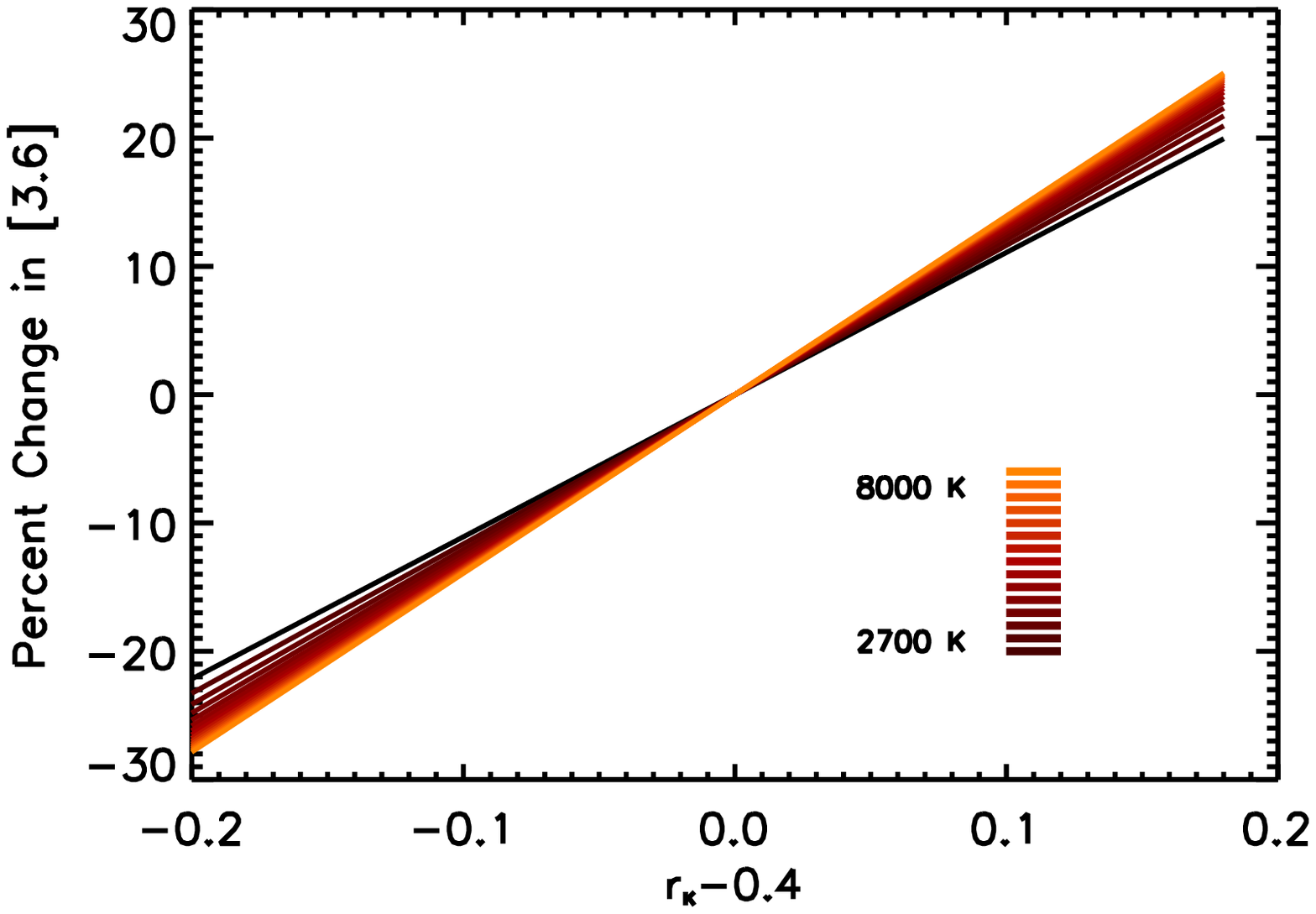}
\includegraphics[scale=.45]{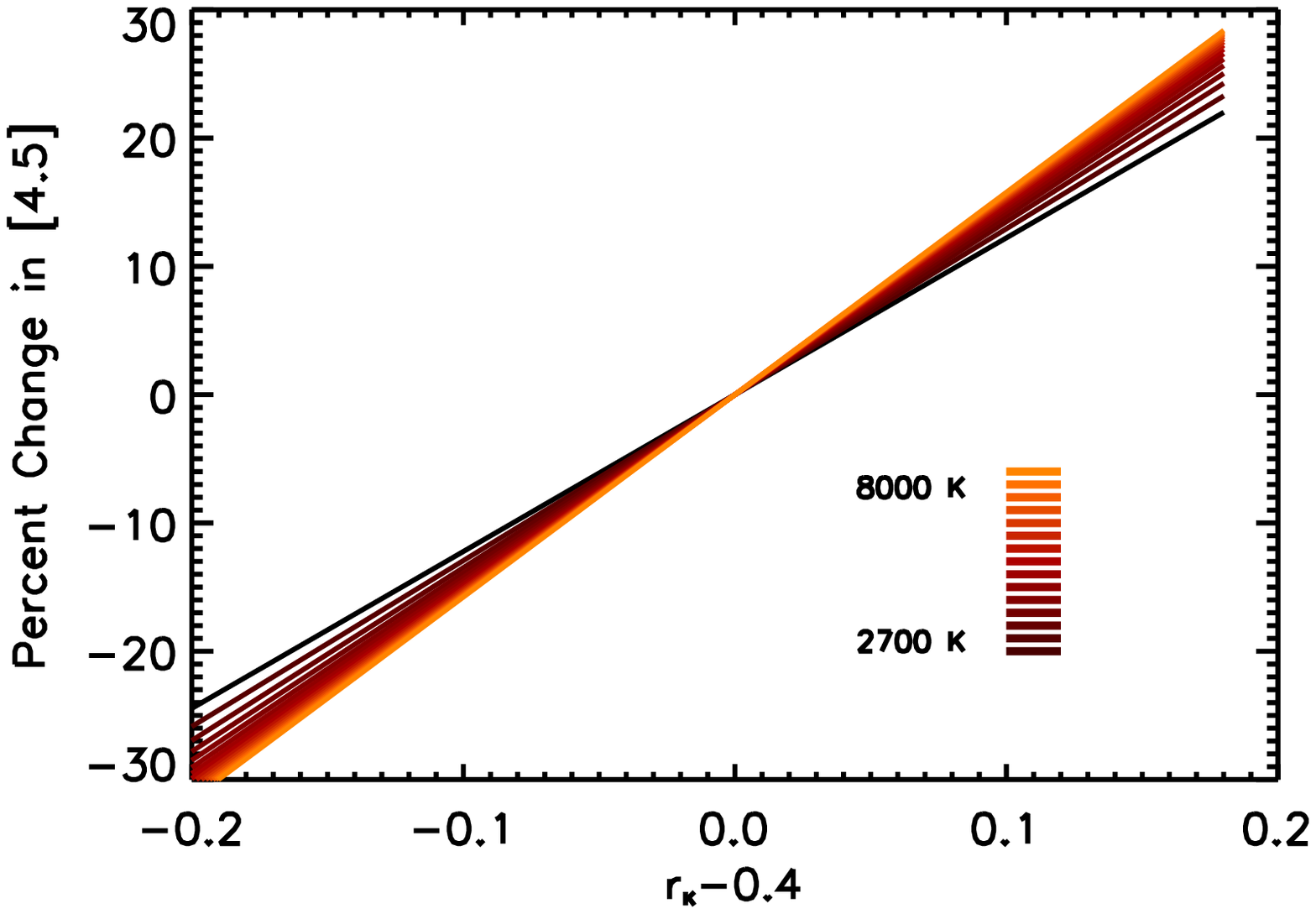}
\caption{The variability in [3.6] (left) and [4.5] (right) for varying levels of discrepancy away from a fiducial excess for stars at various effective temperatures. As the effective temperature of the star decreases the size of the fluctuations decreases for a given change in the excess. This demonstrates that the observed correlation between variability strength and effective temperature is not a physical effect, and is simply due to the different relative strengths of the stellar and disk flux as the peak of the stellar SED moves to redder wavelengths.\label{teff_ir_corr}}
\end{figure*}

\begin{figure*}
\center
\includegraphics[scale=.32]{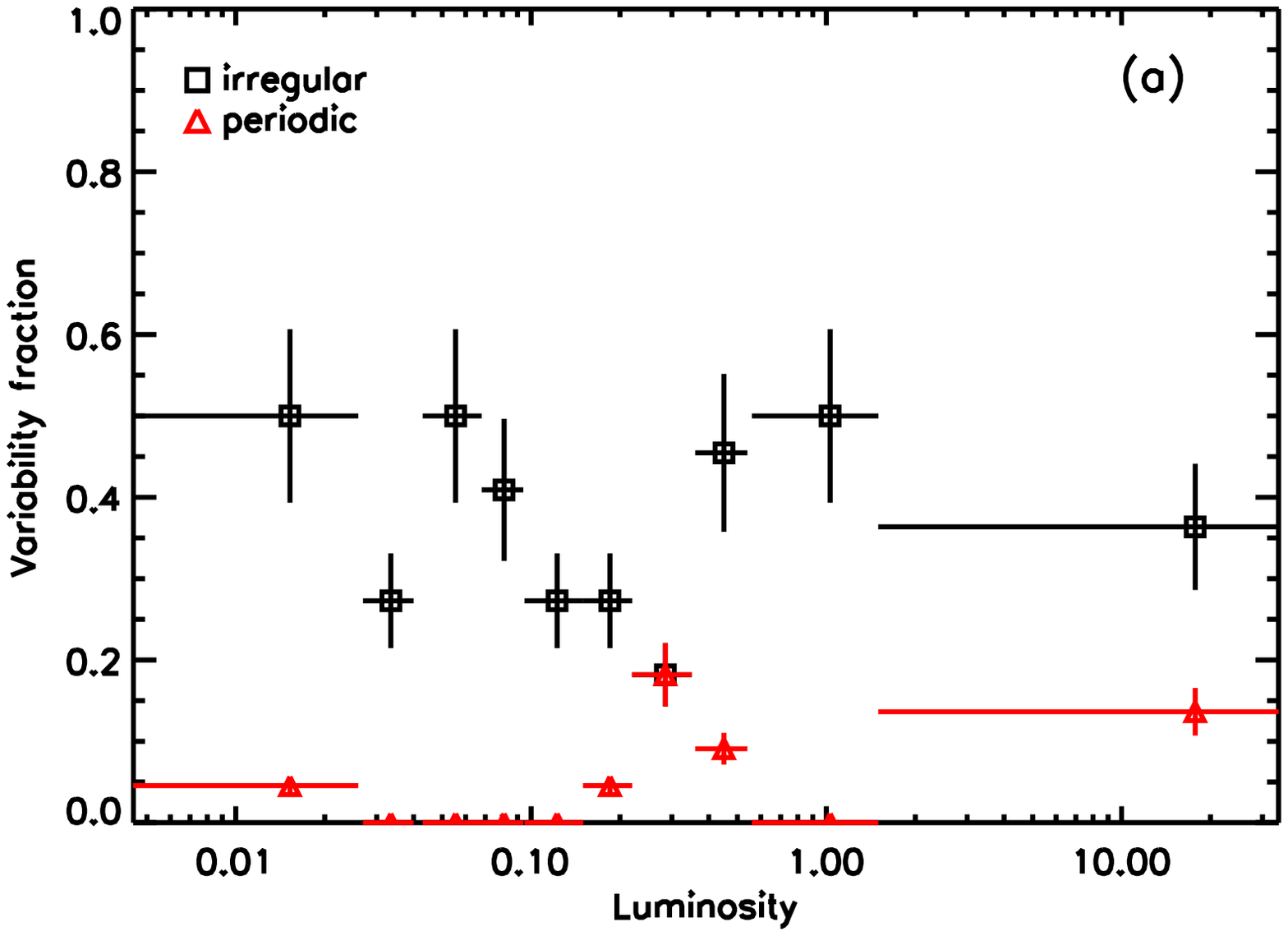}
\includegraphics[scale=.32]{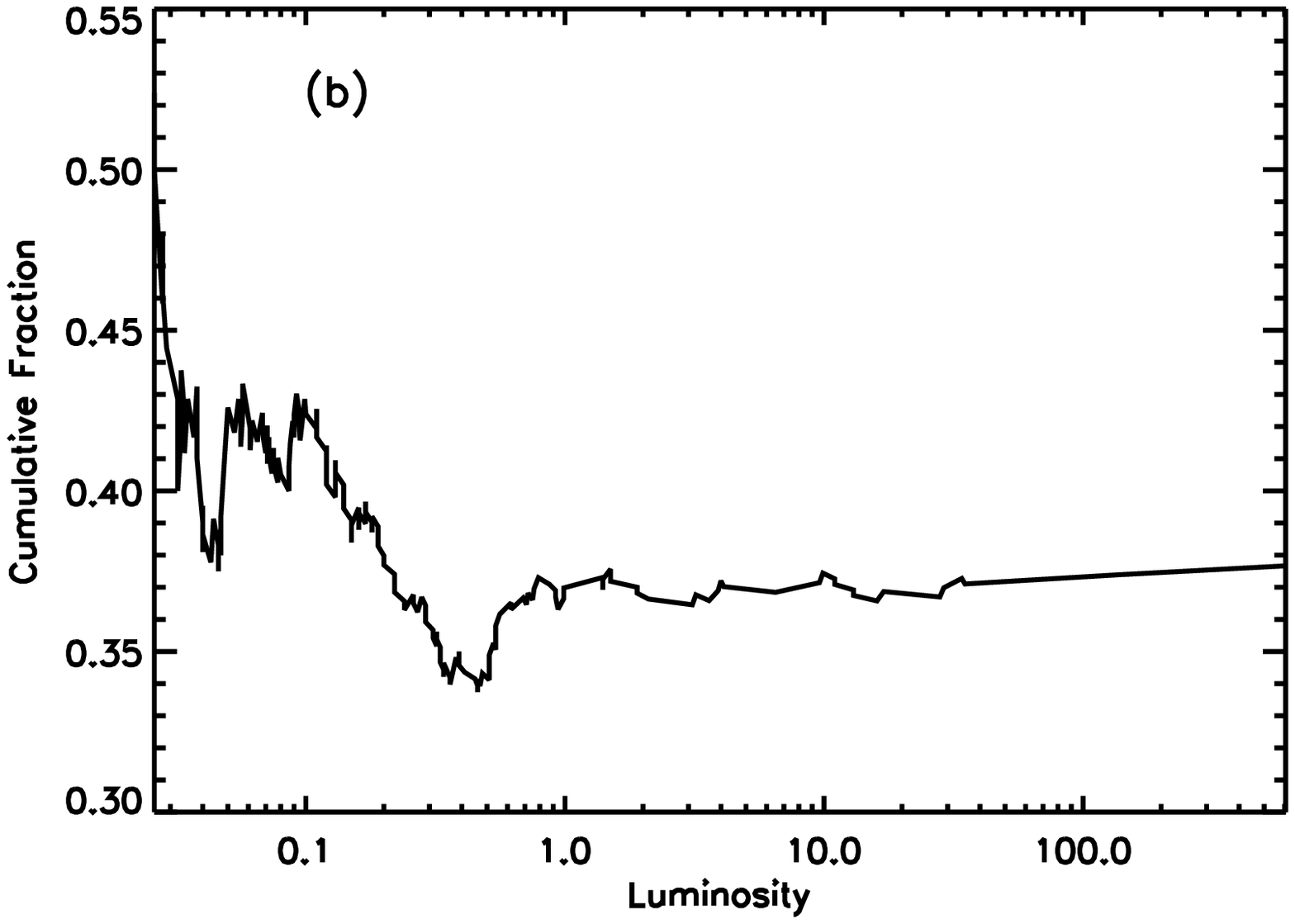}
\includegraphics[scale=.32]{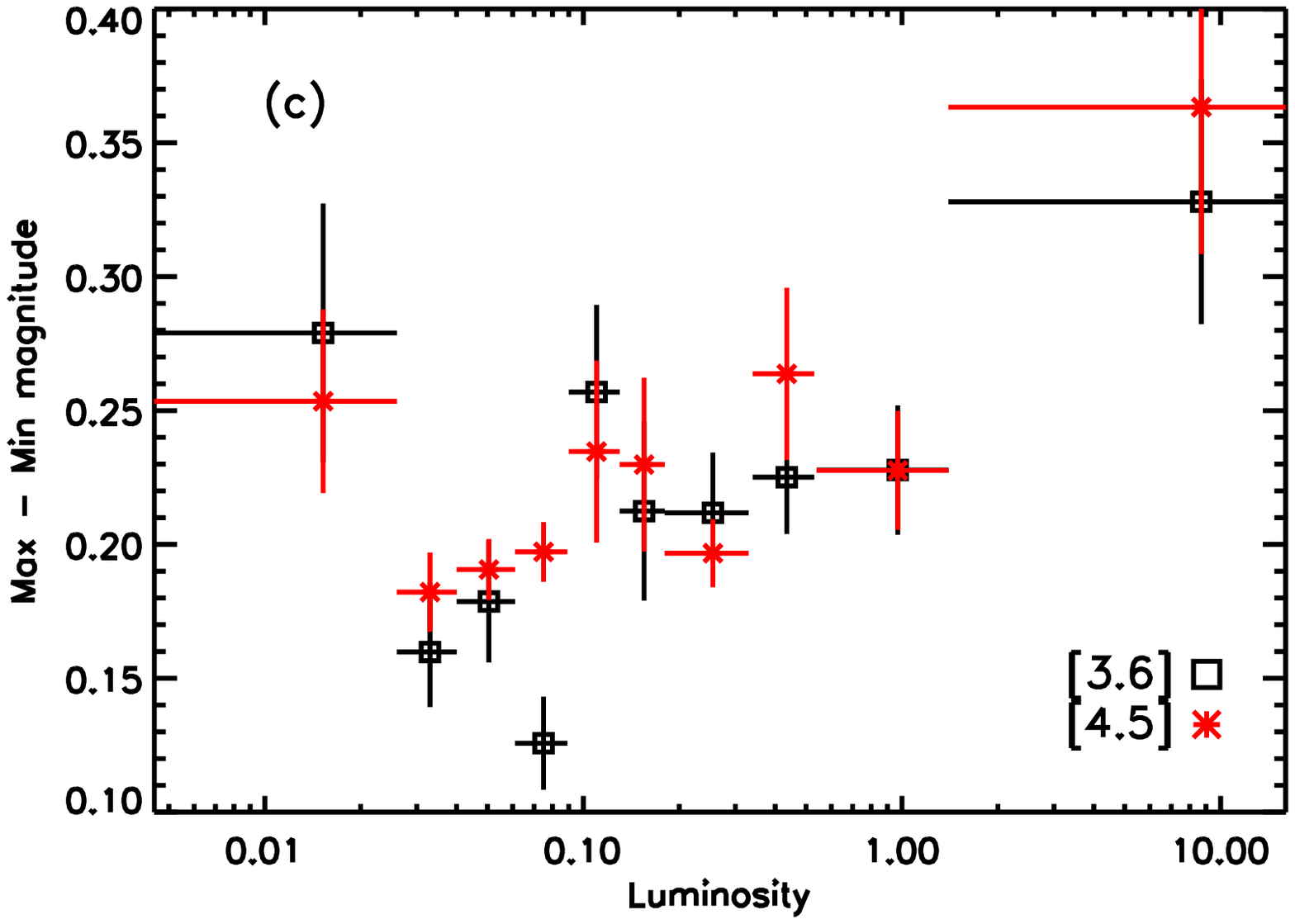}
\includegraphics[scale=.32]{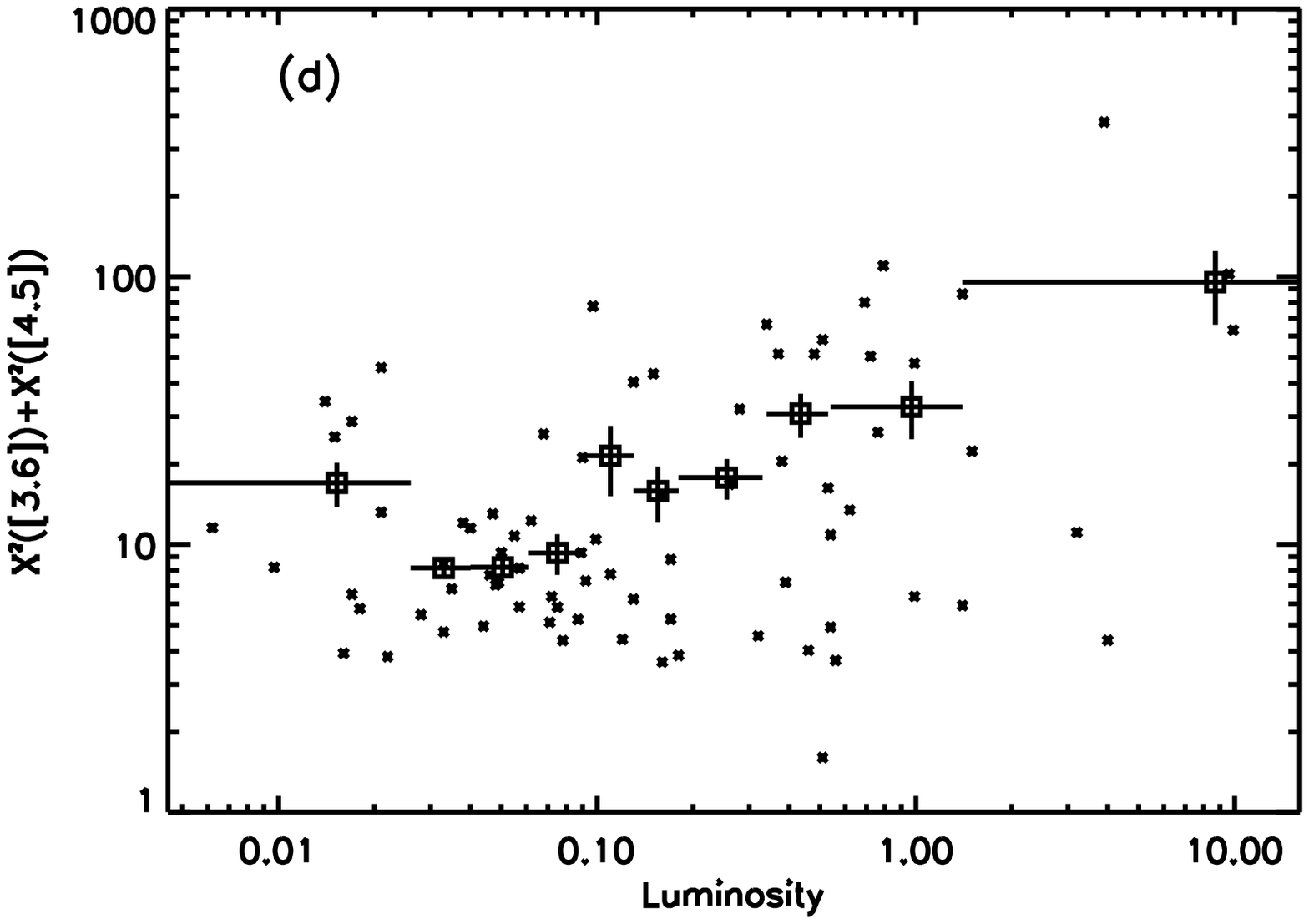}
\caption{(a) Variable fraction as a function of L$_*$ (223 members). (b) Cumulative fraction of irregular variables. (c) Mean peak to peak fluctuations among irregular variables (99 members). (d) Size of the fluctuations. No significant trend is seen between L$_*$ and the measures of variability (the size trend goes away when including T$_{eff}$). \label{var_lstar}}
\end{figure*}

\begin{figure*}
\center
\includegraphics[scale=.32]{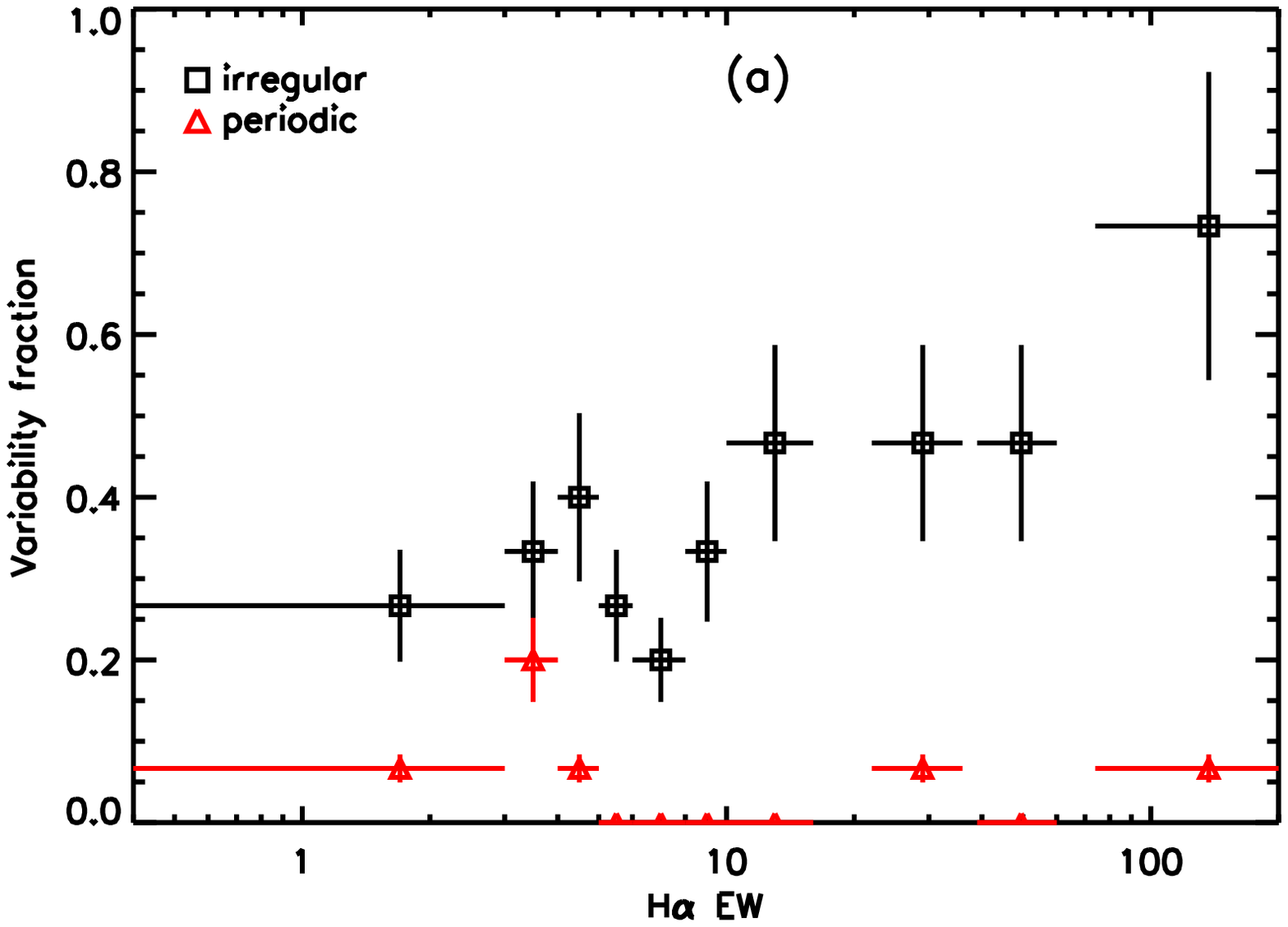}
\includegraphics[scale=.32]{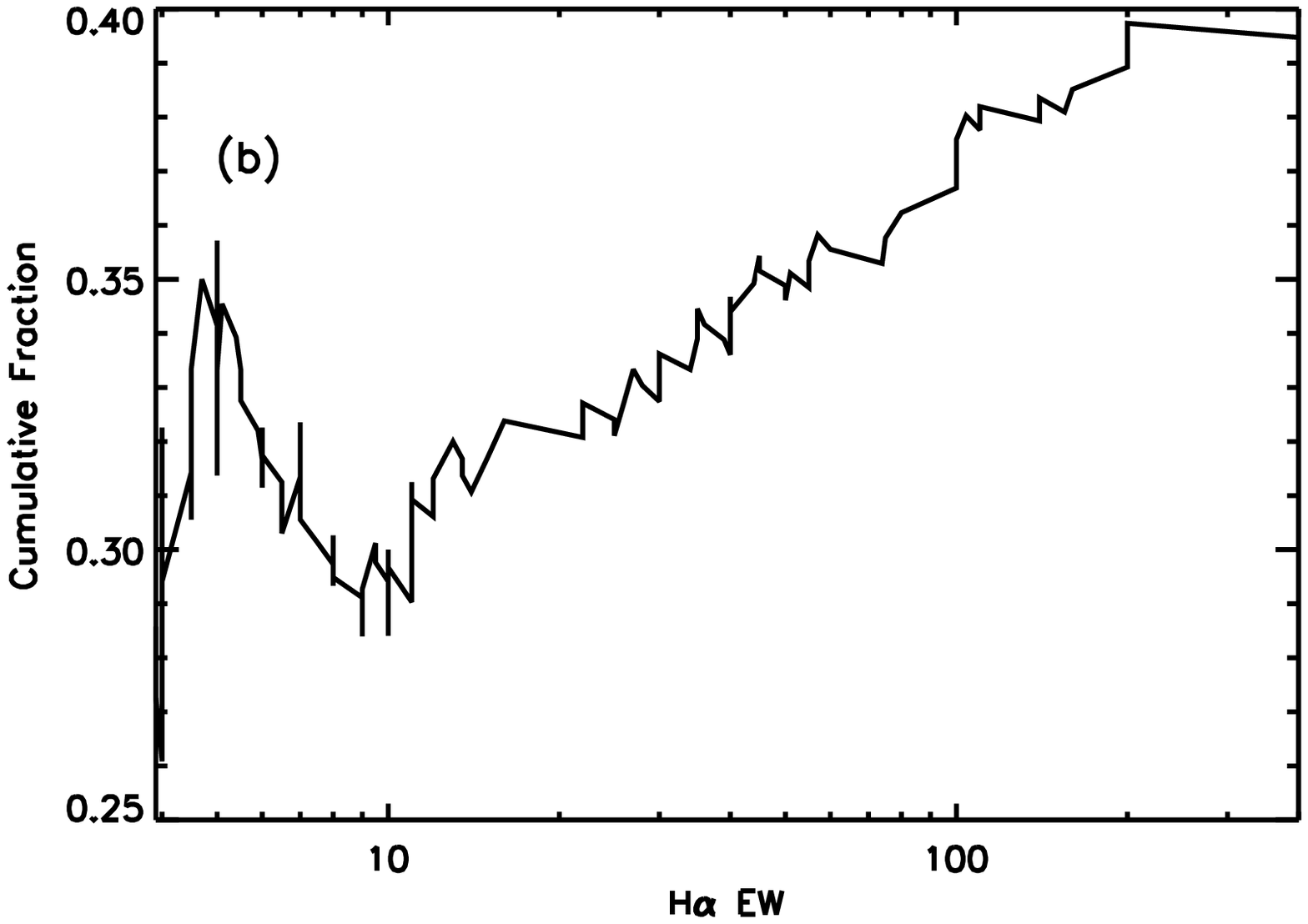}
\includegraphics[scale=.32]{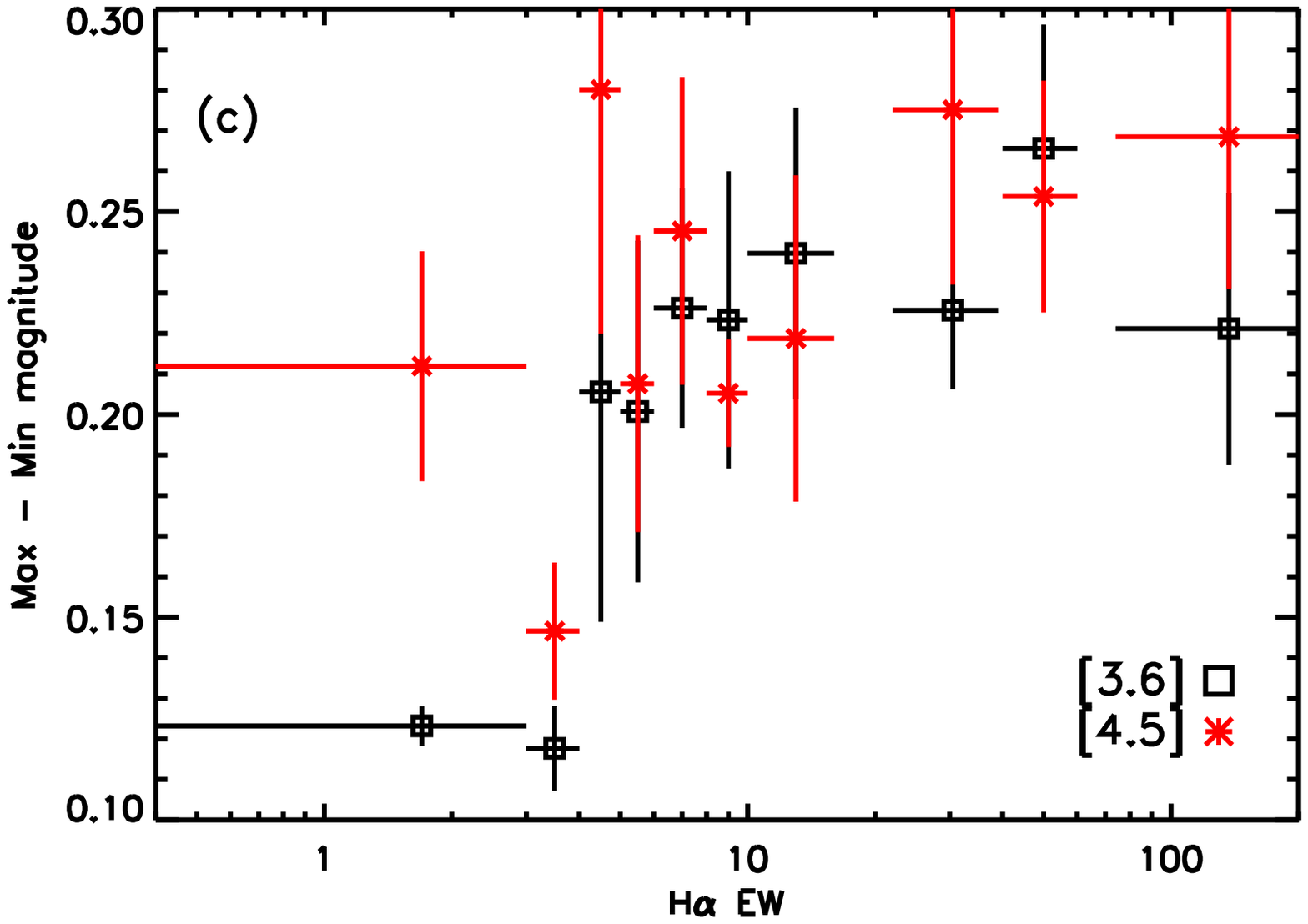}
\includegraphics[scale=.32]{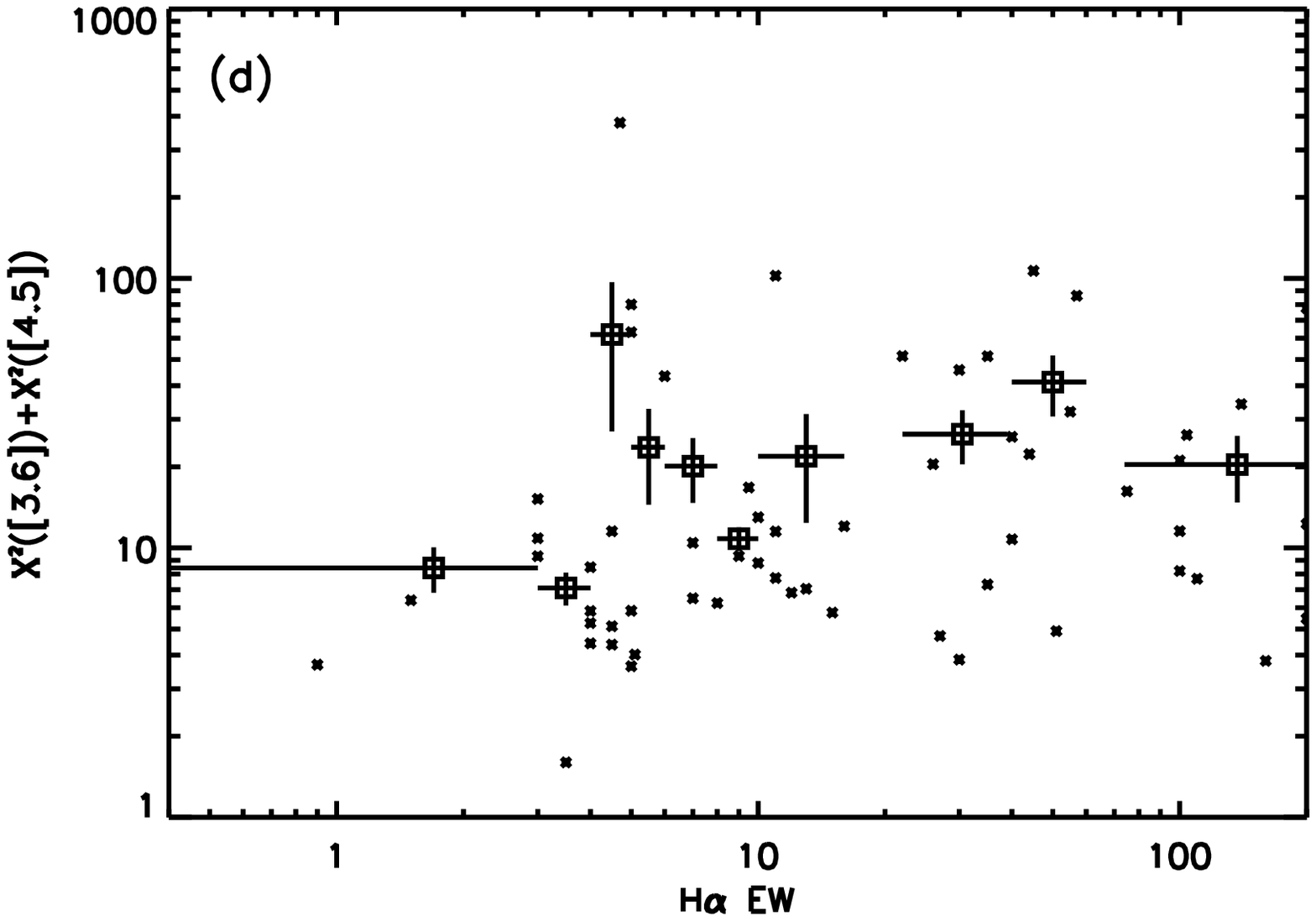}
\includegraphics[scale=.32]{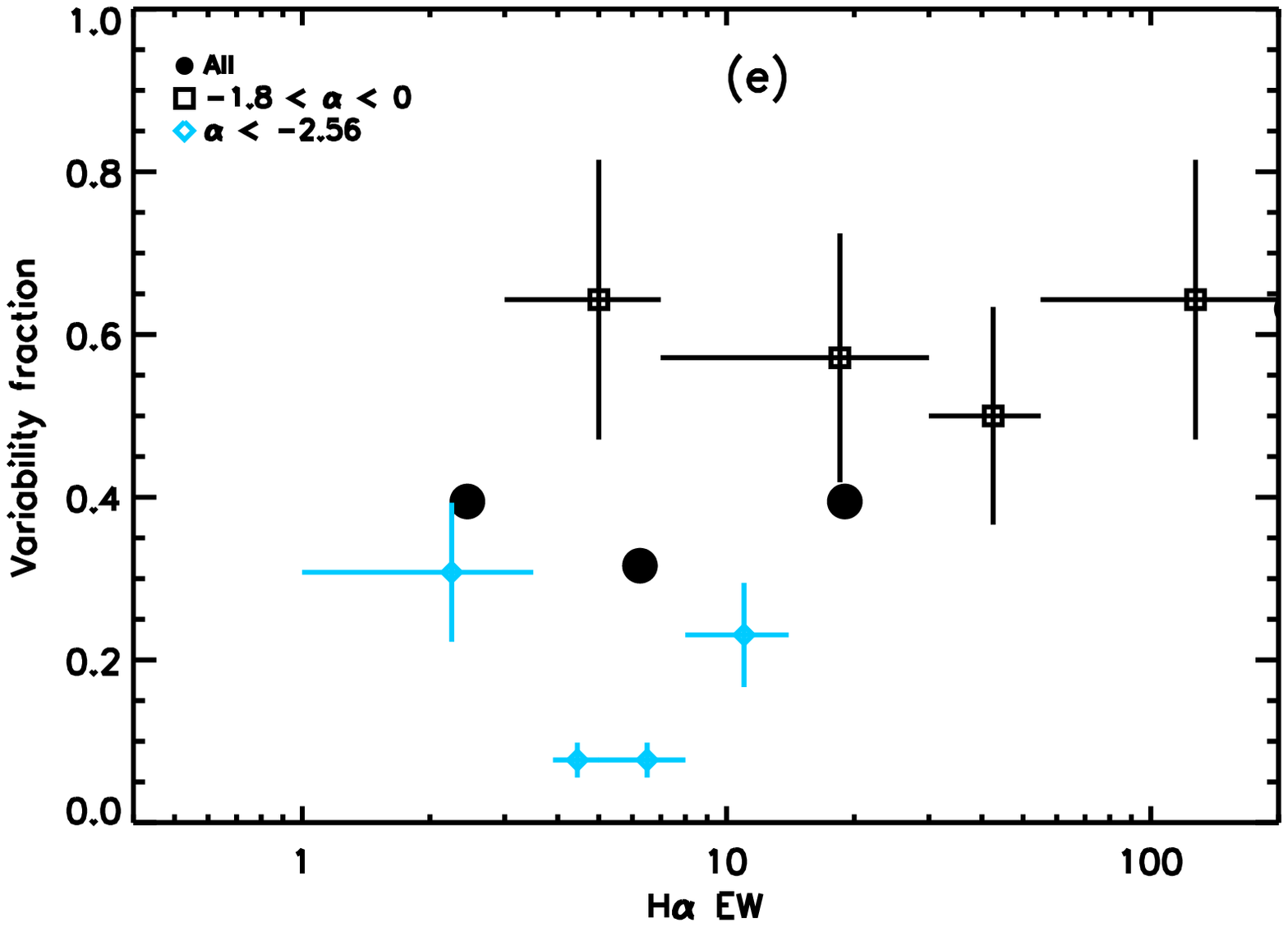}
\includegraphics[scale=.32]{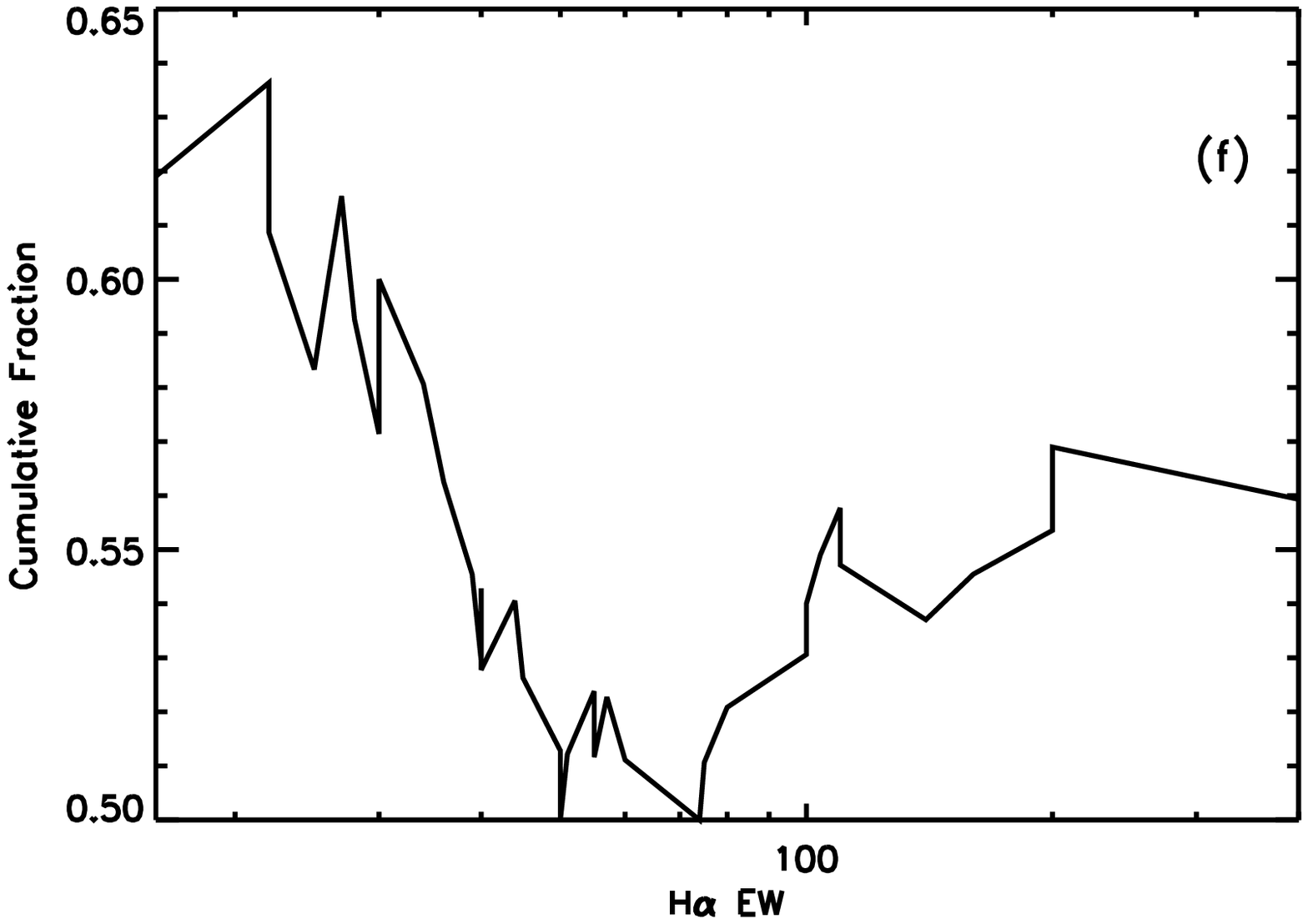}
\includegraphics[scale=.32]{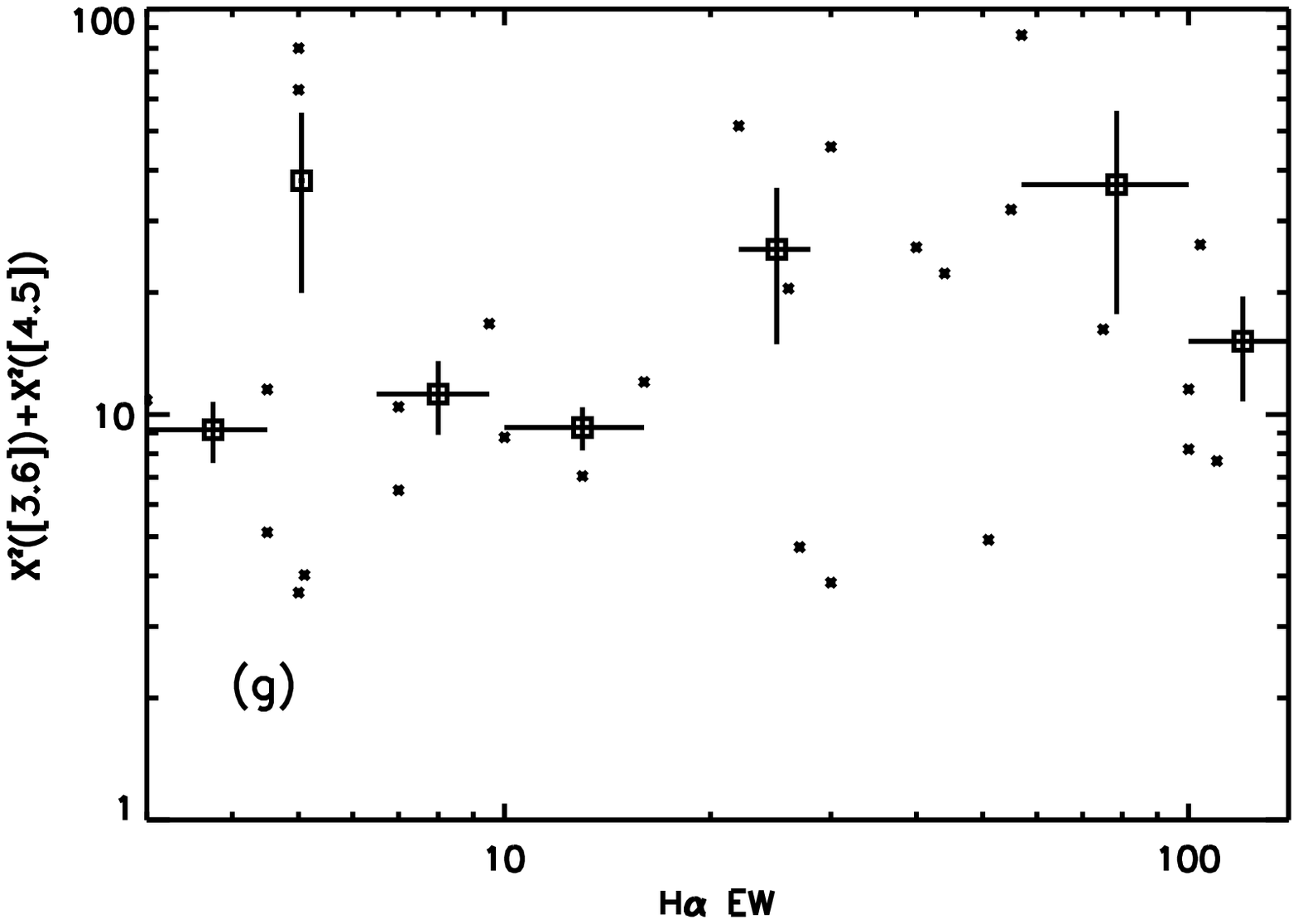}
\caption{(a) Variable fraction as a function of H$\alpha$ EW in units of \AA\ (154 members). (b) Cumulative fraction. (c) Mean peak to peak fluctuations for the irregular variables. (d) Size of the fluctuations (72 members). (e) Variable fraction split into class II (59 members) and class III (56 members)sources. (f) Cumulative fraction of irregular class II variables. (g) Size of the fluctuations among class II stars. While there is an increase in f$_{var}$ with H$\alpha$ emission, this trend disappears when correcting for the trend with $\alpha_{IRAC}$.\label{var_halpha}}
\end{figure*}

\begin{figure*}
\center
\includegraphics[scale=.32]{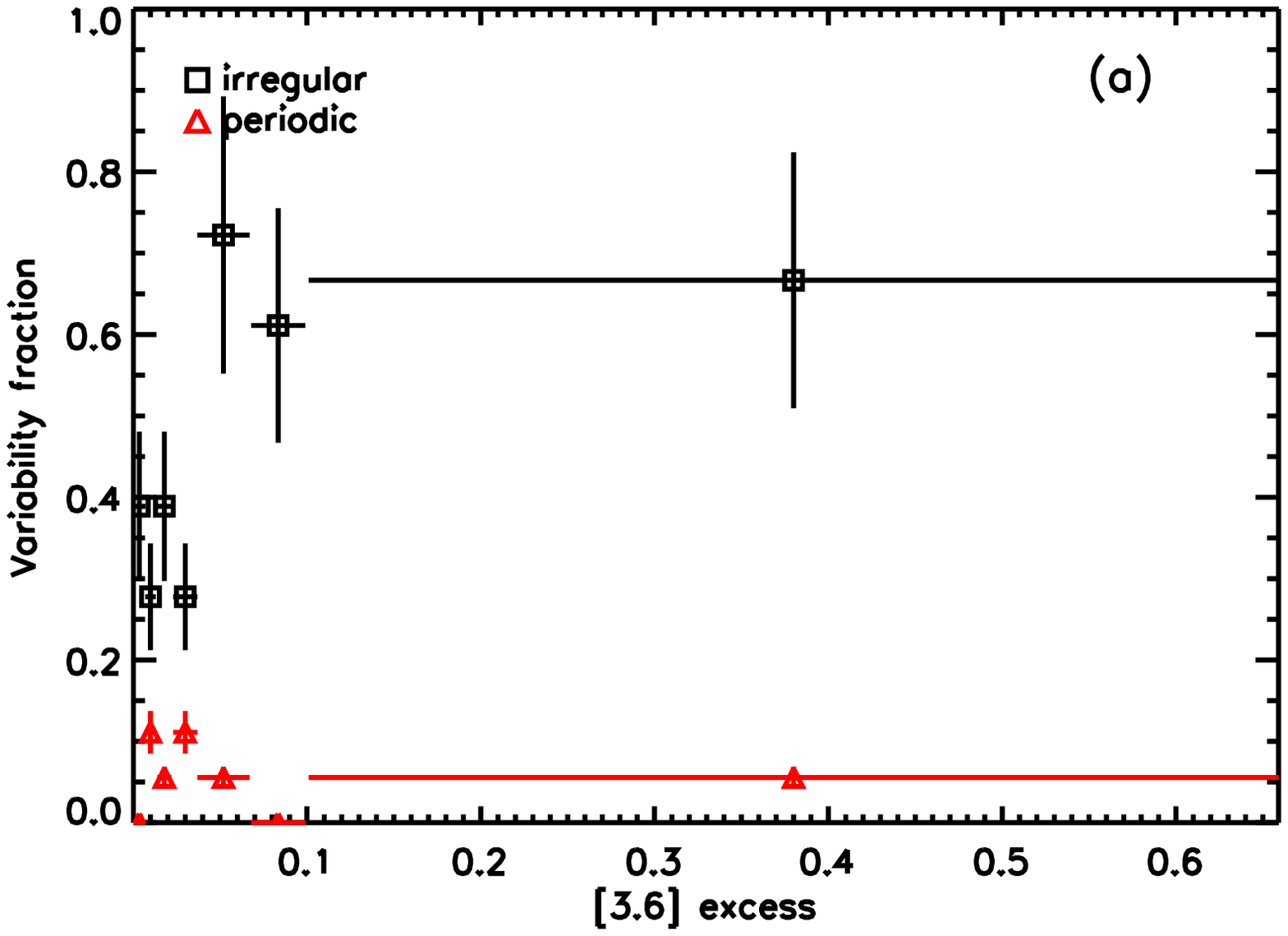}
\includegraphics[scale=.32]{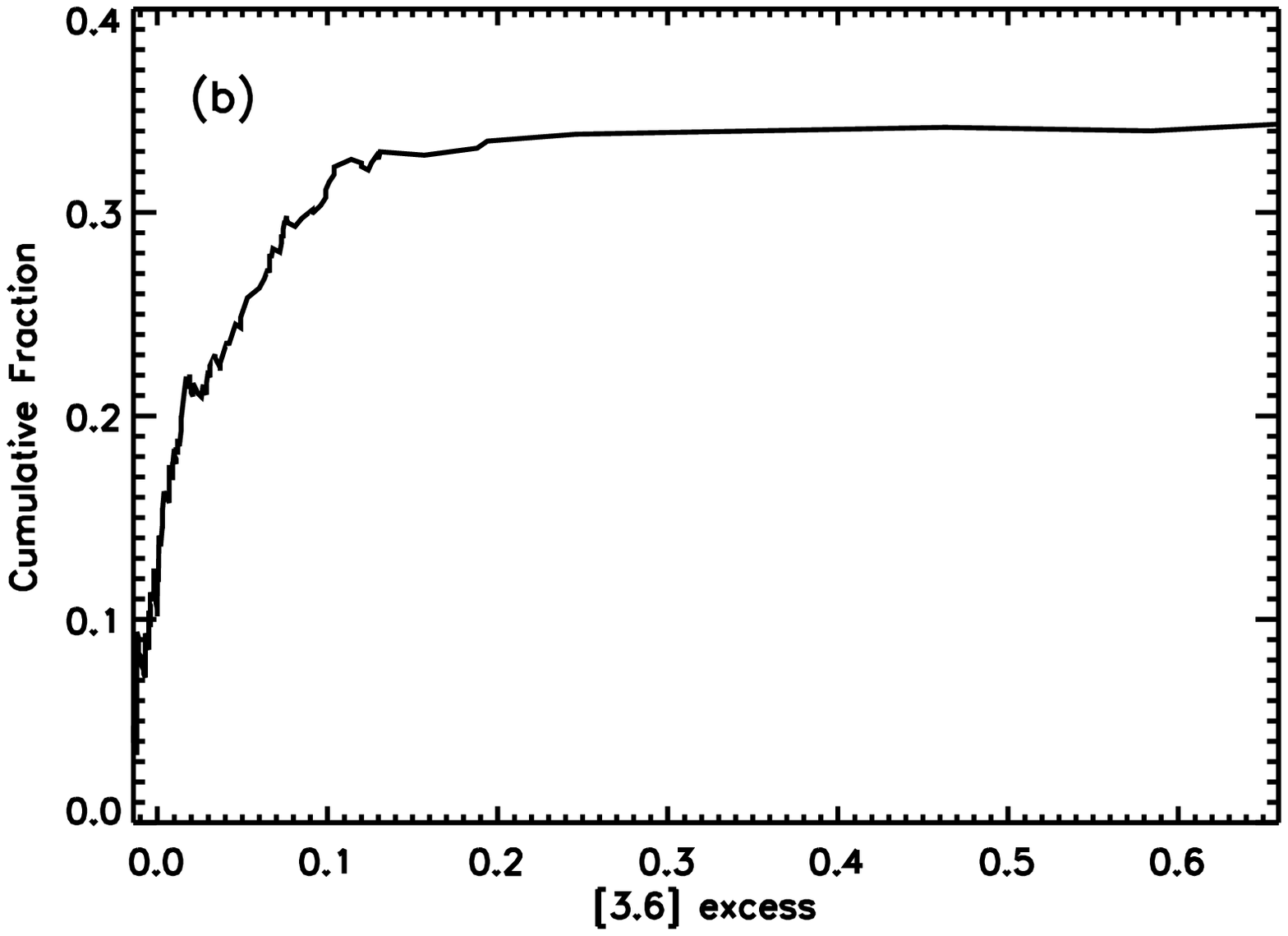}
\includegraphics[scale=.32]{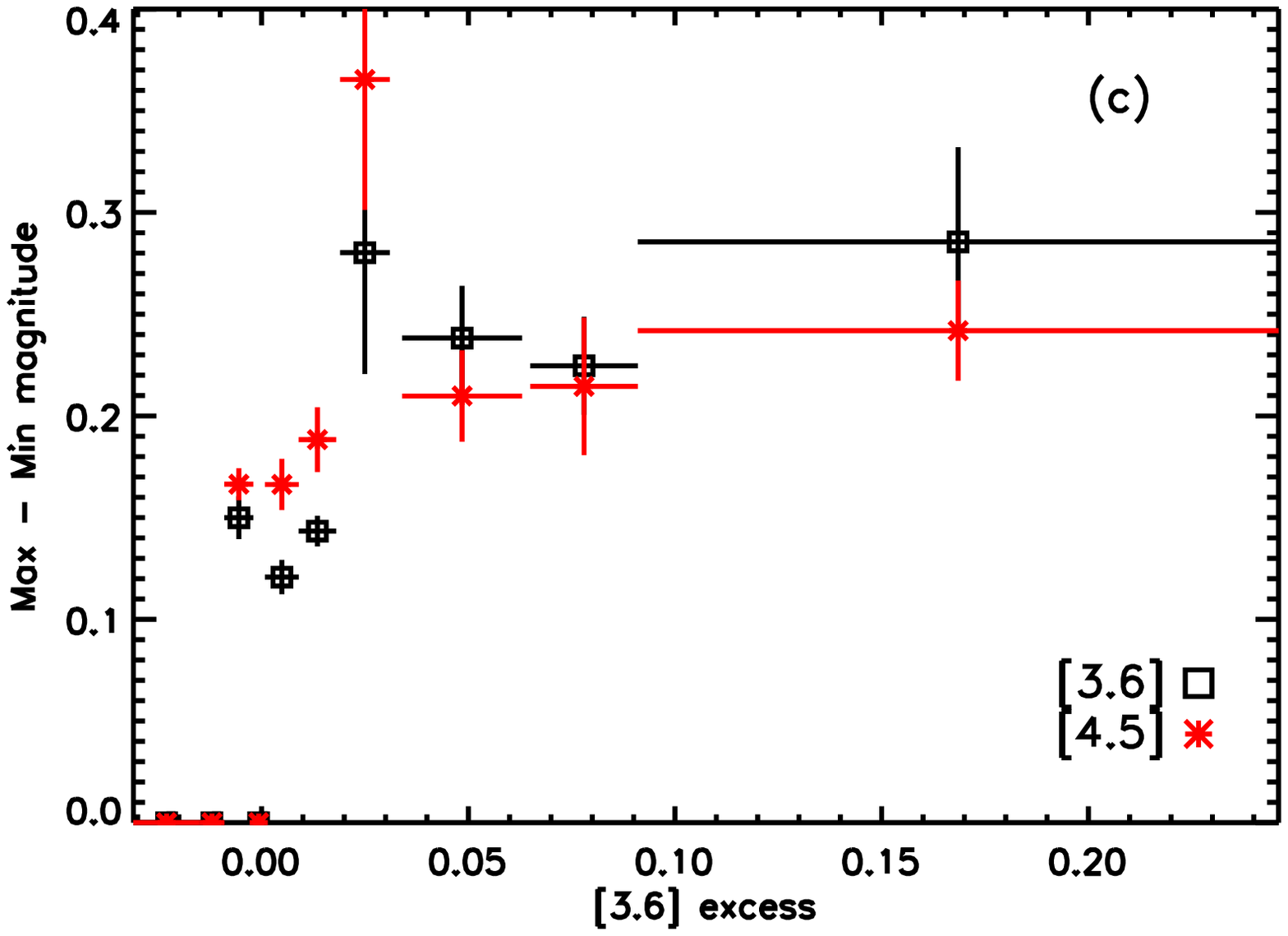}
\includegraphics[scale=.32]{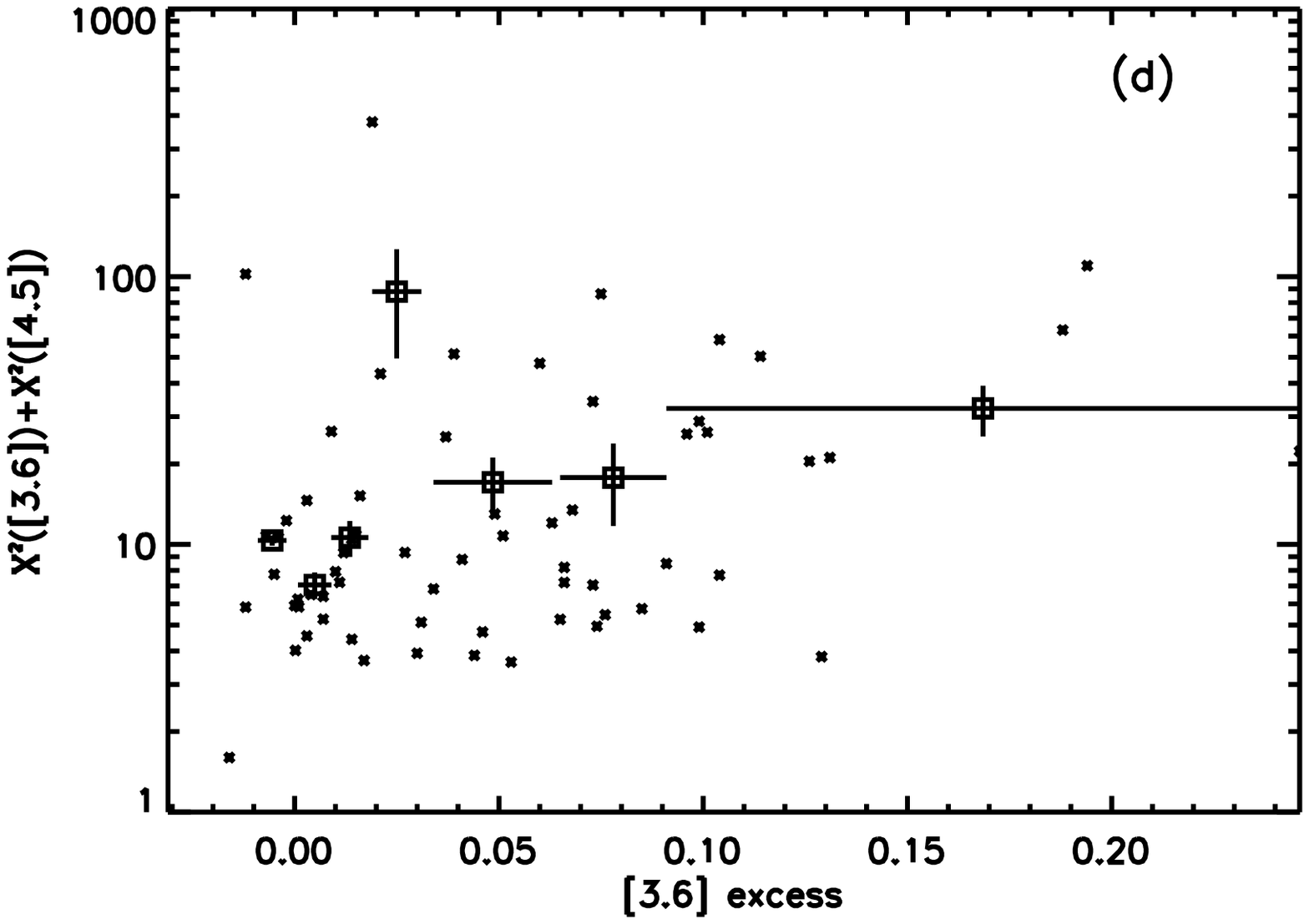}
\includegraphics[scale=.32]{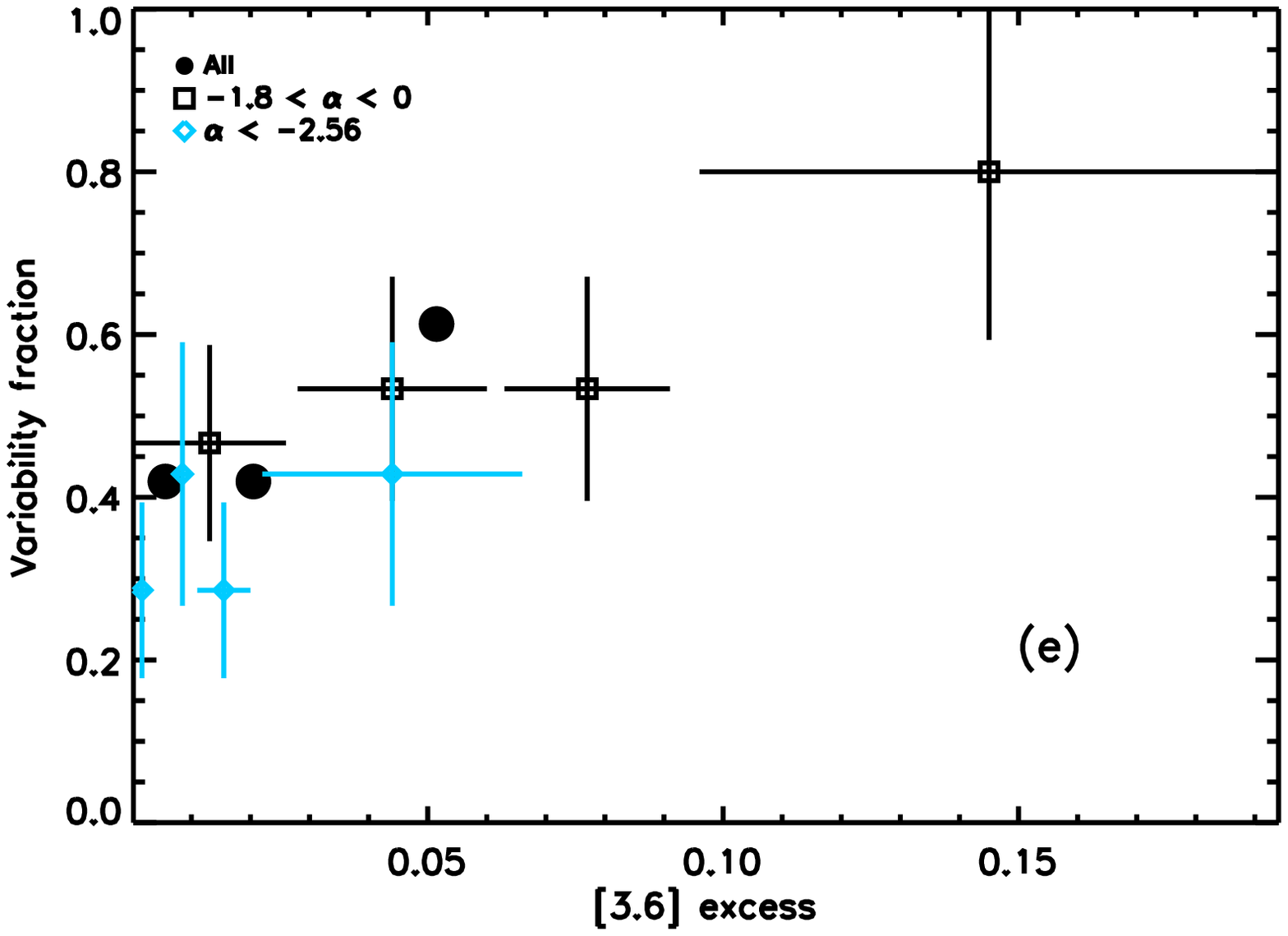}
\includegraphics[scale=.32]{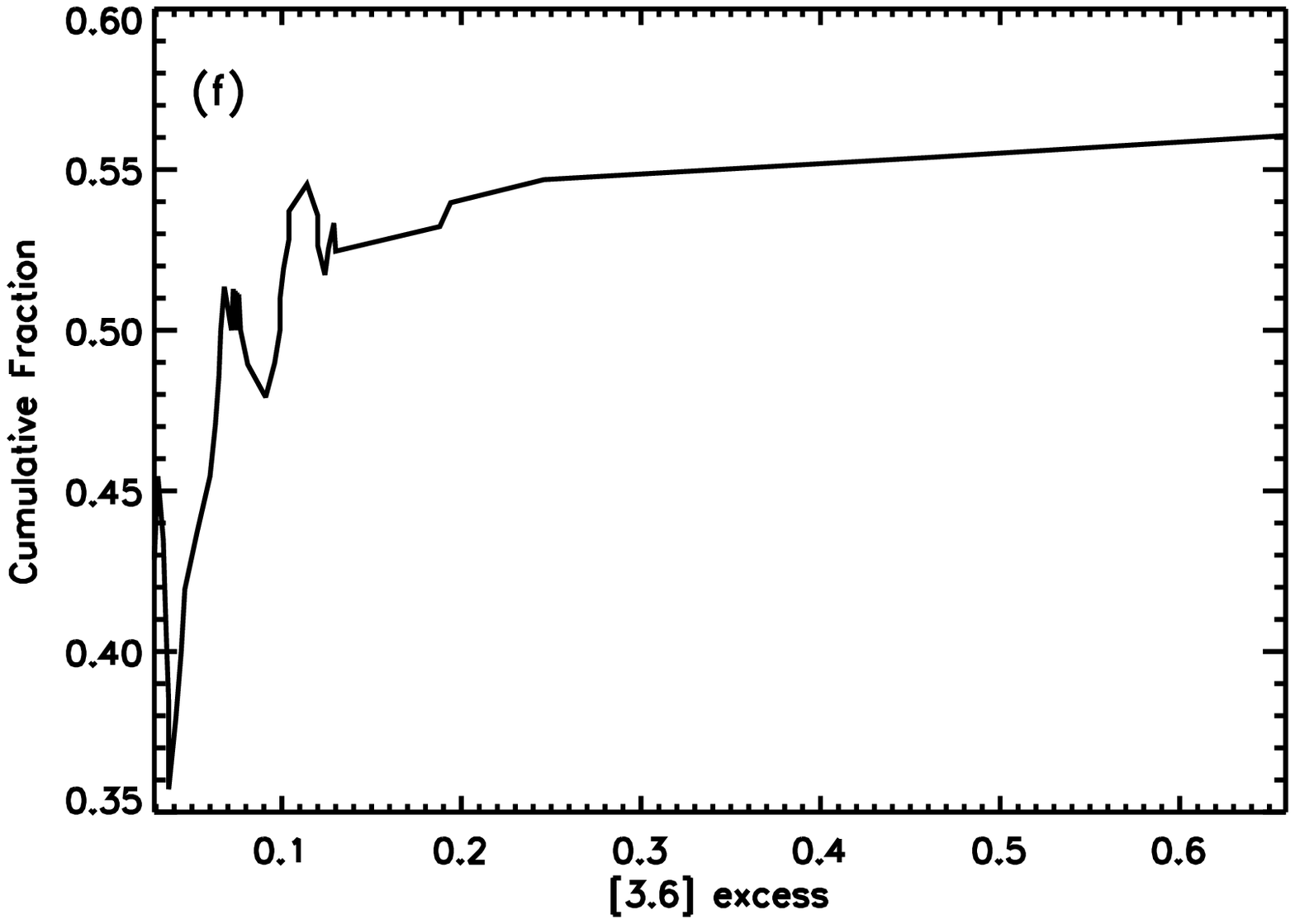}
\includegraphics[scale=.32]{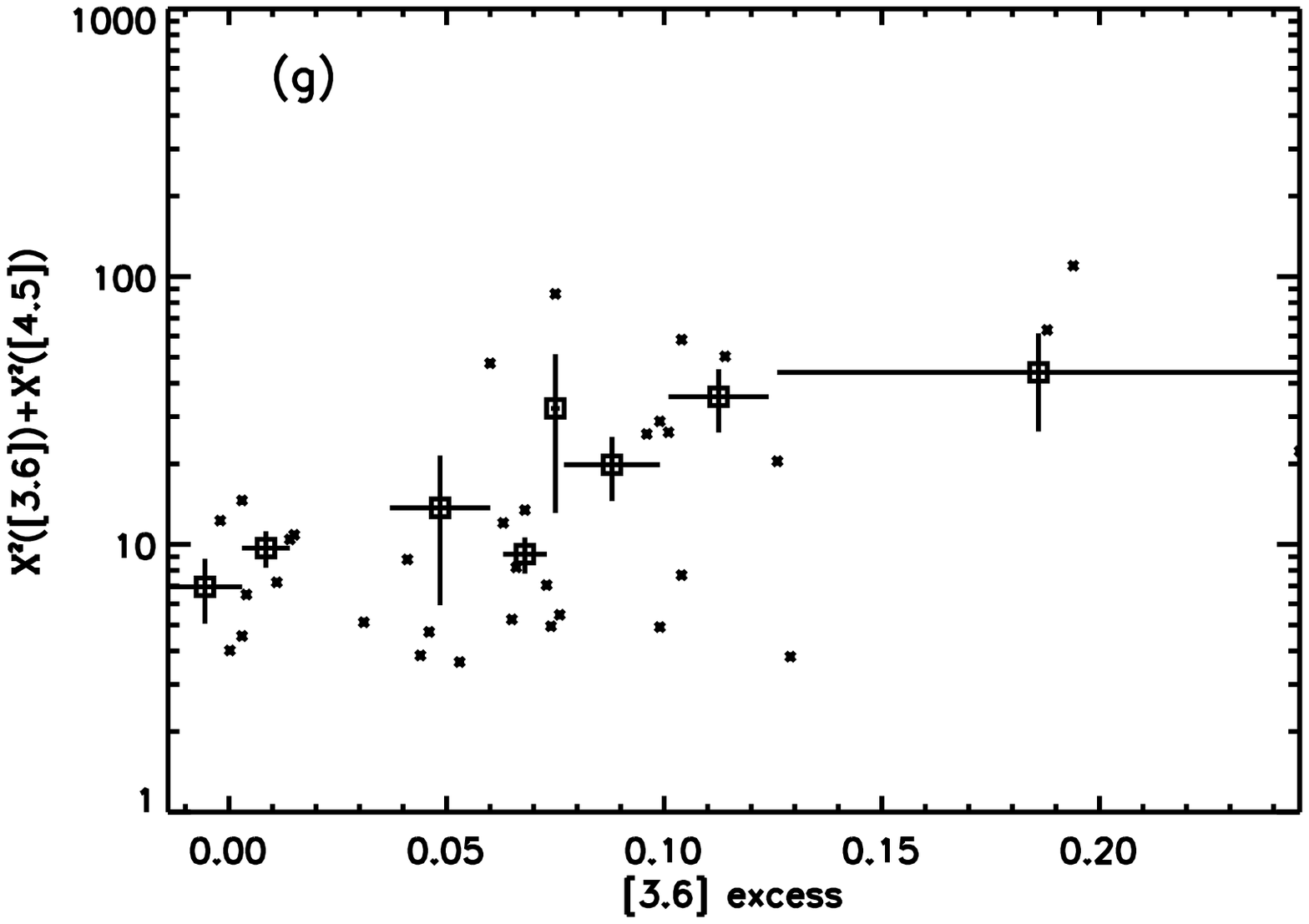}
\caption{(a) Variable fraction as a function of [3.6] excess (F(3.6)$_{obs}$/F(J)$_{obs}$-F(3.6)$_{phot}$/F(J)$_{phot}$) (126 members). (b) Cumulative fraction. (c) Mean peak to peak fluctuations for the irregular variables. (d) Size of the fluctuations (70 members). (e) Variable fraction split into class II (63 members) and class III (30 members) sources. (f) Cumulative fraction of irregular class II variables. (g) Size of the fluctuations among class II stars. While there is an increase in f$_{var}$ and $\chi^2_{\nu}$ with [3.6] excess, this trend is severely diminished when correcting for the trend with $\alpha_{IRAC}$. \label{var_excess}}
\end{figure*}

\begin{figure*}
\center
\includegraphics[scale=.32]{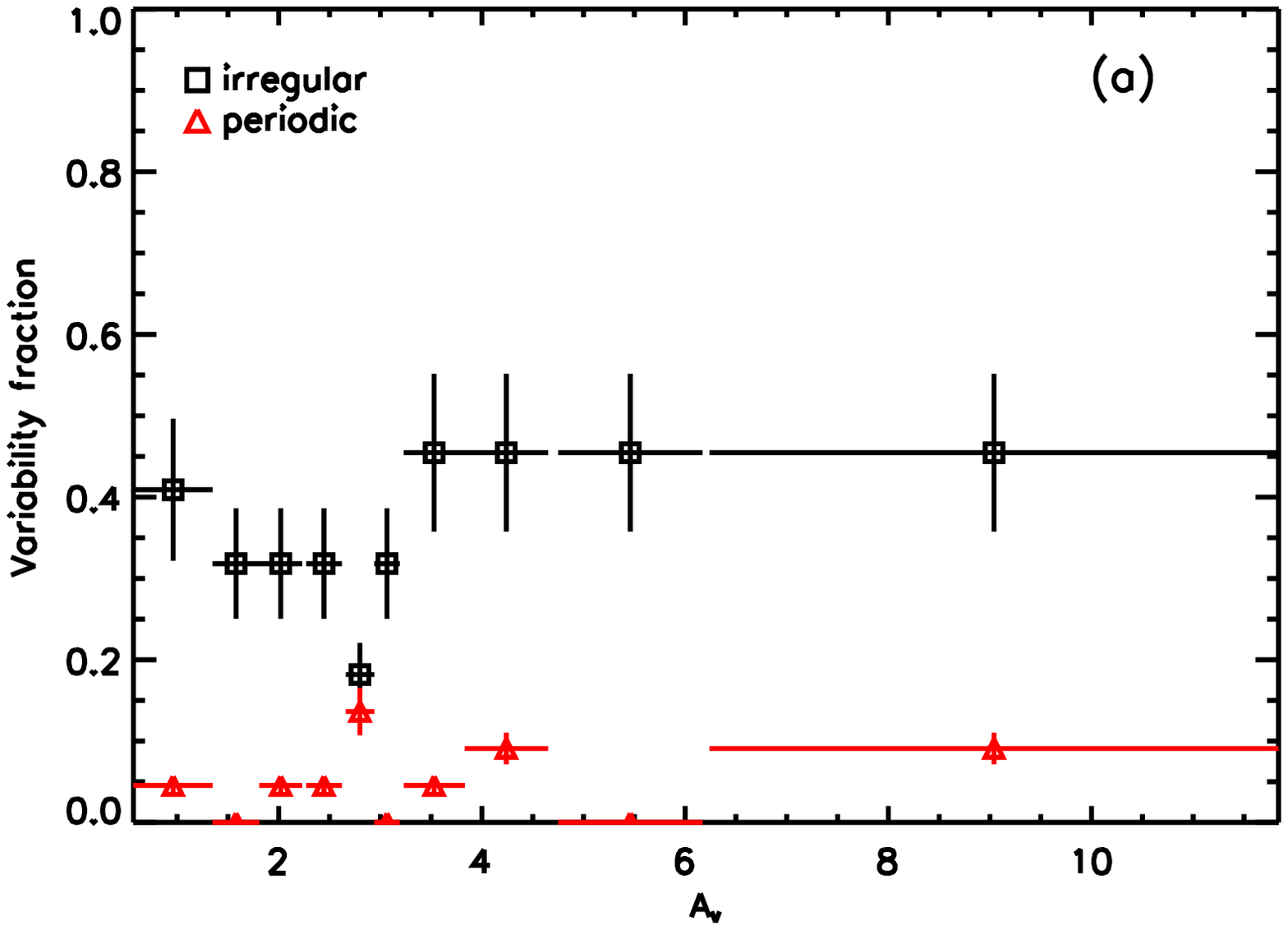}
\includegraphics[scale=.32]{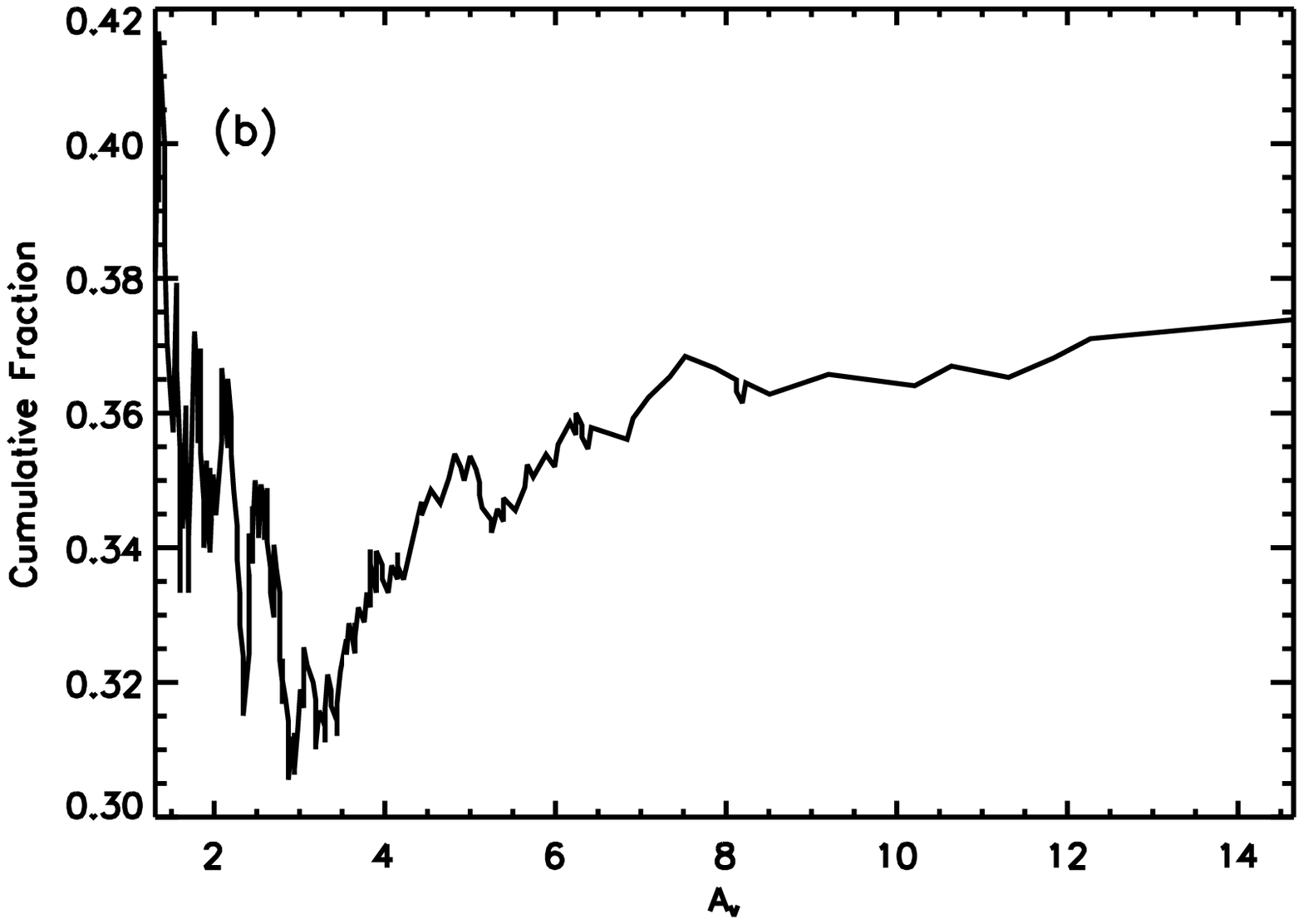}
\includegraphics[scale=.32]{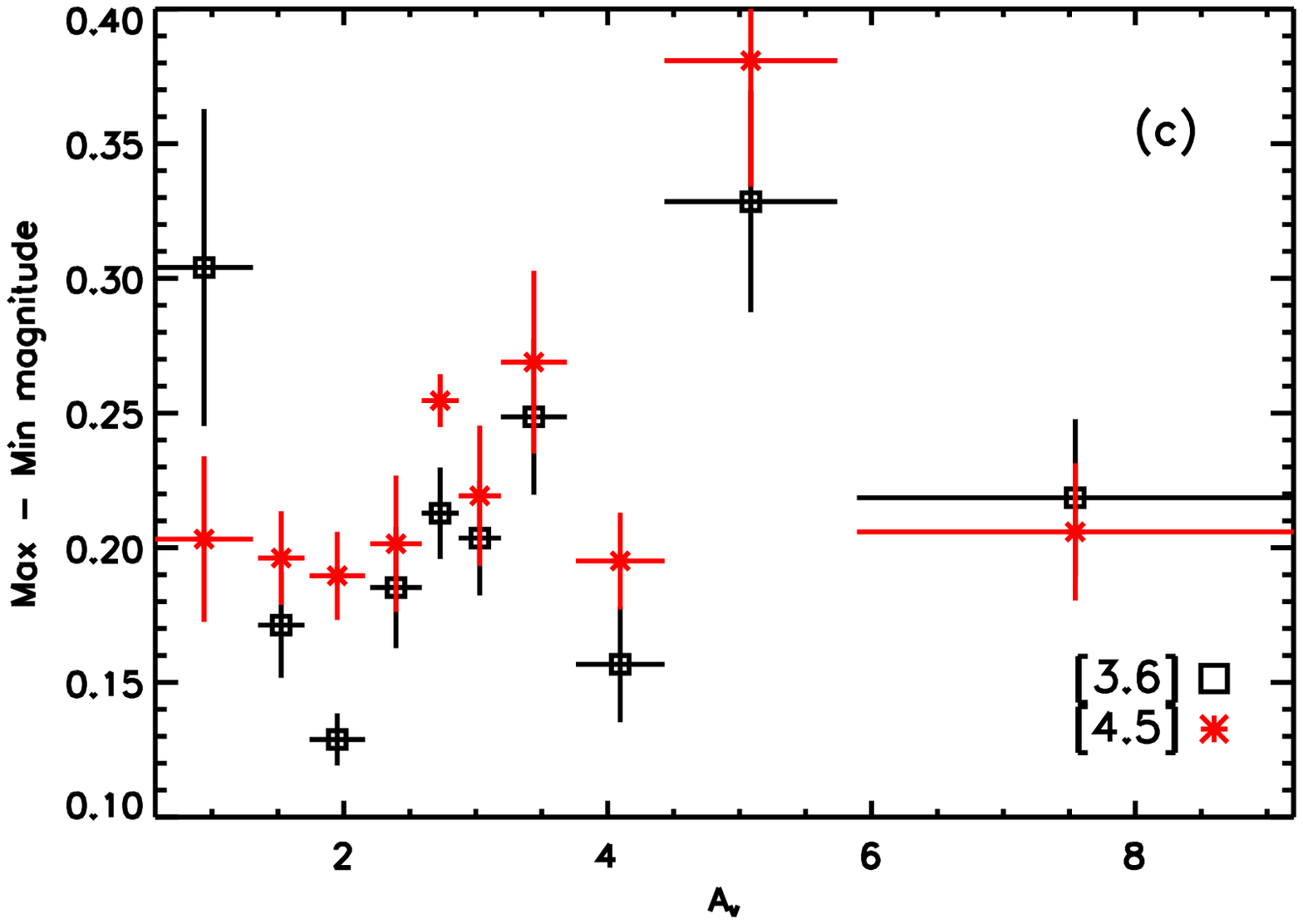}
\includegraphics[scale=.32]{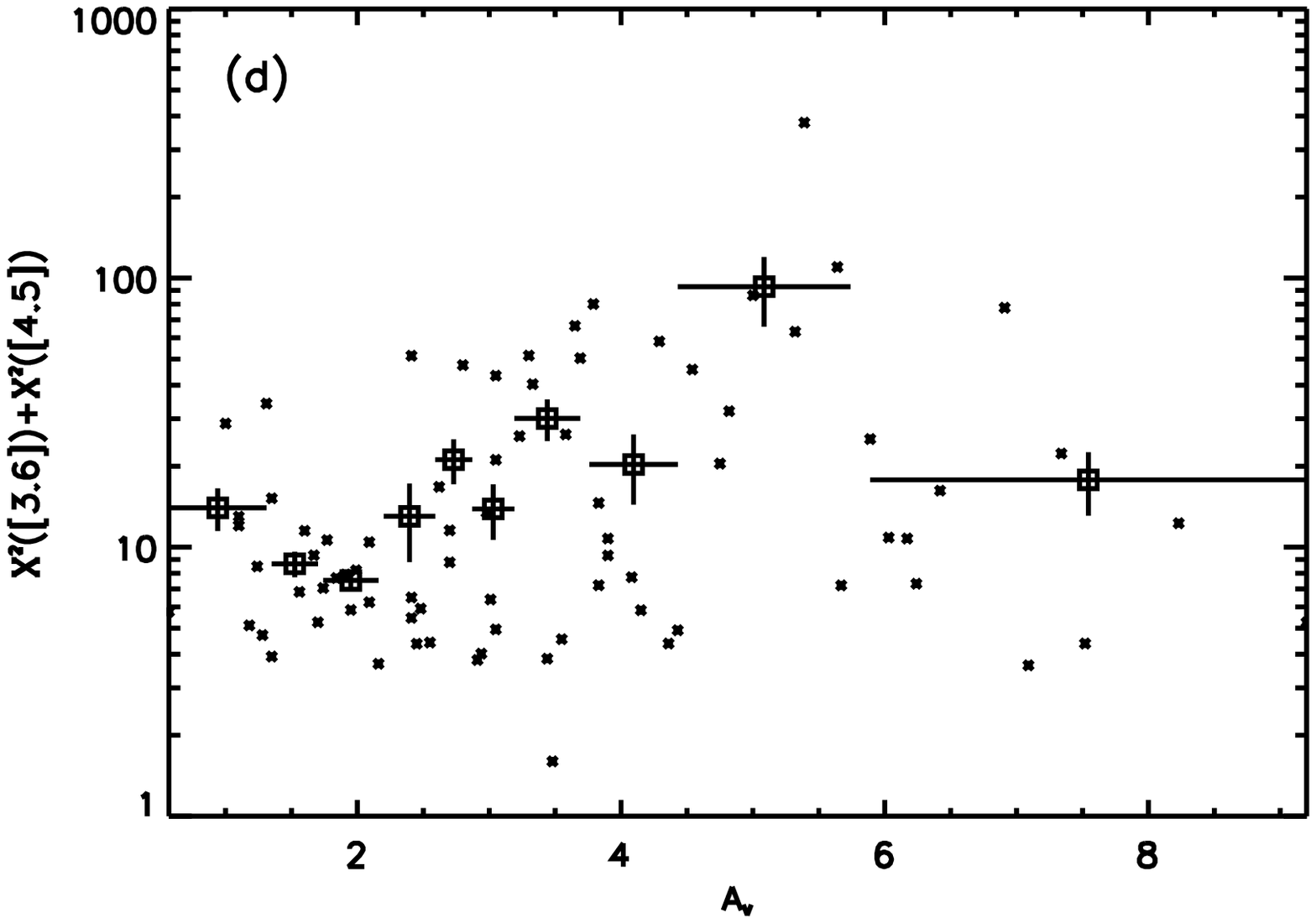}
\caption{(a) Variable fraction as a function of A$_V$ (222 members). (b) Cumulative fraction. (c) Mean peak to peak fluctuations for the irregular variables. (d) Size of the fluctuations (98 members). There is no strong trend between A$_V$ of a star and infrared variability. \label{var_av}}
\end{figure*}

\begin{figure*}
\center
\includegraphics[scale=.32]{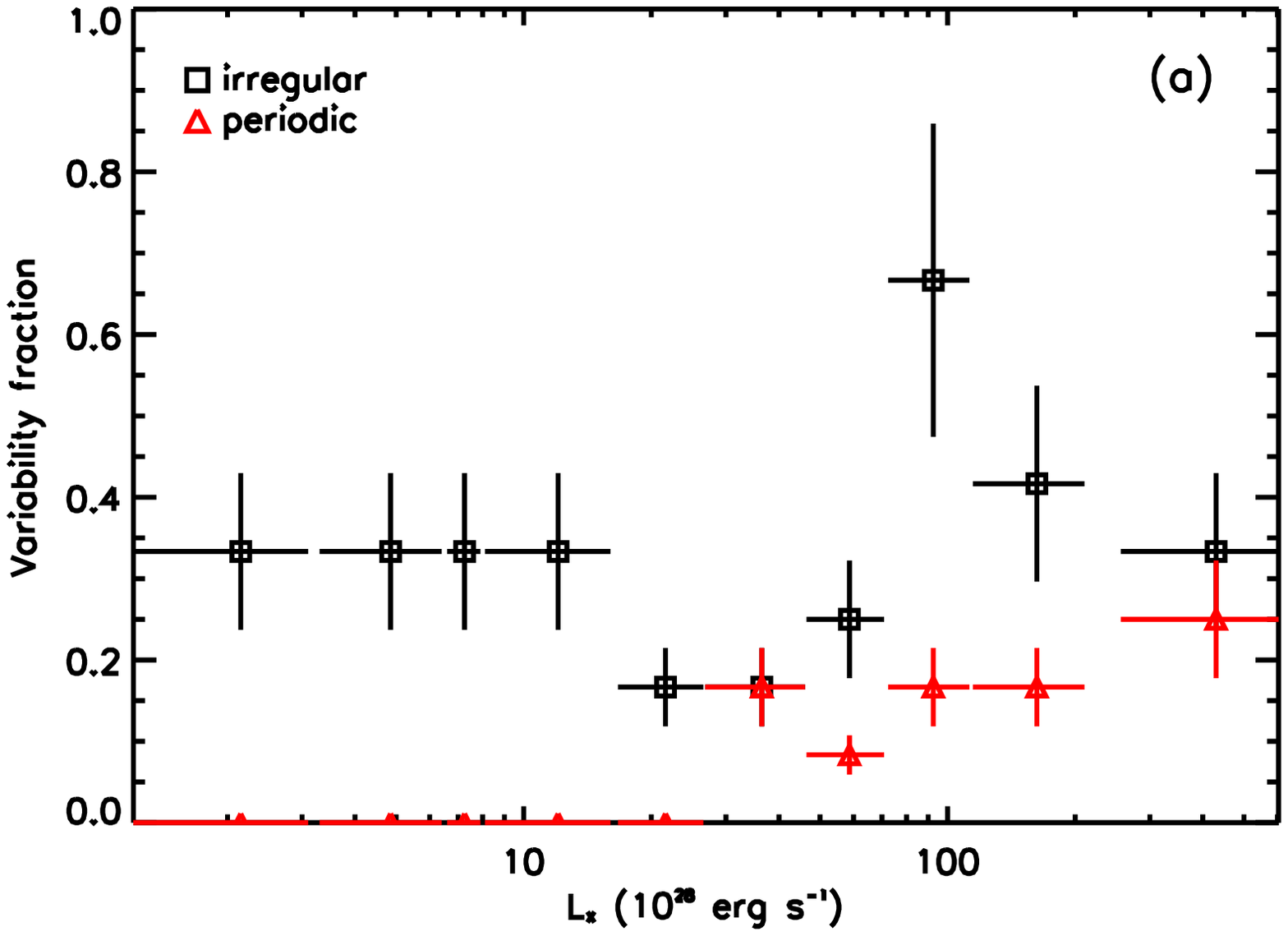}
\includegraphics[scale=.32]{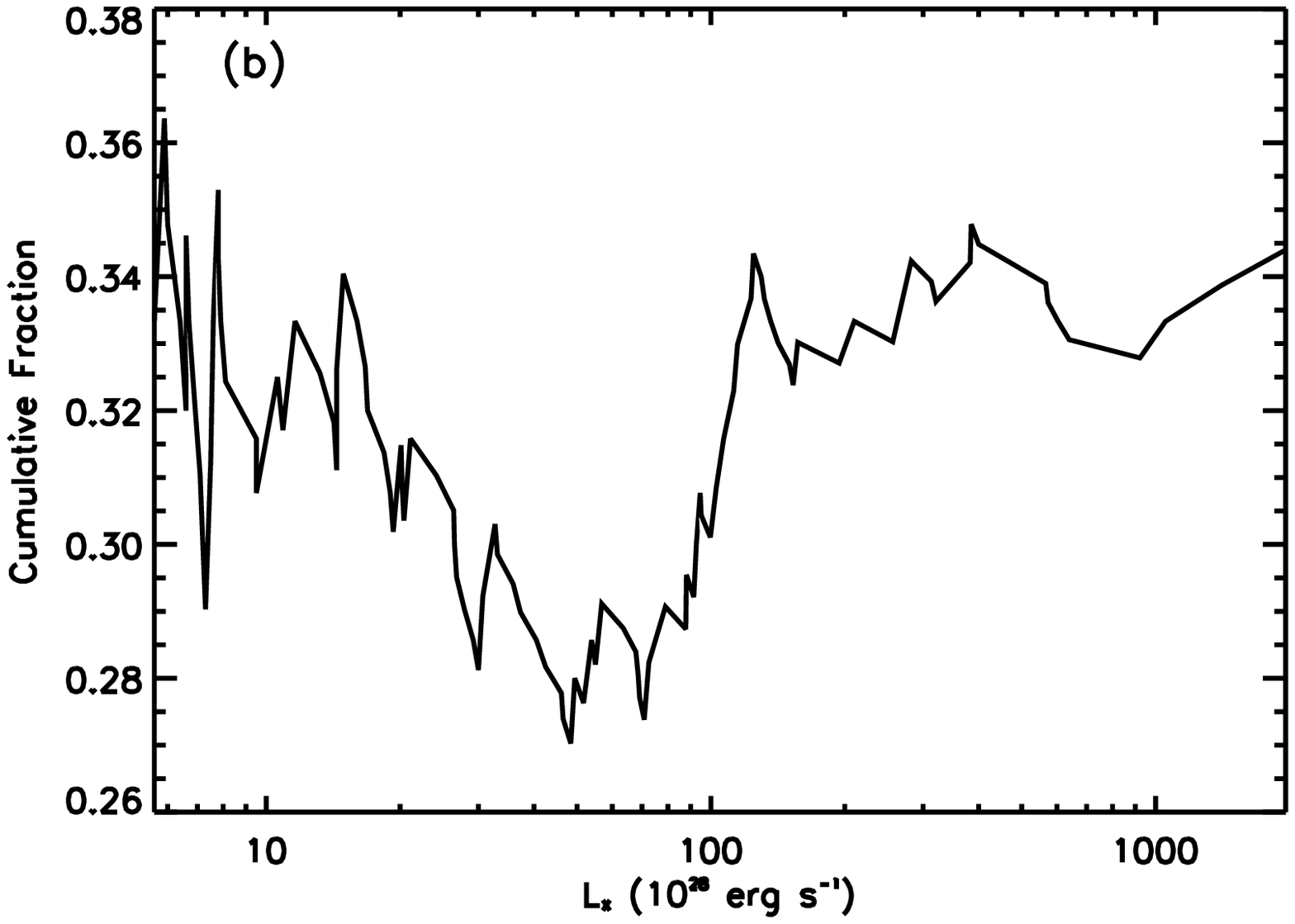}
\includegraphics[scale=.32]{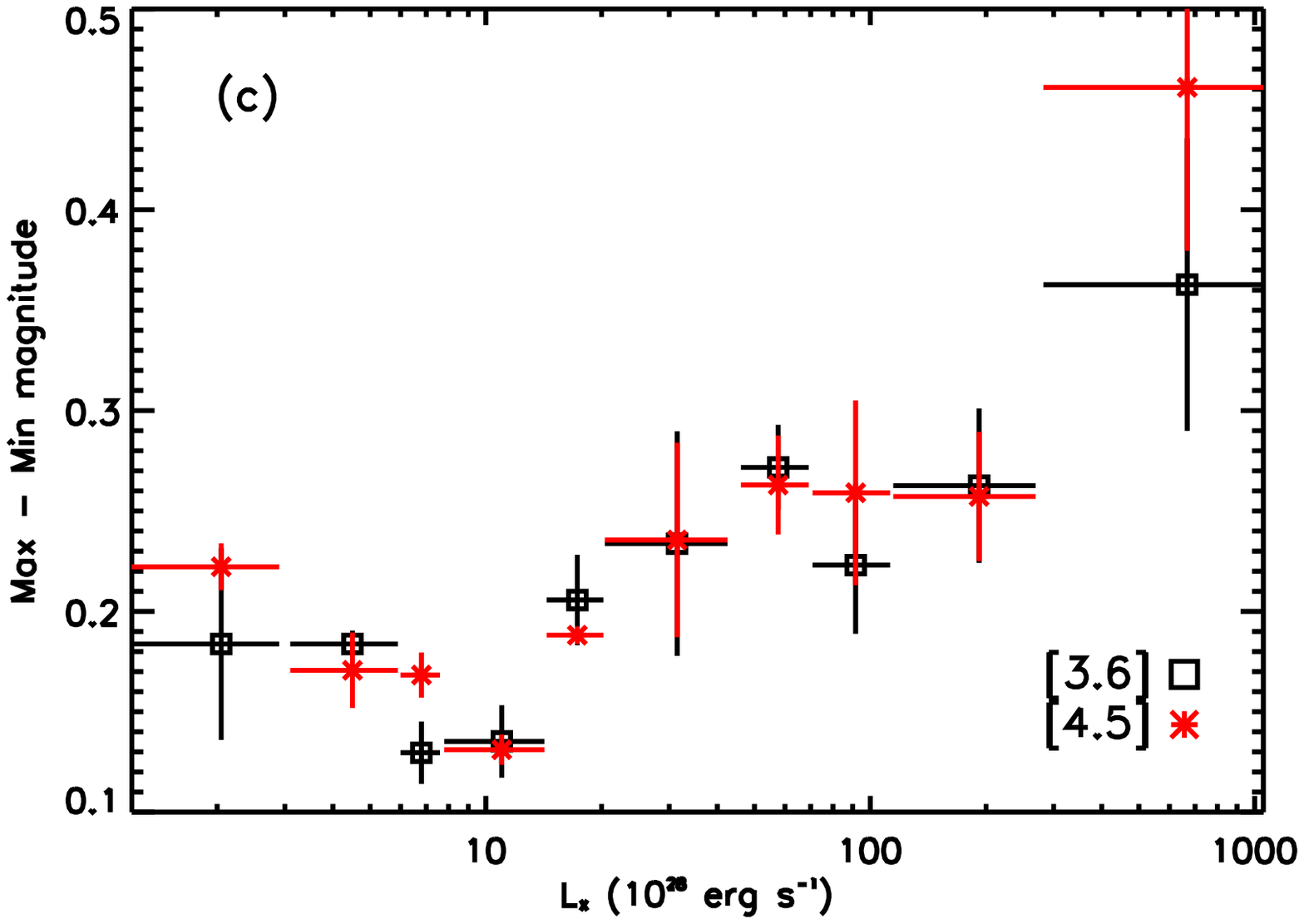}
\includegraphics[scale=.32]{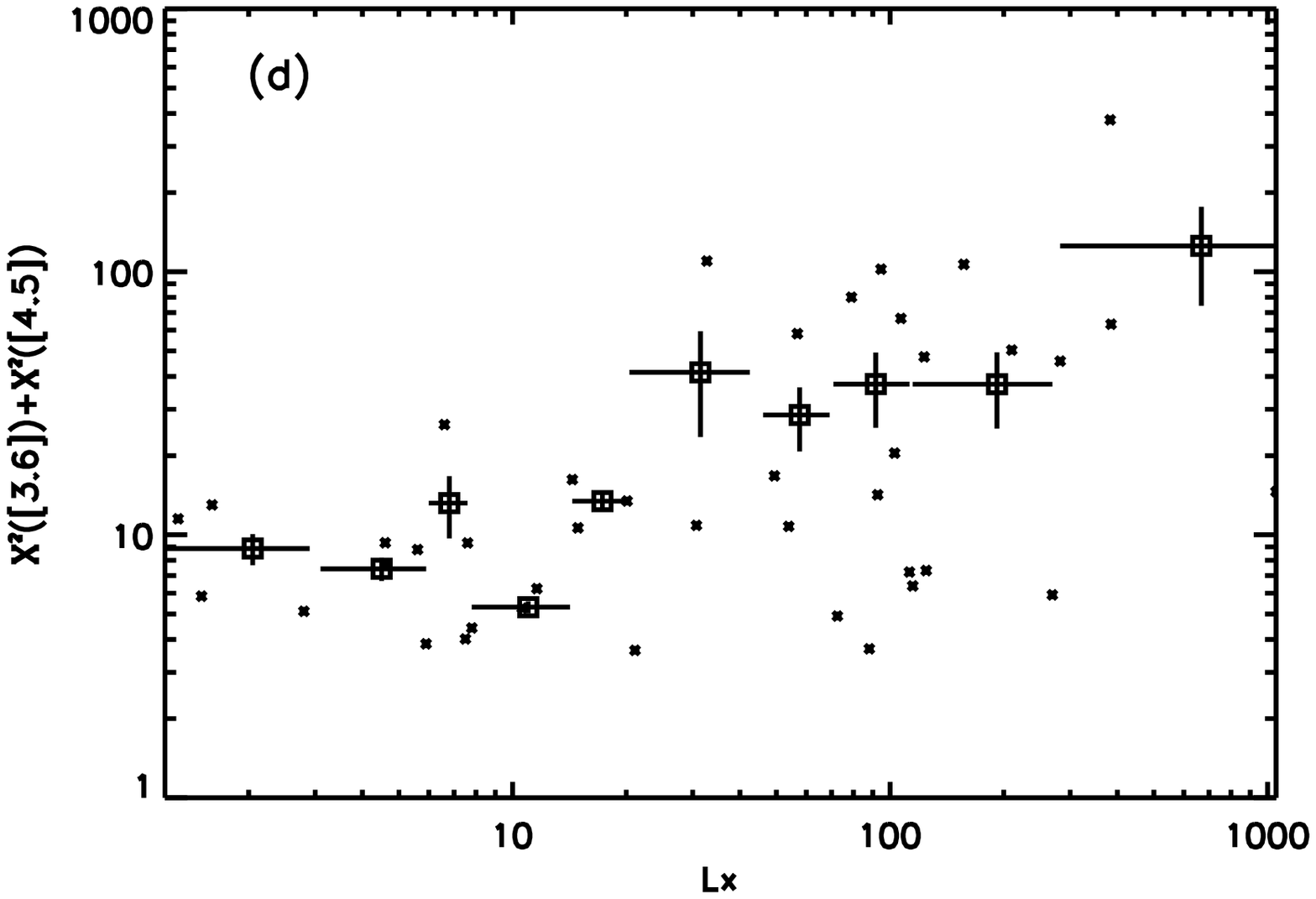}
\includegraphics[scale=.32]{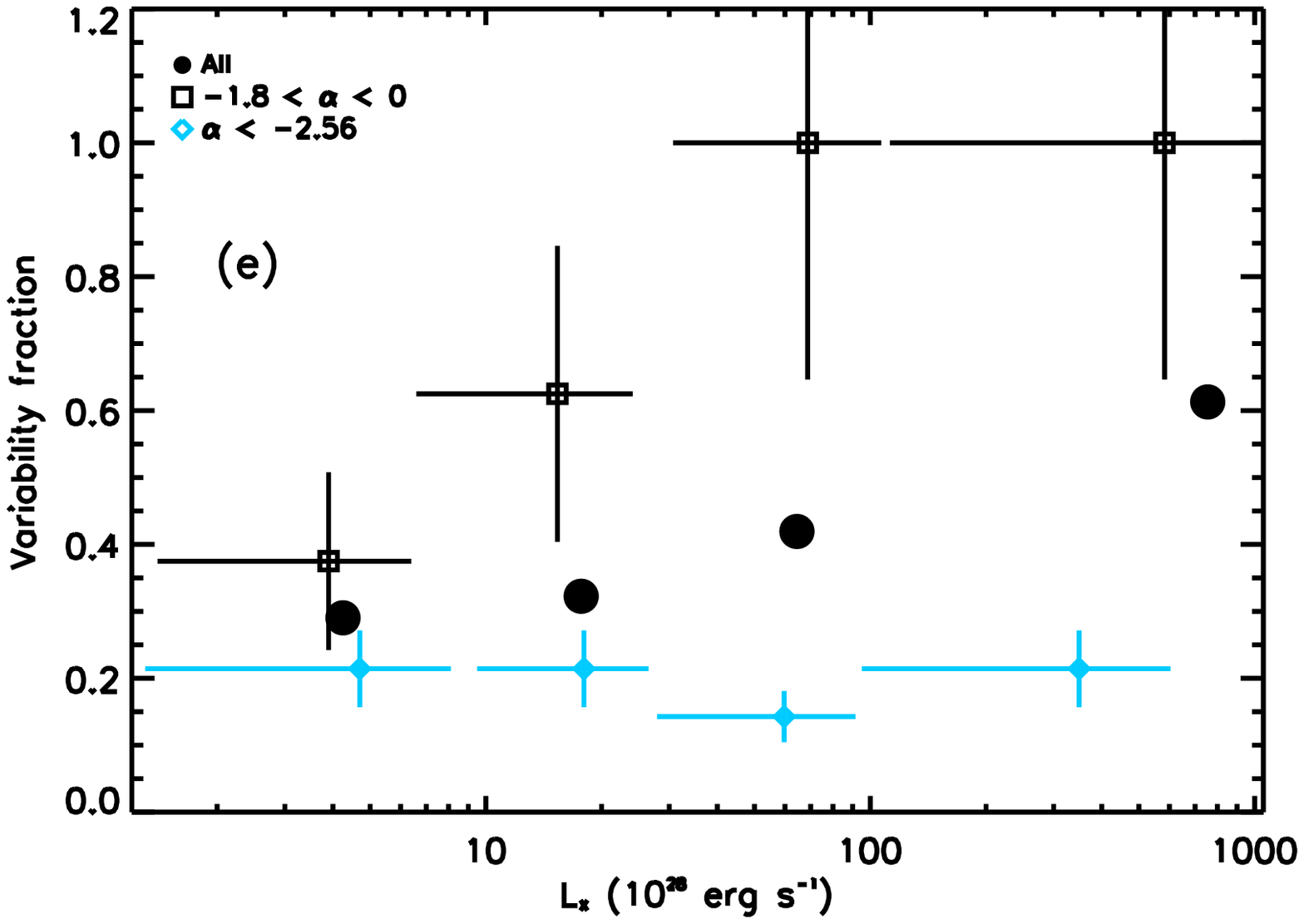}
\includegraphics[scale=.32]{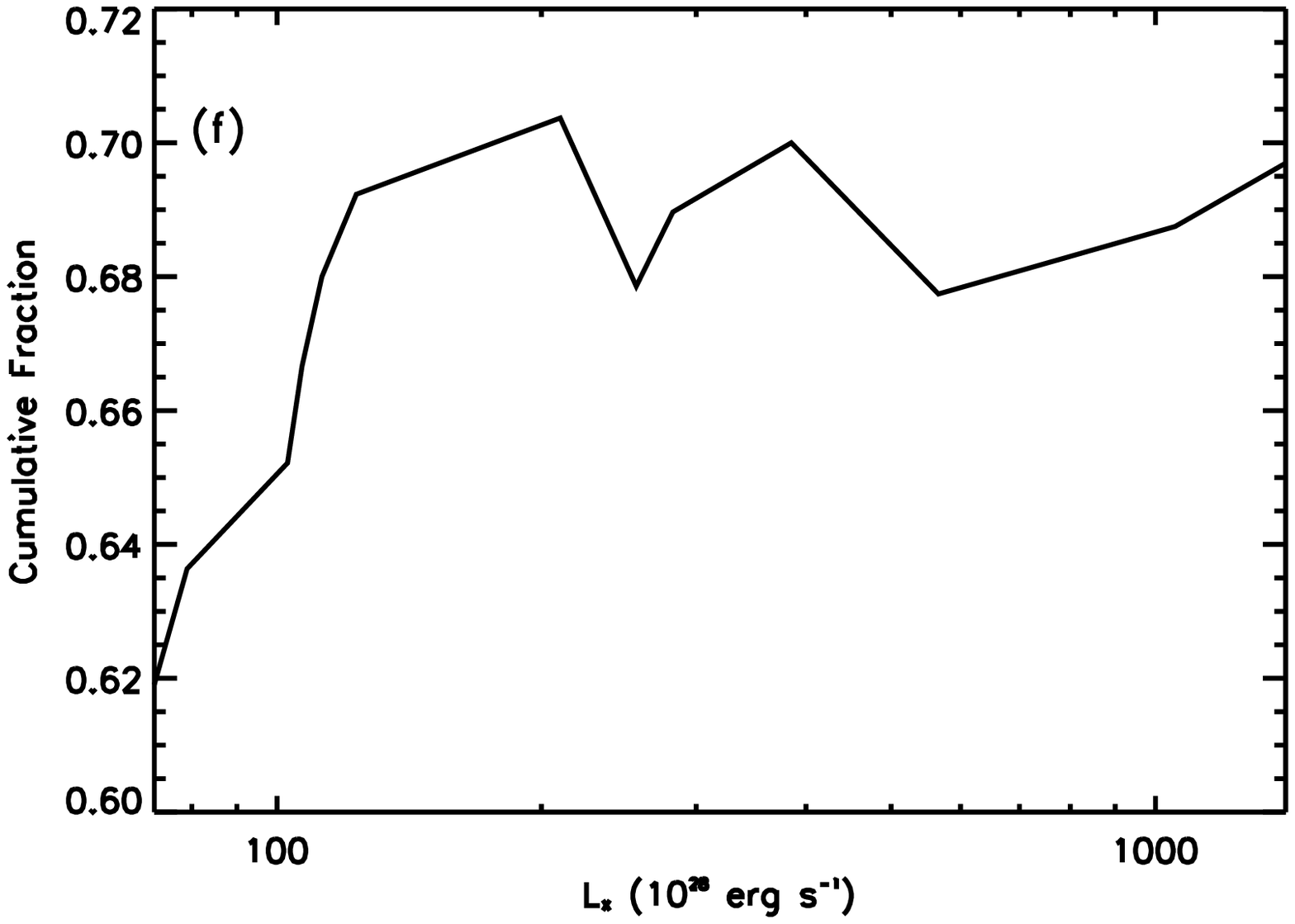}
\includegraphics[scale=.32]{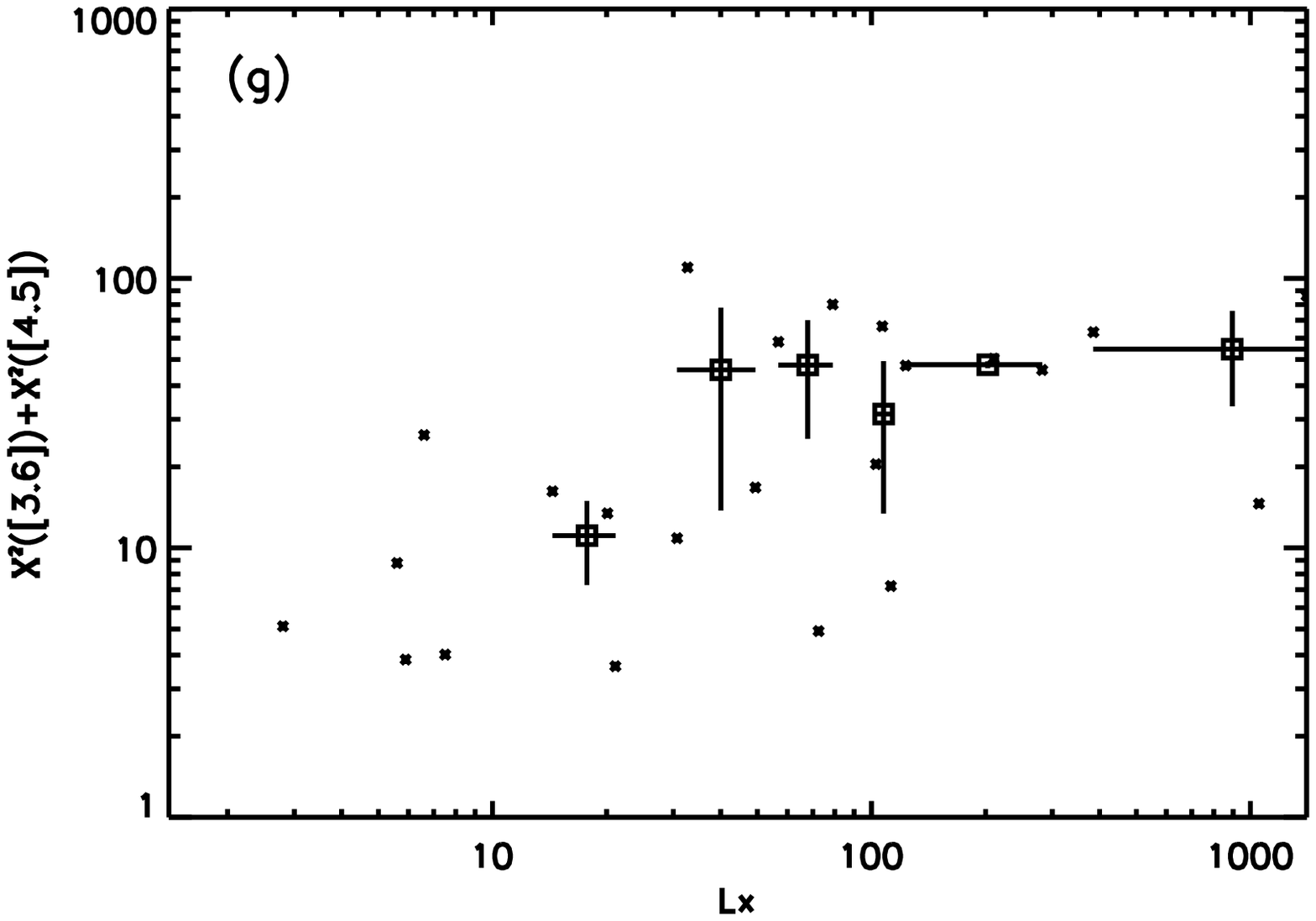}
\caption{(a) Variable fraction as a function of X-ray luminosity in units of 10$^{28}$ erg s$^{-1}$ (154 members). (b) Cumulative fraction. (c) Mean peak to peak fluctuations for the irregular variables. (d) Size of the fluctuations (58 members). (e) Variable fraction split into class II (33 members) and class III sources (60 members). (f) Cumulative fraction of irregular class II variables. (g) Size of the fluctuations among class II stars. When looking at all of the cluster members together there is no trend between f$_{var}$ and X-ray luminosity (first two panels) but when restricting the analysis to just the class II sources (middle panels) there is a strong trend between variability and X-ray luminosity. \label{var_lxray}}
\end{figure*}

This analysis can be extended to the other system parameters, but we must be careful to avoid degeneracies with the shape of the infrared SED. For example, the H$\alpha$ line strength broadly traces the accretion rate onto the star \citep{nat04} and will be correlated with the infrared excess. This may result in a statistically significant trend between variable fraction and H$\alpha$ EW that may simply be a reflection of the underlying trend with infrared SED shape. To avoid this degeneracy we can perform linear regression between the logit or $\chi^2_{\nu}$ and a linear combination of system parameters. This amounts to looking for a trend between H$\alpha$ EW and f$_{var}$ or $\chi^2_{\nu}$ at a given value of $\alpha_{IRAC}$. Using a linear combination as a simple statistical model to relate f$_{var}$ or $\chi^2_{\nu}$ and system parameters almost certainly underestimates the complex relationship between different parameters but it is useful as a first order approximation to determine the most important factors in the system. The F-statistic is then used to determine the significance of adding another variable to our statistical model. While we do consider sources across all SED classes when performing this statistical analysis, we only show the class II sources (-1.8$<\alpha_{IRAC}<$0) in the plots below (Figure~\ref{var_pos}-\ref{var_lxray}) for clarity. 

One possibility is that the correlation with H$\alpha$ EW, or some other parameter, is more fundamental than the correlation with SED shape and that we should instead look for a trend between $\alpha_{IRAC}$ and variability after controlling for the correlation with H$\alpha$ EW instead of the other way around. We have chosen to use the correlation with $\alpha_{IRAC}$ as the more fundamental trend because it is the strongest correlation in our data, it has been seen by other groups, and requiring an infrared detection puts fewer limits on the number of sources we can study than requiring e.g. an H$\alpha$ EW measurement. With the rest of the system parameters, we first examine if there is a significant correlation with variability, and then examine whether or not this trend persists after controlling for the underlying correlation with SED shape. 

Interestingly we find a significant trend between the position within the cluster and the variable fraction (Fig~\ref{var_pos}a,b) for the 249 cluster members with accurate positions and measurable variability. As before, f$_{var}$ is defined as the fraction of stars with $\chi^2_{\nu}>3$ in either [3.6] [4.5] or S$>0.45$, separating out stars with evidence for periodicity, in bins whose boundaries are set so that each bin contains the same number of cluster members. The variable fraction decreases from 70\%\ near HH 211 to 20\%\ on the far side of the cluster (Fig~\ref{var_pos}a). This trend is still significant (F=9.5) even when including the underlying correlation with $\alpha_{IRAC}$ (Figure~\ref{var_pos}d,e). The trend between f$_{var}$ and position is not due to a change in the relative fraction of class I/class II/class III between different regions of the cluster. There is a change in systems of a given SED shape that causes the variability to diminish from one side of the cluster to the other. Even when examining only the class II sources, f$_{var}$ drops from 70\%\ to 50\%\ across the cluster. This can be seen in Figure~\ref{map_fvar} where we show a map of the stars with disks ($\alpha_{IRAC}>-1.8$) within the cluster. In annuli that increase from 5 to 15 arcminutes from HH 211 the variable fraction decreases from 76\%\ down to 52\%. Over this SED range f$_{var}$ is fairly constant as a function of SED shape(Fig~\ref{var_airac}a), but there is still a correlation with position. There must be some change in the stars with disks that causes them to be less variable the farther they are from the dense, embedded region in the south-east. There is a hint of a trend between the size of the variability and the location within the cluster (Fig~\ref{var_pos}c), although this is not as significant.

We also find a significant correlation between the strength of the variability and the effective temperature of the star (among 235 cluster members) with higher T$_{eff}$ stars showing larger fluctuations (Figure~\ref{var_teff}c,d), even when taking into account the underlying trend with $\alpha_{IRAC}$ (F=7.7, Fig~\ref{var_teff}e). Stars earlier than K0 have fluctuations of $\sim0.3$ mag, while stars later than M2 have fluctuations of $\sim0.15$ mag (Fig~\ref{var_teff}c).There is no corresponding trend between the logit and T$_{eff}$ (Fig~\ref{var_teff}a,b). It is likely that the observed relation is an observational bias related to the fact that the fraction of stellar flux in the [3.6] and [4.5] bands increases as the effective temperature decreases. The increased fraction of stellar flux will dilute the signal from the disk, making it appear as though the fluctuations are smaller. We can demonstrate this effect by considering a stellar SED modeled with a single temperature blackbody underneath disk emission at T$_{dust}$=1500K. The disk is chosen to have a single temperature because the disk emission at [3.6] and [4.5] is dominated by flux from the inner wall at the dust destruction radius \citep{muz03}. The fiducial disk flux level is set so that r$_K$, the ratio of disk flux to stellar flux at K-band, is 0.4 and we vary the excess about this fiducial value for effective temperatures ranging from 8000 K (A7) down to 2700 K ($\sim$M6). In this simple model we implicitly assume that inner rim scale height does not depend on stellar mass. Figure~\ref{teff_ir_corr} shows that stars with smaller effective temperature have smaller variations in their [3.6] and [4.5] magnitude for a given change in the disk flux, consistent with the effect that we observe in the data. Because this trend is an observational bias, we do not account for it while searching for a physical cause of the variability.

For the other parameters that we consider (L$_*$, H$\alpha$ EW, [3.6] excess, A$_V$) we find no statistically significant evidence for a correlation with the logit or the $\chi^2_{\nu}$, especially after taking into account the underlying trend with $\alpha_{IRAC}$ (Figures~\ref{var_lstar},\ref{var_halpha},\ref{var_excess},\ref{var_av}). We find that, besides the trend between T$_{eff}$ and size of the fluctuations, variability does not depend strongly on the base stellar properties of T$_{eff}$ and L$_{*}$. This suggests that there is little dependence of infrared variability on stellar mass from G0 to M6 ($\sim2.5-0.1$M$_{\odot}$) The large scatter in f$_{var}$ and $\chi^2_{\nu}$ still leave some room for small trends with these parameters, but we can rule out large trends ($\Delta f_{var}>20\%$, $\Delta[3.6],[4.5]>0.2$). The main limitation in our ability to measure the dependence of variability on these parameters is due to uncertainties in their measurements, which can smear out any small trends. It may also be possible that observational biases, such as the inability to detect H$\alpha$ in deeply embedded class I sources, restrict our ability to detect any real trends in the data. For the rest of our analysis we focus on the trends strong enough for us to detect, but note that we cannot completely rule out correlations with the other parameters.

One potential bias in our analysis is the fact that the source of the variability is likely different between class II and class III sources. The variability in class II sources is most likely related to the structure of the disk and therefore might be correlated with different system parameters than the variability of the class III sources, which is most likely driven by star spots. Our method of regression between the logit or $\chi^2_{\nu}$ and a linear combination of system parameters includes cluster members of all SED shapes and a lack of a trend among class III sources may dilute a real trend among class II sources. To eliminate this bias we repeat the analysis performed above but only looking at the class II sources. We use the same definitions of f$_{var}$ and variability strength as before, we simply change the subset of stars that are examined. We find many of the same correlations as before; $\alpha_{IRAC}$ is correlated with both f$_{var}$ and the variability size, position is still strongly correlated with f$_{var}$ and T$_{eff}$ is still strongly correlated with the size of the fluctuations. Interestingly we find that f$_{var}$ is marginally correlated with X-ray luminosity and the strength of the variability is strongly correlated with X-ray luminosity (Figure~\ref{var_lxray}e,f,g). This was not the case when including the class III sources (Fig~\ref{var_lxray}a,b), indicating that X-ray luminosity is only related to the variability of the disk.  As the X-ray luminosity increases from 10$^{28}$ to 10$^{30}$ erg cm$^{-2}$ s$^{-1}$ the variable fraction increases from 40\%\ to 100\% and the size of the fluctuations increases from 0.15 to 0.4 mag. For the class III sources f$_{var}$ is roughly constant with X-ray luminosity. The same trend can be seen using X-ray fluxes from the recent analysis of \citet{ste12} who present a stack of X-ray data from multiple Chandra observations that is able to detect an additional 30 cluster members.

\subsubsection{Other Trends}

The extensive optical monitoring of this cluster allow us to compare the presence of optical variability, in particular irregular fluctuations, with the infrared variability. Of the 16 stars marked as irregular optical variables by \citet{coh04}, all of which have an infrared excess, 88\%\ are variable in the infrared, which is higher than when considering the entire sample. Of the stars that do not show significant irregular optical variability, 33\%\ of the strong disks are variable, 30\%\ of the evolved disks are variable and 21\%\ of the non-excess sources are variable in the infrared. For the strong disks the variable fraction is significantly smaller than the entire sample($\sim60\%$) indicating that the irregular optical variability and the infrared variability are strongly related.

\begin{figure}
\center
\includegraphics[scale=.45]{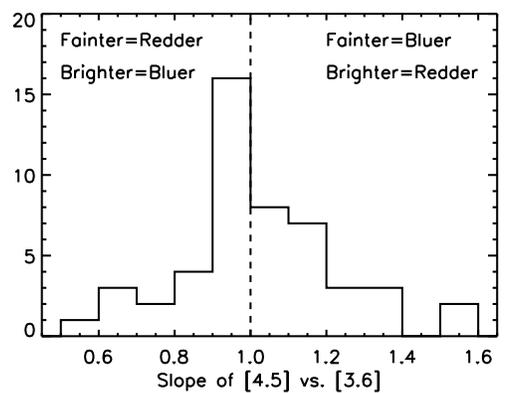}
\caption{Slope of [4.5] vs [3.6] for those stars where the uncertainty in the line fit to [4.5] vs [3.6] is less than 20\%. Most stars show no change in color, although a small handful either get redder as they get fainter or get bluer as they get fainter. \label{slopes}}
\end{figure}

We also find that the fluctuations at [3.6] and [4.5] are highly correlated. This is not surprising given that the disk flux at both wavelengths is dominated by emission from the puffed up wall at the dust sublimation radius. To quantify any variations in color we calculate the slope of the data in a plot of [3.6] vs. [4.5], taking into account the uncertainty in the measurements and any intrinsic scatter in the relation between [3.6] and [4.5] \citep{kel07}. Slopes greater than 1 corresponds to the star getting bluer as it gets fainter (or redder as it gets brighter) while a slope of less that 1 corresponds to the system getting redder as it gets fainter (or bluer as it get brighter). Figure~\ref{slopes} shows the slopes for the star with the most highly correlated fluxes. Of these stars, 76\%\ have a slope within 3$\sigma$ of 1, indicating no change in color, while the rest exhibit significant changes in color. Four sources (LRLL 21, 31, 67, 90; see Table~\ref{tab_cluster} for positions and other ancillary information) get redder as they get brighter, with slopes larger than 1. This is most likely due to the larger stellar flux in [3.6] versus [4.5] diluting some of the change in [3.6] due to the change in disk flux when compared to the change in [4.5]. The eight sources that get bluer as they get brighter (LRLL 13, 75, 245, 276, 435, 1889, 40182, 54361) all have a strong infrared excess ($\alpha_{IRAC}>-1.0$) and possibly have an envelope contributing scattered light to the [3.6] and [4.5] bands. If the variability consists of a short pulse, then as the light echo passes through the envelope the level of scattered light will increase, as is seen in LRLL 54361 (Muzerolle et al. submitted).

\subsection{Periodic Stars}
We detect 13 stars that show periodic behavior in their infrared variability, based on the SigSpec analysis of either [3.6] or [4.5] (Table~\ref{table_period}). Another 14 stars have weak evidence for periodicity based on SigSpec but are likely periodic because either the same period is derived in both [3.6] and [4.5] or the infrared period matches the optical period. That is, about 25\%\ of the variable stars are likely to be periodic; the fluctuations in the majority of cases are dominated by short-term aperiodic variations. When the [3.6] and [4.5] period do not agree we consider the period with the strongest significance as the actual period. Of the 13 certain periodic stars, the periods range from 2 to 22 days, as expected given our sensitivities. Eight of the stars are periodic in the optical, and of these five (LRLL 64, 69, 87, 116, 125; see Table~\ref{tab_cluster} for positions and other ancillary information) have infrared periods that match their optical periods suggesting that the same physical process, most likely a cool spot rotating across the star, is driving the infrared and optical variability. All of these stars have a weak excess  ($\alpha_{IRAC}<$-2.3), consistent with the [3.6] and [4.5] fluxes being dominated by photospheric emission as well as fluctuations small enough ($\sigma<.050$) to be consistent with cool spots. Of the three stars whose optical and infrared periods differ, LRLL 29 has an infrared period less than the optical period while LRLL 36 and 90 and infrared periods much longer than the optical period. LRLL 29 is a class III source ($\alpha_{IRAC}=-2.73$) with a small fluctuation ($\sigma=0.03$) and we are likely seeing a cool spot at a different rotation period than previously measured. For these last two sources, both have strong infrared excesses that would mask out stellar fluctuations and the variability is most likely due to a long-lived structure within the disk. The periods (14.7 days for LRLL 36 and 20 days for LRLL 90) corresponds to material at $\sim0.1$ AU from the central star. Of the other sources that do not show strong periodic fluctuations in the optical, LRLL 94 does not have an excess, and we are likely seeing rotating spots, while the other three sources (LRLL 15, 186, 435) all have a strong excess and the periodicity is likely due to rotating structure within the disk. The star LRLL 54361 is notable not only because it is a periodic class I source, but also because it is the strongest variable in our sample, with fluctuations up to 2.5 magnitudes in both [3.6] and [4.5]. We refer to Muzerolle et al. submitted for more details on the observations, and possible physical cause, of the variability in this source.

\begin{deluxetable*}{cccccc}
\tablewidth{0pt}
\tablecaption{Periodic Stars\label{table_period}}
\tablehead{\colhead{Star ID}&\colhead{[3.6] period}&\colhead{[3.6] significance}&\colhead{[4.5] period}&\colhead{[4.5] significance}&\colhead{Optical Period}}
\startdata
\cutinhead{Periodic stars}
LRLL 15 & 15.5 & 4.4 & 14.8 & 4.9 & not periodic\\
LRLL 29 & 2.2 & 2.8 & 2.2 & 5.4 & 10.8\\
LRLL 36 & 14.7 & 4.9 & 14.7 & 5.54 & 5.1\\
LRLL 64 & 8.3 & 6.1 & \ldots & \ldots & 8.4\\
LRLL 69 & 8.9 & 6.4 & 15.5 & 3.6 & 9.1\\
LRLL 87 & 12.9 & 5.9 & \ldots & \ldots & 14\\
LRLL 90 & 19.8 & 4.2 & 20.6 & 4.3 & 2.2\\
LRLL 94 & 5.6 & 4.4 & 5.5 & 4.9 & \ldots\\
LRLL 116 & 7.6 & 6.4 & 4.2 & 2.6 & 7.0\\
LRLL 125 & 2.5 & 2.3 & 8.4 & 5.6 & 8.0\\
LRLL 186 & 8.3 & 4.5 & 8.3 & 5.9 & not periodic\\
LRLL 435 & 13.9 & 5.4 & 13.6 & 4.3 & \ldots\\
LRLL 54361 & 21.5\tablenotemark{a} & 6.7 & 21.4\tablenotemark{a} & 6.7 & \ldots\\
\cutinhead{Likely Periodic stars}
LRLL 41 & 12.5 & 3.4 & 12.3 & 3.6 & 2.8\\
LRLL 47 & 4.9 & 2.7 & 4.9 & 2.9 & 4.857\\
LRLL 53 & 3.0 & 3.2 & 2.9 & 3.6 & 3.0\\
LRLL 60A & 3.7 & 4.1 & 3.8 & 2.2 & 6.340\\
LRLL 79 & 1.9 & 3.4 & 1.9 & 3.3 & 1.9\\
LRLL 85 & 2.7 & 4.4 & \ldots & \ldots & 2.7\\
LRLL 88 & 5.2 & 3.3 & 2.5 & 1.4 & 5.5\\
LRLL 93 & 4.3 & 3.5 & \ldots & \ldots & 4.5\\
LRLL 149 & 2.5 & 2.1 & 2.5 & 4.8 & 2.5\\
LRLL 178 & 6.8 & 3.4 & \ldots & \ldots & 6.9\\
LRLL 182 & 2.7 & 3.3 & 2.6 & 2.5 & 2.7\\
LRLL 228 & 4.8 & 4.0 & 4.8 & 3.6 & not periodic\\
LRLL 276 & 4.6 & 3.9 & 4.6 & 1.8 & \ldots\\
LRLL 40182 & 2.6 & 3.2 & 2.6 & 3.7 & \ldots\\
\enddata
\tablenotetext{a}{The period presented here is derived using only the fall 2009 Spitzer data. Analysis of all available infrared data indicates a period closer to 25 days (Muzerolle et al. submitted)}
\tablecomments{The significance corresponds to the log of the probability that the peak is due to random fluctuations. A value of 4 corresponds to a 1 in 10$^4$ chance that the peak in the SigSpec analysis is due to noise.}
\end{deluxetable*}

\subsection{'Dippers'}
\citet{mor11} identified a subsample of infrared variables in Orion that exhibit brief dips in their light curves. These flux drops of a few tenths of a magnitude are seen in both [3.6] and [4.5] as well as contemporaneous optical and near-infrared data, and last for typically 1-2 days. In some cases they are periodic, recurring on weekly timescales, but sometimes only one event is seen. The fact that the dips are more extreme at shorter wavelengths indicates that they are most likely due to extinction events in which the star is temporarily obscured by a clump of dust. The rapid timescale indicates that the clump of dust resides in the disk, close to the inner edge. These extinction events are similar to those seen around AA Tau \citep{bou07} and the class of object known as UXors \citep{wat98}. 

We do not see any sources that we can confidently classify as dippers in our sample, but this may be due to our sparse observing cadence. We can simulate our sensitivity to dippers by randomly generating light curves with dips whose characteristics match those seen by \citet{mor11} and running these artificial light curves through our algorithms for detecting variables. Our simulated cluster has the same [3.6] and [4.5] magnitude distribution as the actual observations in order to match our measured sensitivities. We randomly select stars to be 'dippers' and assign them a dip with a strength between 0.05 and 0.2 magnitudes and a duration between 0.5 and 2 days. We repeat this simulation 1000 times for a given 'dipper' fraction and determine how many of these artificial clusters have more than one detected dipper. Our 99\%\ confidence limit corresponds to the dipper fraction for which fewer than 1\%\ of the artificial clusters have more than 1 dipper. In this way we can put an upper limit of 5\%\ on the dipper fraction. This is roughly consistent with the fraction derived by \citet{mor11}.

\section{What is happening to the disk to cause this variability?}

Among the IC 348 cluster members we find that many of the stars show infrared variability. This variability is more common among stars with a strong infrared excess, as well as stars with a large X-ray luminosity and stars located in the south-western ridge. Stars with irregular optical fluctuations are much more likely to be infrared variables than those without such fluctuations. Very few of the variables are periodic, or appear to have flux changes consistent with rapid extinction events. The [3.6] and [4.5] flux is highly correlated for the majority of stars, suggesting a common origin for their variability. We now discuss what could be causing this variability.  

\subsection{Variable Structure}
Our observations at 3.6 and 4.5\micron\ are dominated by emission from the puffed-up wall at the dust sublimation radius that marks the very inner edge of the dusty disk.  The flux from this ring is proportional to its emitting area times the blackbody function evaluated at the dust sublimation temperature. Determining the source of the variability reduces to a determination of why either the emitting area or temperature of the inner wall would change on day to week long timescales.

Since the dust in the inner wall is already at the sublimation temperature, any increase in temperature will cause the dust grains to evaporate. As an example, the timescale for a molten piece of magnesium-iron silicate to evaporate is given by:

\begin{equation}
t_{evap}=\frac{\rho_c R_c}{P_v(T)}\left(\frac{2\pi kT}{m}\right)^{1/2}
\end{equation}

\noindent where $P_v$ is the vapor pressure ($\approx 1$dyne cm$^{-2}$), $R_c$ is the radius of the molten droplet, $\rho_c$ is the density of the dust grain, $m$ is its mass and T is its temperature. For T$\approx$1700 K, $R_c\approx$1\micron, $\rho_c\approx$2 g cm$^{-3}$ the evaporation time is less than a few hours \citep{shu96}. The vapor pressure is in fact highly temperature dependent, increasing by two orders of magnitude from 1500 to 2000 K \citep{nut06}. This timescale is effectively instantaneous relative to our observing cadence, allowing us to assume that if the heating at a location on the disk changes it will rapidly equilibrate thermally. Assuming that the condensation timescale is similar to the evaporation timescale then as the heating of the disk decreases the inner disk will rapidly fill in smaller radii with dust until it reaches 1500 K.

That is, on our observing cadence the dust should always be near 1500 K, consistent with recent observations of the temperature of the inner rim as its strength fluctuates \citep{fla11,fla12}. This only leaves a change in emitting area as the possible source of the variability. We consider a simple scenario in which some fraction of the inner wall, $f_d$, undergoes a fractional change in its emitting area, quantified as $f_{\Omega}$. The old disk emission depends on the fiducial emitting area, while the new disk emission depends on both $f_d$ and $f_{\Omega}$.
\begin{eqnarray}
F_{old}\propto R_0 H_0B_{\lambda}(T_d)\\
F_{new}\propto((1-f_d)R_0 H_0+f_df_{\Omega}R_0 H_0)B_{\lambda}(T_d)\\
F_{new}\propto((1-f_d)+f_df_{\Omega})F_{old}
\end{eqnarray}
The parameters $R_0$ and $H_0$ are the fiducial radius and scale height of the inner disk while $B_{\lambda}(T_d)$ is the blackbody function evaluated at the dust sublimation temperature. The change in disk emission depends on both the fractional change level of emission and the amount of the disk that is subject to this perturbation. By focusing on changes in disk emission, rather than the absolute level of disk flux, we can ignore complicating factors such as the dust grain size and composition (which set the fiducial values of R$_0$ and H$_0$ \citep{dul10}). While $f_d$ and $f_{\Omega}$ are highly degenerate, we can vary $f_{\Omega}$ assuming a value for $f_d$ until the difference between F$_{new}$ and F$_{old}$ matches the observations, to put loose constraints on the typical size of the structural fluctuation. As noted earlier, the size of the [3.6] and [4.5] fluctuations depends on the effective temperature of the star, since the [3.6] and [4.5] fluxes arise partly from the photosphere; we account for this effect by using the known effective temperatures of the variable class II sources to simulate the observed fluctuations. For each star in our sample we start with blackbody emission at the stellar effective temperate plus a blackbody representing the dust at 1500 K with a flux ratio of the disk and star emission at K band of F$_{disk}$/F$_{star}$=0.4. We assume that $f_d$ and $f_{\Omega}$ are normally distributed about a mean value with some small dispersion and draw random values for each star from these distributions. New disk fluxes at 3.6 and 4.5\micron\ are calculated and added to the stellar flux at these wavelengths. The ratios between the old and new 3.6 and 4.5\micron\ stellar+disk fluxes are converted to magnitudes and compared to the observed distribution of max-min magnitudes in each band. This simulation allows us to estimate appropriate values of $f_{\Omega}$ for a given value of $f_d$. 

As expected, $f_d$ and $f_{\Omega}$ are highly anti-correlated. If the entire inner disk is subject to a structural perturbation ($f_d=1$) then $f_{\Omega}\sim1.25$, a 25\%\ increase in emitting area. This increase in emitting area could correspond to an increase in the radius of the inner disk, an increase in the scale height or some combination of the two. When only 1\%\ of the inner wall undergoes a structural variation ($f_d=0.01$), then $f_{\Omega}\sim$25. If the radius of the inner wall stayed fixed, then this would imply that the scale height of 1\%\ of the inner disk increases by over an order of magnitude. If this change in emitting area seems extreme, then instead a substantial portion of the disk must be undergoing a structural perturbation.  If only a few tens of percent of the inner wall fluctuates ($f_d=0.1-0.3$) then the emitting area must change by a factor of 2-3. 

If 'dippers' and UXors exhibit structural perturbations similar to those observed here then we can use their observations to help constrain f$_d$ and f$_{\Omega}$. The short-lived extinction events in these sources are likely due to clouds above the midplane near the inner edge of the disk. In our nomenclature one could say that f$_d$ of the disk has had its scale height increased by f$_{\Omega}$. The ratio of the duration of the extinction event to the period of the event corresponds to f$_d$. AA Tau \citep{bou07} exhibits an extinction event that occurs every 8 days and lasts $\sim$1.5 days corresponding to $f_d\sim0.2$. Based on the observations of \citet{mor11} typical extinction events last around 1 day with periods between 2 and 10 days, corresponding to f$_d\sim0.1-0.5$. 

One key distinction between the dippers and our observations is that the dippers can be explained by a long-lived structural perturbation that rotates with the rest of the disk while our observations require a rapid change in the structure. A non-axisymmetric perturbation, such as a warp, rotating with the disk produces infrared fluctuations on the order of only 1-2\%\ \citep{fla10}, much smaller than our observations. To explain the variations we see, f$_{\Omega}$ must represent how much either the radius or height of the inner disk changes on weekly to monthly timescales. The fact that the extinction events in AA Tau are seen in observations separated by many years \citep{bou07} indicates that the warp causing the extinction is fairly stable. However, not all of the dippers in Orion exhibit periodic behavior \citep{mor11}, suggesting that at least some of these structural perturbations are short-lived and may be related to the infrared variability. The dipper timescale then gives some idea of the size of the typical disk structure that might vary, although more observations are needed to directly connect these two phenomena.

One way to increase the emitting area is to increase the amount of dust in the inner disk. Increasing the surface density will increase the optical depth causing the $\tau=1$ surface, which defines the size of the emitting region, to occur at a larger height above the midplane. The accretion flow can increase the surface density by dragging dust in and out of the disk, and it does so on a viscous timescale. At 0.1 AU the viscous timescale is on the order of thousands to tens of thousands of years, much longer than the weekly fluctuations that we observe, making it unlikely that a change in the accretion flow is causing an observable change in surface density. A disk wind could also pull material out of the disk, decreasing the surface density, causing the disk to appear to shrink. Near the base of the wind, at the surface of the disk, the speed of the wind is sub-sonic \citep{bla82}, which implies a timescale to remove material that is longer than the thermal timescale. The thermal timescale is on the order of years to decades, implying that a wind could not remove dust fast enough to match the observed fluctuations. Given the long timescales for the accumulation/removal of dust by both accretion and a wind it is unlikely that we are observing a large change in dust mass. 

A number of other theories, besides a variable accretion flow through the disk and a variable wind, can produce fluctuations in disk structure on daily to weekly timescales \citep[see figure 12 in][]{fla11}. \citet{tur10} and more recently \citet{hir11}, find that in regions of the disk unstable to magneto-rotational instability (MRI) the disk magnetic field can become buoyant and lift out of the midplane. The magnetic fields drags dust along with it, causing rapid fluctuations in the scale height of the disk. \citet{ke12} model irradiation by X-ray flares and find that the the increased ionization caused by an X-ray flare will lead to dust being accelerated up along the stellar magnetic field lines away from the midplane. This would lead to changes in dust scale height that are directly correlated with large X-ray flares. If the coupling between the stellar magnetic field and the disk is moderate then the stellar magnetic field will periodically inflate, possibly leading to the creation of a warp in the disk \citep{goo99}. A companion circling the primary star on an orbit that is inclined with respect to the disk can create a warp as it drags material out of the midplane \citep{fra10}. Another possibility, which we consider in detail below, is variable illumination by a hot spot and the accompanying change in structure as it sublimates part of the inner wall.

\subsection{Variable Illumination from a Stellar Hot Spot}
Variable illumination from a hot spot on the surface of the star is one possible mechanism for explaining the observed infrared variability \citep{mor09}. As gas flows along the stellar magnetic fields lines it creates a shock as it strikes the surface of the star. The emission from this shocked-material is strongest at shorter wavelengths due to its high temperature and can be approximated with blackbody emission at 6000-10000 K \citep{cal98}. The increased luminosity from this hot spot is responsible for the large irregular optical fluctuations seen among actively accreting stars \citep{her94}. The lack of periodicity in the optical suggests that the distribution of spots on the stellar surface is asymmetric and that the spots do not live for multiple rotation periods. A rapid change in the covering fraction of these hot spots would lead to rapid variations in the illumination of the disk. Here we examine the effect of variable illumination on the disk structure to determine if it can plausibly explain our observations. 

We restrict ourselves to an examination of the emission from the directly illuminated puffed-up wall at the sublimation radius that defines the inner edge of the disk, since the 3.6 and 4.5\micron\ excess flux is dominated by emission from this region. As discussed earlier, if the inner wall is heated above 1500 K the dust will rapidly evaporate and the wall will push outward until it reaches a radius defined by the new sublimation radius. This sudden loss of dust over the newly illuminated region of the disk may have an effect on the dynamics of the system. While the gas dominates the mass, and hence the dynamics, the dust dominates the opacity and is responsible for setting the disk temperature and the gas may be affected by this loss of opacity. Without the dust, the gas in this azimuthal region of the disk will cool off and collapse to the midplane on a thermal timescale. The cooling timescale is roughly the thermal energy (=$nkT$) divided by the cooling rate (=$\Lambda$). For the temperatures in the inner disk, the cooling rate is $\Lambda/n_e n_p\approx10^{-26}$ erg cm$^3$ s$^{-1}$. Assuming the surface density is 1000 g cm$^{-2}$, the height of the disk is 1 AU, and the disk is made entirely of hydrogen molecules, we obtain n$\sim10^{13}$cm$^{-3}$. Assuming the ionization fraction is $\sim$10$^{-12}$, to convert from n$_h$ to n$_e$ gives a cooling timescale of tens of thousands of years. This is much longer than the lifetime of the hot spot, and as a result, the gas will not instantly collapse to the midplane when the dust disappears. If the condensation timescale is not substantially different from the evaporation timescale, then once the hot spot rotates out of view the dust will quickly fill in the disk back to the original sublimation radius.

The fact that the dust can sublimate fairly quickly and that the gas, which drives the dynamics of the disk, responds more slowly means that a hot spot rotating across the surface of the star would sweep across the inner wall like a lighthouse, creating a small region that has a higher sublimation radius than the rest of the wall. This kink in the inner wall (Fig~\ref{schematic}) could lead to a change in the strength of the infrared emission. Given that hot spots are not very long lived, the size of the kink will constantly fluctuate on timescales of days to weeks, similar to the infrared variability timescales.

The next question is whether the changes induced by variable illumination are large enough to produce the observed infrared fluctuations. \citet{mor09} decided that a hot spot alone could not produce large enough infrared variations. They had to include a warp in the disk that rotated in step with the hot spot to create fluctuations of $\sim0.1$ mag. Here we consider the effect of the changing sublimation radius in response to the hot spot to see if this structural perturbation is large enough to match the observations. As discussed earlier, the emission from the rim is proportional to its emitting area.
\begin{equation}
F_{rim}\propto R_{rim}H_{rim}B_{\nu}(T_{d})=\frac{H_{rim}}{R_{rim}}R^2_{rim}B_{\nu}(T_{d})
\end{equation}
 If we assume that H/R stays roughly constant as the disk is pushed outward, then the flux from the rim is proportional to R$^2$. The radius of the inner disk roughly scales as the square root of the stellar luminosity \citep{dul10}, and in turn the inner rim flux scales directly with the stellar luminosity. A 10,000 K spot covering 1\%\ of the stellar surface of a 4,000K star would produce stellar luminosity fluctuations close to 30\% and a corresponding 30\% increase in disk flux. This is roughly consistent with the size of the infrared fluctuations that we have observed.

We now estimate the distribution of infrared variability among the cluster members in the hot spot model. The observed changes in [3.6] and [4.5] will depend on the size of the hot spot and on the temperature of the star. A hot spot covering 1\%\ of a M dwarf will have a more significant effect on the luminosity striking the disk than if the same hot spot covered 1\%\ of a G star because of the larger temperature contrast. Also, an M star has a stellar effective temperature closer to the dust temperature, leading to smaller changes in [3.6] and [4.5] for a given change in disk flux.  Our estimate is based on the known distribution of effective temperatures for the cluster members, while also making some simple assumptions about the typical size of a hot spot, its temperature, how much of the disk is heated by the hot spot and the strength of the excess. We start with the effective temperatures of the variable stars with $-1.8<\alpha_{IRAC}<0$.  For the disk we use a 1500 K blackbody that has F$_{disk}$/F$_{star}$=0.4 at K-band, and sum the stellar and disk flux at 3.6 and 4.5\micron\ to determine the fiducial system flux. We randomly draw a hot spot size from a normal distribution with a given mean and standard deviation and add a spot of this size with T=8000 K to the stellar flux. We assume that only one-third of the disk is illuminated by the hot spot (f$_d$=1/3); the flux from the other two-thirds of the disk stays constant.The illuminated region of the disk has its flux increased by the same factor as the change in stellar luminosity for the given spot size and stellar effective temperature (f$_{\Omega}=\Delta$L$_*$/L$_*$). We then add this new dust flux to the new stellar flux at 3.6 and 4.5\micron\ and calculate the ratio of this new flux to the fiducial flux. We convert the flux ratios at [3.6] and [4.5] into magnitude differences and compare the distribution of these differences to the observations. We also compare our model with observed I-band fluctuations in the irregular variables in IC348 \citep{coh04}. Including the optical data, while not available for all of the infrared variables, helps constrain the size of the star spot independent of our assumptions about the disk. Matching to the observed fluctuations at all wavelengths allows us to determine if a consistent set of hot spot sizes can reproduce the observed distribution of fluctuations. Given that the size of each hot spot is random, many simulations are repeated in order to determine how often the distribution of magnitude fluctuations matches the observations. The uncertainty in the assumptions (ie. hot spot temperature, how much of the disk is heated by the hot spot, the exact response of disk flux to a change in stellar flux) prevent us from using this simulation to accurately constrain the hot spot sizes, but it can be used to check for plausible solutions. 

\begin{figure*}
\center
\includegraphics[scale=.7]{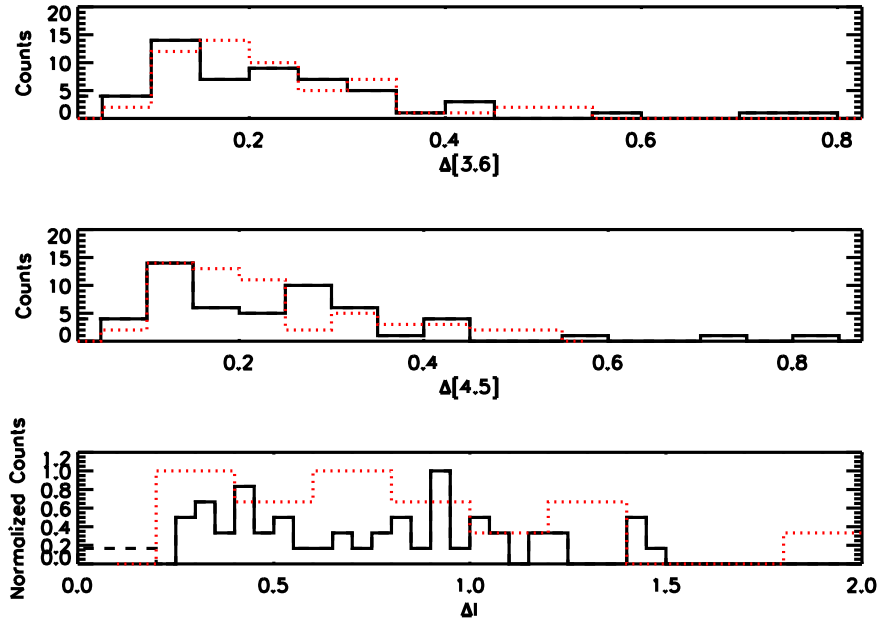}
\caption{Comparison of magnitude fluctuations derived from out 10,000 K hot spot model (black solid line) and the observed fluctuations for the variable class II sources (red dotted line). We show, from top to bottom, the max-min [3.6], [4.5] and I band magnitude. For the I band data the number of counts have been normalized since we use only the 15 stars observed as irregular variables by \citet{coh04}, but include the entire simulated sample ($\sim$50 stars). Dashed lines include the sources that would lie below our infrared detection limit. Adding 10,000 K hot spots with an average size of 2.5\%\ of the stellar surface, and accounting for the change in inner radius due to the additional heating from these hot spots, is able to reproduce the observed infrared fluctuations. Similar results are found when using an 8,000 K hot spot with an average size of 3\%\ of the stellar surface. \label{hot_spot_model}}
\end{figure*}

We find that hot spots that are 8000 K, with a mean size of 3$\pm$1\%\ of the stellar surface are able to reproduce the observed fluctuations in [3.6] and [4.5] as well as I band. We can also fit the observations with a 10,000K hot spot covering on average 2.3$\pm$0.7\%\ of the surface and heating 15\%\ of the disk (Fig~\ref{hot_spot_model}). Both of these scenarios produce infrared fluctuations of $\sim$0.2 mag as well as optical fluctuations of 0.5-1.5 mag, similar to the observations. This suggests that heating from a hot spot is a reasonable model for the fluctuations. The structural perturbations required by \citet{mor09} can naturally be explained by the increase in disk radius as dust rapidly sublimates in response to the increased illumination. We do assume that H/R stays constant as the inner edge of the disk is pushed outward, but this assumption can be relaxed if f$_d$ is allowed to vary. If H/R drops by a factor of 2, then f$_d$ needs to increase by a factor of 2 from 1/3 (15\%) to 2/3 (30\%) for an 8,000 (10,000) K hot spot to fit the observations. The change in H/R and f$_d$ are highly degenerate, preventing us from accurately constraining either of them. Once the material behind the inner wall becomes directly illuminated by the star it will heat up, but its gas pressure scale height will not change substantially because the thermal timescale is on the order of years and decades. The observed height will simply be the fiducial height. The region directly behind the inner wall lies in the shadow of the wall and will be heated by radial diffusion, which is difficult to model \citep{dul10}. Our assumption of constant H/R is most likely not accurate, but is simply used as a first order approximation of reality. The fact that these first order results are reasonable suggest that this model warrants further investigation, including a better treatment of the height of the disk.

\subsubsection{How well does the hot spot model fit the trends we see in our data?}

We have found that infrared variability is strongly correlated with large irregular optical fluctuations, position within the cluster and X-ray luminosity.  Here we examine how those correlations could be explained with illumination of the disk by variable hot spots.

{\it Optical Variability:} Large irregular optical fluctuations are directly attributed to the presence of hot spots on the stellar surface. The strong influence of hot spots on disk structure is consistent with our finding that almost 90\%\ of the class II sources with large irregular optical fluctuations are also variable in the infrared while only 30\%\ of the class II sources without irregular optical fluctuations are variable in the infrared. While the optical data were taken six years before our infrared data, it is plausible that the size of hot spot fluctuations does not change over the course of a few years. While the short-lifetime of hot spots may make it difficult to search for a direct correlation between optical and infrared fluctuations, contemporaneous observations may show a correlation between the size of the infrared and optical fluctuations.

{\it The South-West Region:} Along the southwestern edge of the main cluster is a region of higher density gas and dust, in which many of the most deeply embedded members of this cluster are located \citep{her08}. The main concentration of sub-mm sources \citep{jor08}, H$_2$ outflows \citep{eis03} and class I and class 0 objects \citep{mue07} are in this southwestern ridge, which has a total mass of $\sim10M_{\odot}$ \citep{taf06}. The presence of stars still embedded in massive infalling envelopes, and the young dynamical age of the outflows \citep{wal06}, indicate that this is a region of very recent star formation. The relative youth of this site may be the most relevant feature for the variability. IC 348 is not home to any massive O stars that could be producing strong photo-ionizing radiation and we do not expect that the ambient radiation field will vary strongly across the cluster. While infrared classes are broadly associated with the age of a source, it is possible that within a single SED class there is a spread in ages. The class II sources in the south-western ridge may be younger that the class II sources in the more exposed cluster center. Their relative youth may be associated with higher average accretion rates. Accretion rates decrease as t$^{-2}$ \citep{har98} and from 1 to 3 Myr, a plausible age spread between regions in IC 348, the average accretion rate will decrease by a factor of six. \citet{cal98} find that the excess from the hot spots around stars with different accretion rate can be explained by simply changing the covering area of the hot spot, with little change in the density and temperature within the spot. Stars with higher accretion rates have more of their surface covered by hot spots. Simulations find that the flow of gas onto the star is highly variable with more severe instabilities for higher accretion rates \citep{kul08} and it is possible that a larger average hot spot size, due to a higher accretion rates, would be associated with larger fluctuations in their coverage. In this way, the higher accretion rate of the younger sources in the southwestern ridge may be associated with larger more variable spots that in turn lead to larger disk structural fluctuations. 

The importance of age would imply that class I sources ($\alpha_{IRAC}>0$) should be more variable that class II sources ($-1.8<\alpha_{IRAC}<0$) since class I sources represent an earlier stage of stellar evolution. In reality we do not see a trend of increasing f$_{var}$ from class II to class I sources but this may partly be due to the fact that $\alpha_{IRAC}$ is an imperfect measure of evolutionary stage. Some nearly edge-on class II sources can appear to have a sharply rising infrared SED, while nearly pole-on class I sources have a spectral shape similar to class II sources \citep{rob07}. This mixing of evolutionary stages will dilute any trend between f$_{var}$ and SED shape among the youngest sources. It is also possible that our limitation of only considering sources with [3.6]$<$14 in our statistical analysis preferentially selects older class I sources with smaller envelopes, and hence less extinction of the central disk emitting at 3\micron, further diluting our results.

{\it X-ray Luminosity:} The connection between X-ray emission and accretion has been difficult to establish, producing conflicting results. Actively accreting stars are known to be on average less X-ray luminous than stars that have no strong ongoing accretion \citep{fei07}. In some stars the soft X-ray component has been associated with low temperature, high-density material in the accretion shock, while the hard X-ray flux arises from hotter, low-density gas in the corona \citep{arg11}.  \citet{sta07} find that optical variability is strongly correlated with X-ray luminosity, which they suggest may be due to larger magnetic topology giving rise to both stronger X-ray emission and larger spots. In terms of our analysis, this implies that stars that are more X-ray luminous have larger spots, giving rise to larger fluctuations in disk illumination and disk structure. The lack of correlation among the class III sources is simply due to the fact that these stars do not have any dust that can reflect the varying spot structure.

\subsubsection{How well does the hot spot model fit other literature results?}

In our analysis, we restricted ourselves to the puffed-up inner rim, but an extended radial region of the disk will be heated to a higher than normal temperature by the hot spot. Longer wavelength flux arising from large radii would rise in concert with the short-wavelength flux observed here, as has been seen in infrared spectroscopy \citep{kos12} and photometry (Megeath et al. accepted) of class II sources. Both of these studies found that for the majority of class II sources the increase in 3-10\micron\ flux was relatively constant with wavelength. These wavelengths are sensitive to the surface layers of the disk, which respond almost instantaneously to an increase in illumination \citep{chi97}, allowing for the disk flux to increase as the illumination increases. 

\begin{figure*}
\center
\includegraphics[scale=.4]{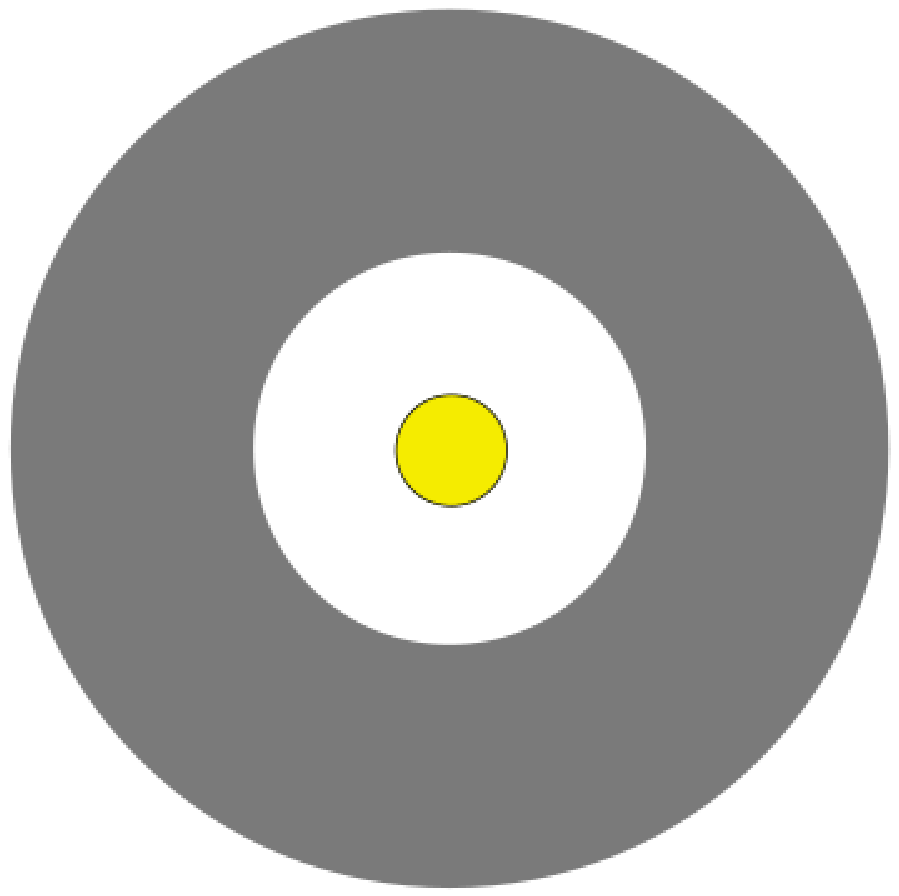}
\includegraphics[scale=.4]{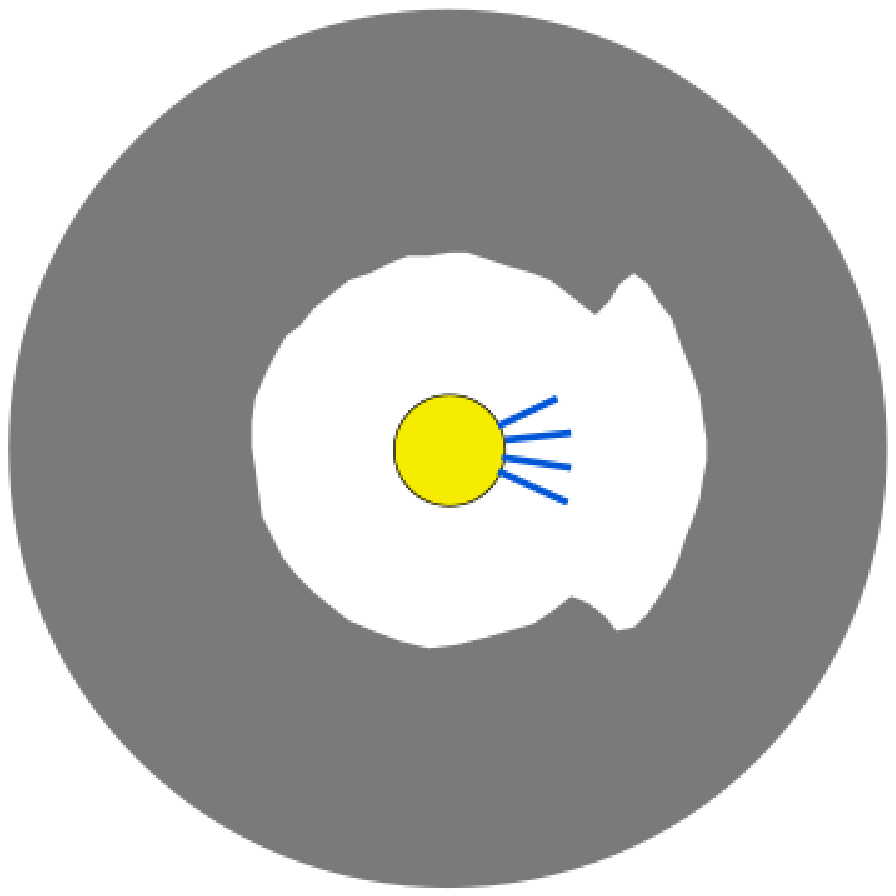}
\includegraphics[scale=.4]{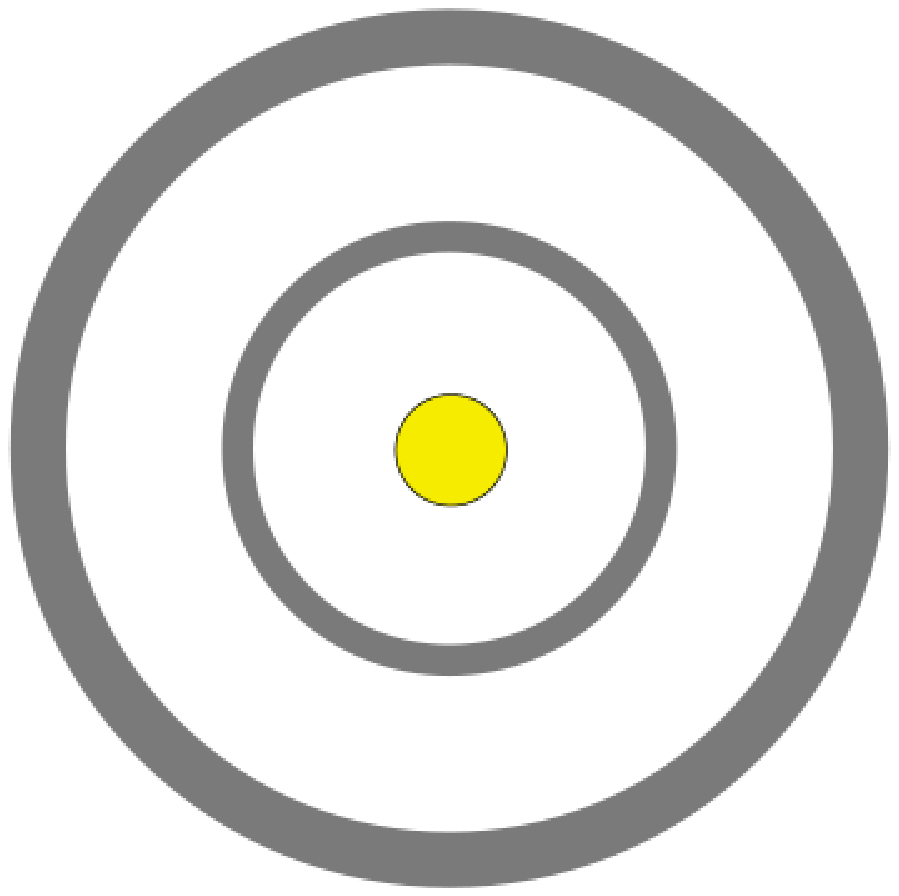}
\includegraphics[scale=.4]{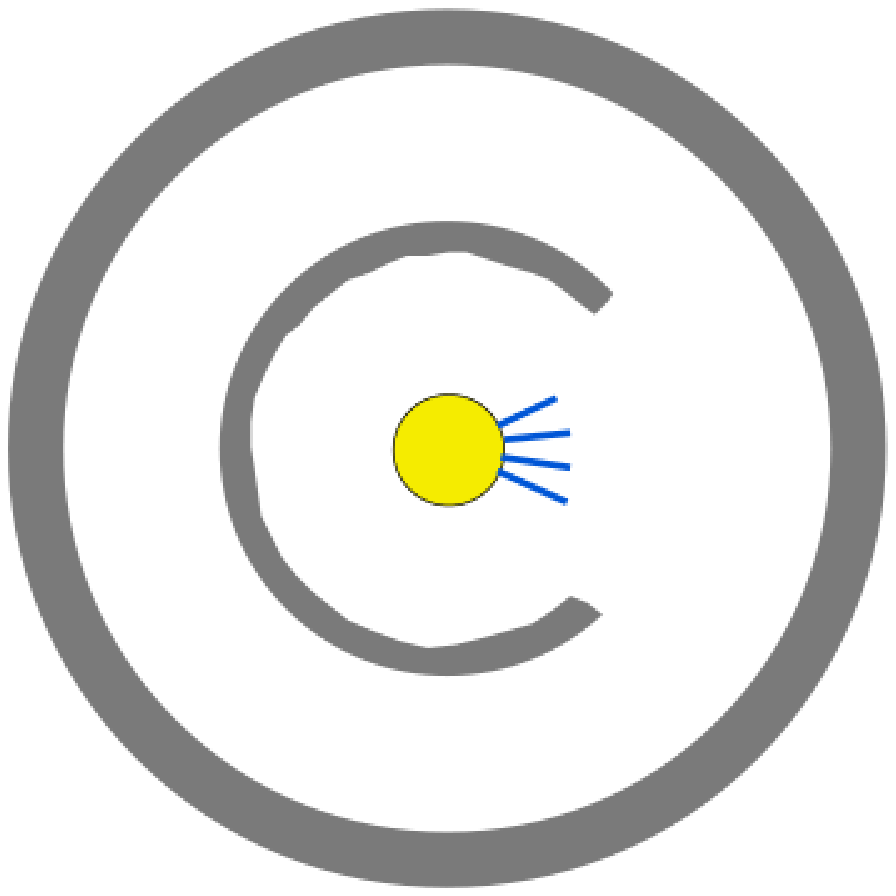}
\caption{Schematic diagrams illustrating how the presence of a hot spot can lead to large infrared fluctuations. When a hot spot appears on the surface of the star it heats up the disk, causing the dust to rapidly sublimate, effectively pushing out the inner edge of the disk (left and middle left). This increase in emitting area causes an increase in infrared flux. In the case of a transition disk, in which there is very little dust behind the wall, instead of creating a dent the hot spot punches a hole in the inner disk (middle right and right). This decreases the short-wavelength flux, while also directly illuminating the outer disk, which increases the long-wavelength flux. Given the small observed accretion luminosities among evolved disks, the inner wall must be very thin, but otherwise it is consistent with the 'seesaw' behavior that is often observed in transition disks. \label{schematic}}
\end{figure*}

This wavelength dependence contrasts with many transition disks that have infrared fluctuations in which the short-wavelength ($\lambda<8\micron$) flux decreases while the long-wavelength ($\lambda>8\micron$) flux increases, and vice-versa \citep{muz09,esp11,fla12}. This 'seesaw' behavior has been explained by changing the scale height of the puffed-up inner wall; as the scale height increases the short-wavelength flux emitted by this part of the disk increases while the long-wavelength flux decreases as the outer disk becomes more shadowed by the inner disk. This model has been successful at matching the observations but it may not be the only explanation. The observations themselves do not necessarily require a change in the scale height; they only require that the covering fraction of the inner disk as seen from the outer disk changes with time. In an axisymmetric model the only way to accomplish this is to change the scale height of the inner disk. Another possibility is that the scale height stays constant, but the azimuthal extent of the inner wall changes. Instead of being a solid ring wrapped around the star, the wall may be a crescent whose size changes with time. In the context of our hot spot model, this could happen as an accretion spot heats and sublimates a portion of the inner wall. If the wall is radially thin and there is no dust behind it, a geometry that is consistent with the SED shapes \citep{esp10}, then instead of putting a dent in the wall the increased luminosity from the hot spot will punch a hole through it (Fig~\ref{schematic}). The outer disk will become directly illuminated, increasing the long wavelength flux while the short wavelength flux decreases because the emitting area of the inner wall has decreased. In this way illumination from variable accretion spots can explain the wavelength dependence of both class II and transition disk sources.

Observations of the accretion luminosity can put limits on the application of this model. We have previously studied a handful of evolved disks using near-infrared hydrogen recombination lines whose flux is proportional to the accretion rate \citep{fla11,fla12}. We found that the accretion rate was highly variable, sometimes dropping below our detection limit, but in general the accretion luminosity was small compared to the stellar luminosity (L$_{acc}$/L$_*<$.06). This would imply that the change in luminosity due to the appearance of a hot spot would be small, creating a small change in inner wall radius. For a 6\%\ increase in luminosity the radius only increases by 6\%. For our model of an accretion spot blowing a hole in the inner wall, the radial depth of the wall must be very small ($\Delta R/R<.06$). SED modeling of pre-transition disks often assumes that the inner wall is infinitesimally thin and is still able to fit the SED \citep{esp10}. While the presence of dust directly behind the puffed-up inner wall is difficult to constrain because it is shadowed by the wall and appears very faint, a very narrow wall is consistent with the observations. In these previous studies, a direct correlation between infrared and accretion rate fluctuations was not seen, but this may be due to the complex geometry of the accretion flow, in combination with our particular line of sight. A hot spot on the far side of the star would produce a large change in infrared excess while not producing a significant change in the line flux. More detailed models, and continued observations, may help to resolve this potential conflict.

One possible exception to our model is LRLL 2.  \citet{esp12} were able to model its SED without a large gap. This star still exhibits a 'seesaw' behavior in its SED, which is difficult to explain with hot spot illumination without a narrow wall and a gap. It is possible that there is in fact a gap in the disk, but it is too small to strongly effect the SED shape. \citet{and11} find that some sources with a gap based on resolved sub-mm observations do not show the traditional SED shape of a pre-transition disk. It is possible that LRLL 2 represents another example of a system that has a radial gap that does not announce itself in the SED.

\subsection{Class III Variables}
For Class III sources, which by definition have little to no dust emission, alternative methods of producing infrared variability must be considered. Out of our entire sample of 108 class III sources, we find 21 possible variable stars. Four sources (LRLL 154, 286, 298, 329) have only a single discrepant data point and are unlikely to be real variables. Four sources (LRLL 64, 69, 87, 125) are periodic in the optical and infrared, with similar periods in both. Periodic fluctuations in the optical are most likely driven by cool spots rotating across the surface of the star, and we are likely seeing the effects of these spots in the infrared. One source, LRLL 314, is periodic in the optical but was not detected as periodic in the infrared, possibly because of the weakness of the signal. Cool spots should produce smaller fluctuations in the infrared than in the optical, and it may be difficult to pick out periodicity given the small size of the fluctuations. One source, LRLL 29, has an infrared period of 2.2 days, but has an optical period of 10.8 days. While some fluctuations in the period are seen \citep{coh04}, these changes are too small to account for the difference reported here. It is possible that period aliasing has led to different measures of the period in the optical and infrared. The remaining ten sources (LRLL 1, 4, 47, 48, 53, 179, 175, 259a, 277, 1939; see Table~\ref{tab_cluster} for positions and other ancillary information) have no evidence for an optical or infrared period. The fluctuations tend to be weak, with min to max differences less than 0.2 magnitudes. None of them are detected at [24], ruling out a cold massive disk that does not produce any excess over the 3 to 8\micron\ range. LRLL 1,4 53 show brief epochs with a small handful of data points fainter than the rest of the sample, similar to the 'dippers' seen by \citet{mor11}. These could be extinction events caused by small clumps of dust, which may not be massive enough to create a measurable excess. Another possibility is that there exists an unseen companion with a circumstellar disk that is variable. An M6 star with a disk whose K-band veiling was close to zero but occasionally increases by a few tenths, would produce [3.6] and [4.5] fluctuations of 0.05 to 0.2, similar to what is observed. While the presence of this dust would create a small excess, increasing $\alpha_{IRAC}$ by 0.15 at the most above the single star value, the infrared SED may still be steep enough to be counted as a diskless source. More detailed followup is needed to confirm the variability as well as determine its origin.

\section{Conclusion}
We have used 38 epochs of [3.6] and [4.5] photometry spread over 40 days and of over 200 low-mass members of the IC 348 cluster to study the structure of protoplanetary disks. We have found that many of the stars are variable in the infrared and that this variability is strongest and most frequent among the stars with the strongest infrared excesses. Fluctuations of a few tenths of a magnitude on weekly timescales are often seen from these young systems. It also appears that the systems in the south-west ridge, which are most likely younger than the rest of the cluster, are more likely to be variable. Those sources with larger X-ray luminosity also seem to be more variable, as well as stars with large irregular optical fluctuations. Relatively few of the stars show periodic fluctuations, most of which are related to stellar fluctuations from cold spots rather than changes in disk structure.

The frequency and size of the variability indicate that the disk is not a static entity. One likely possibility is that there are large structural changes in the disk, due to dust and gas being moved on a dynamical timescale, which at the inner edge of the disk is on the order of a week for many of these stars. We also find that dust can quickly sublimate when heated above 1500 K, which may result in large changes in the structure of the disk. Both the movement of material, and the rapid sublimation of dust, will be manifest as large structural changes in the emitting area of the inner disk. This increase in emitting area must be substantial in order to produce the observed fluctuations; increasing the scale height of 30\%\ of the disk by a factor of 2 could reproduce the observations. A broad range of scale height increases can also explain the measurements if appropriate adjustments are made in the amount of the disk undergoing a perturbation. Multiple theories for structural perturbations exist in the literature, including buoyant disk magnetic fields lifting dust out of the midplane, X-ray flares accelerating dust out of the midplane, and perturbations by the stellar magnetic field. Our observations suggest that these theories must be capable of structural changes on large physical scale. We explore the effect of variable illumination by a hot spot in additional detail. A 10,000 K hot spot covering on average 2.5\%\ of the stellar surface and illuminating 15\%\ of the inner wall can reproduce the size of the fluctuations and is consistent with the observed timescale. Systems that are younger and more X-ray luminous are more likely to have variable accretion, which would explain the ancillary trends that have been seen. While this model is dependent on many simplifying assumptions, it does indicate that variable illumination warrants further consideration.

This study of infrared variability, as well as others, points to a new picture of the disk. It is no longer a static system that slowly evolves, but it is a bubbling, boiling, wrinkled, dented, warped mass of gas and dust. These asymmetric structures cover a substantial portion of the disk and appear/disappear over the course of a month. While their source is still uncertain, it is clear that they are common and large. The structures are too small to be resolved with current instruments, but with infrared variability we have learned about their presence. Future resolved observations should frequently find non-axisymmetric structure that may have a large impact on the growth and migration of planetesimals within the disk.

\begin{figure}
\includegraphics{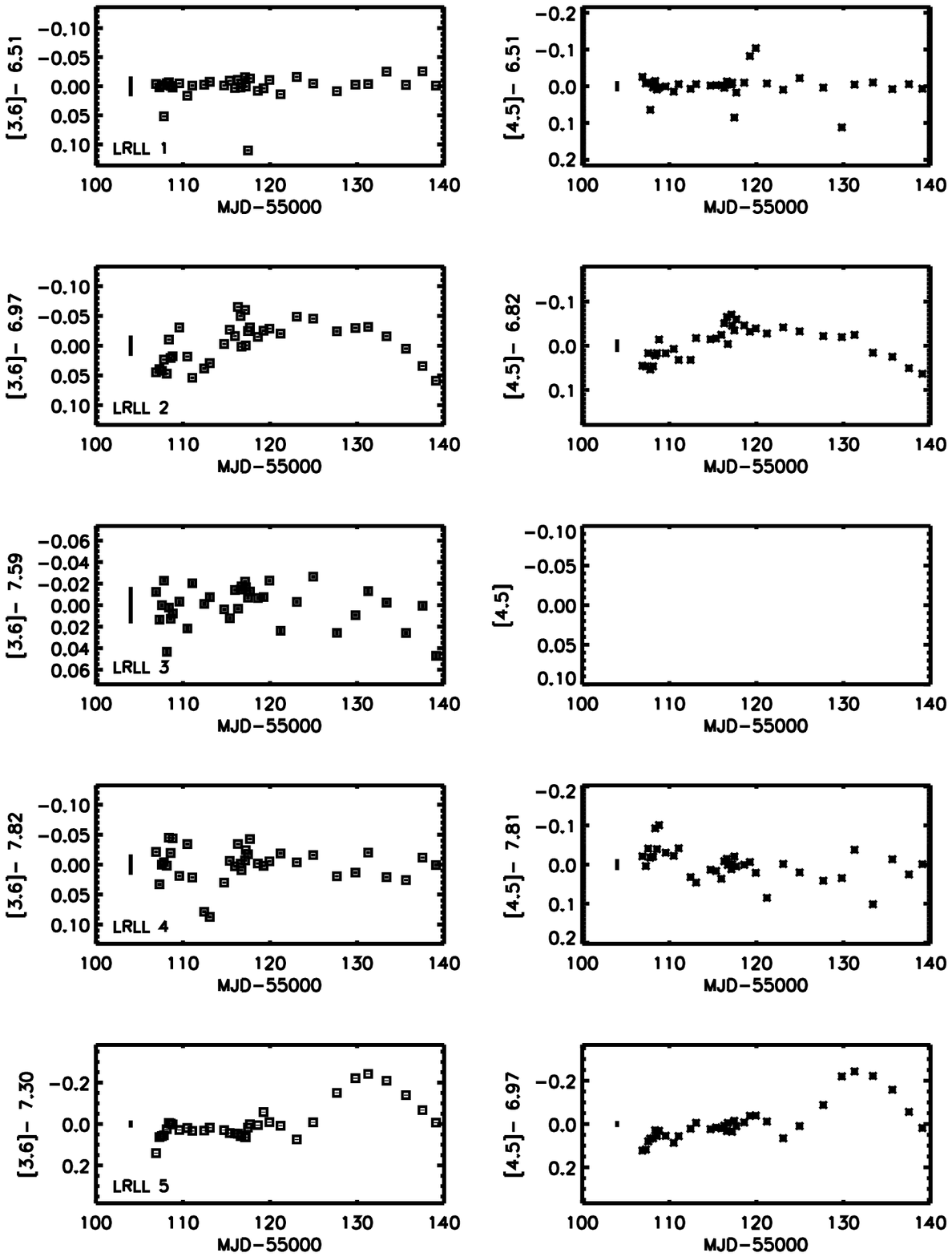}
\caption{[3.6] (left) and [4.5] (right) light curves for cluster members. Error bars on individual points have been included, as well as the error bars including the systematic uncertainty (vertical line, description of the derivation in the text). Only data points with an uncertainty of less than 0.5 mag are shown. Magnitudes have been normalized to the mean magnitudes, which are marked in the y-axis label. We show one set of light curves here, while the rest are available as an online figure.}
\end{figure}

\clearpage

\begin{deluxetable}{cccccccccccc}
\tabletypesize{\scriptsize}
%\footnotesize
\tablewidth{0pt}
\tablecaption{Cluster Member Properties\label{tab_cluster}}

\tablehead{ \colhead{LRLL} & \colhead{RA} & \colhead{Dec} & \colhead{T$_{eff}$} & \colhead{L$_*$} & \colhead{H$\alpha$ EW\tablenotemark{a}} & \colhead{Optical} & \colhead{A$_V$} & \colhead{L$_X$\tablenotemark{c}} & \colhead{[3.6]} & \colhead{$\alpha$\tablenotemark{d}} & \colhead{$\alpha_{IRAC}$\tablenotemark{e}}\\ 
 \colhead{\#} & \colhead{(J2000)} & \colhead{(J2000)} & \colhead{} & \colhead{} & \colhead{} & \colhead{Period\tablenotemark{b}} & \colhead{} & \colhead{(10$^{28}$} & \colhead{excess}& \colhead{} & \colhead{}\\
 \colhead{} & \colhead{(degrees)} & \colhead{(degrees)} & \colhead{(K)} & \colhead{(L$_{\odot}$)} & \colhead{($\AA$)} & \colhead{(days)} & \colhead{} & \colhead{ erg/s)} & \colhead{} & \colhead{} & \colhead{}} 
\startdata
1& 56.1425 & 32.1629 & 15400 & 605.000 & -10.000 & -10.000 & 1.910 & -10.000 & 0.010 & -2.76 & -2.80 \\
2& 56.1474 & 32.1679 & 8970 & 137.000 & -10.000 & 0.000 & 3.830 & 1053.000 & 0.003  & -1.41 & -1.47\\
3& 56.2110 & 32.3185 & 9520 & 135.000 & -10.000 & 6.097 & 4.470 & -10.000 & -0.012  & -2.77 & -2.85\\
4& 56.1300 & 32.1061 & 7200 & 34.000 & -10.000 & 0.000 & 1.770 & 14.900 & -0.004 & -2.69 & -2.72\\
5& 56.1084 & 32.0751 & 5520 & 9.900 & 5.000 & 6.536 & 5.320 & 385.000 & 0.188  & -1.14 & -1.24\\
6& 56.1539 & 32.1126 & 5830 & 17.000 & -10.000 & 1.659 & 3.900 & 1961.000 & -0.007  & -2.02 & -2.08\\
7& 56.0353 & 32.1213 & 9520 & 35.000 & -10.000 & -10.000 & 0.920 & -10.000 & -0.002  & -2.70 & -2.72\\
8& 56.0381 & 32.1193 & 8970 & 28.000 & -10.000 & -10.000 & 0.960 & -10.000 & -0.001  & -2.38 & -2.40\\
9& 56.1632 & 32.1551 & 5520 & 11.000 & -10.000 & 2.600 & 5.530 & 573.000 & -0.015  & -2.76 & -2.86\\
10& 56.1028 & 32.1708 & 6890 & 13.000 & -10.000 & 0.000 & 1.910 & 13.200 & -0.004  & -2.86 & -2.89\\
\enddata
\tablecomments{The full version of this table is available in machine-readable format. Here we show a small portion for illustration. The full table includes columns for LRLL \#, RA, Dec, T$_{eff}$, L$_*$, H$\alpha$EW, Optical Period, A$_V$, L$_X$, [3.6] excess, $\alpha$, $\alpha_{IRAC}$, HH 211 distance, Var? and Reference. Values of -10 or -100 correspond to a non-detection. We do not include sources  8024, 8042, 8078, 9042, 9060, 30190, 30191, 30192 from \citet{lad06} because they are duplicates of sources 24A, 42A, 78A, 42B, 60A, 92,1,9 respectively}
\tablenotetext{a}{Here we actually report the negative of the H$\alpha$ EW, with positive numbers corresponding to emission. H$\alpha$ EW is not reported when it is in absorption, and a value of -10 corresponds to a non-detection.}
\tablenotetext{b}{Optical period is taken from \citet{coh04,cie06} and references therein. Values of 0 correspond to stars that were detected in these surveys, but no period structure was detected, while values of -10 refer to stars that were not reported as detections in these surveys.}
\tablenotetext{c}{X-ray luminosity taken from \citet{pre02}}
\tablenotetext{d}{The slope of the observed infrared SED, defined as $\lambda$F$_{\lambda}\propto\lambda^{\alpha}$ over the IRAC (3-8\micron) wavelengths.}
\tablenotetext{e}{The slope of the dereddened infrared SED. Unless otherwise stated, this is the value of the infrared SED that is used in the text.}
\tablenotetext{f}{Var?: y = Variable, n = not variable, y1 = Variable, but only detected at [3.6], n1 = not variable, but only detected at [3.6], p = periodic, b = a nearby source prevents us from accurately assessing variability, $\ldots$ = source was not detected}
\tablenotetext{g}{Reference key: luh03 = \citet{luh03}, mue07 = \citet{mue07}, lad06 = \citet{lad06}, muz06 = \citet{muz06}, luh05 = \citet{luh05}}
\end{deluxetable}

 \begin{deluxetable}{cccccccccccc}
%\rotate
\tabletypesize{\scriptsize}
\tablecaption{[3.6] photometry\label{ch1_phot}}
\tablehead{\colhead{} & \colhead{RA (J2000)} & \colhead{Dec (J2000)} & \colhead{55106.9} & \colhead{} & \colhead{55107.3} & \colhead{} & \colhead{55107.8} & \colhead{} & \colhead{55108.1} & \colhead{} \\ 
\colhead{} & \colhead{(Degrees)} & \colhead{(Degrees)} & \colhead{([3.6])} & \colhead{([3.6] unc)} & \colhead{([3.6])} & \colhead{([3.6] unc)} & \colhead{([3.6])} & \colhead{([3.6] unc)} & \colhead{([3.6])} & \colhead{([3.6] unc)} } 

\startdata
& 56.1425 & 32.1629 & 6.510 & 0.0021 & 6.516 & 0.0017 & 6.512 & 0.0021 & 6.566 & 0.0020 \\
 & 56.1474 & 32.1679 & 7.015 & 0.0018 & 7.009 & 0.0016 & 7.012 & 0.0021 & 6.993 & 0.0019 \\
 & 56.2110 & 32.3185 & 7.575 & 0.0019 & 7.601 & 0.0021 & 7.587 & 0.0019 & 7.564 & 0.0020 \\
 \enddata
 \tablecomments{The full version of this table is available in machine-readable format. Here we show a small portion for illustration. Column headings refer to the MJD at which the data was taken. Magnitudes less than zero, with uncertainties of 10, correspond to a non-detection.}
 \end{deluxetable}
  
 \begin{deluxetable}{cccccccccccc}
 %\rotate
 \tabletypesize{\scriptsize}
 \tablecaption{[4.5] photometry\label{ch2_phot}}
 \tablehead{\colhead{} & \colhead{RA (J2000)} & \colhead{Dec (J2000)} & \colhead{55106.9} & \colhead{} & \colhead{55107.3} & \colhead{} & \colhead{55107.8} & \colhead{} & \colhead{55108.1} & \colhead{}  \\  & \colhead{(Degrees)} & \colhead{(Degrees)} & \colhead{([4.5])} & \colhead{([4.5] unc)} & \colhead{([4.5])} & \colhead{([4.5] unc)} & \colhead{([4.5])} & \colhead{([4.5] unc)} & \colhead{([4.5])} & \colhead{([4.5] unc)} }

\startdata
 & 56.1425 & 32.1629 & 6.489 & 0.0022 & 6.507 & 0.0021 & 6.504 & 0.0023 & 6.578 & 0.0022  \\
 & 56.1474 & 32.1679 & 6.862 & 0.0022 & 6.863 & 0.0022 & 6.833 & 0.0020 & 6.870 & 0.0022  \\
 & 56.2110 & 32.3185 & -100.000 & 10.0000 & -100.000 & 10.0000 & -100.000 & 10.0000 & -100.000 & 10.0000 \\
 \enddata
  \tablecomments{The full version of this table is available in machine-readable format. Here we show a small portion for illustration. Column headings refer to the MJD at which the data was taken. Magnitudes less than zero, with uncertainties of 10, correspond to a non-detection.}
 \end{deluxetable}


\begin{thebibliography}{}
\bibitem[Alencar et al.(2010)]{ale10} Alencar, S.H.P., et al. 2010, \aap, 519, 88
\bibitem[Andrews et al.(2011)]{and11} Andrews, S.M., Wilner, D.J., Espaillat, C., Hughes, A.M., Dullemond, C.P., McClure, M.K., Chunhua, Q., \&\ Brown, J.M. 2011, \apj, 732, 42
\bibitem[Argiroffi et al.(2011)]{arg11} Argiroffi, C., et al. 2011, \aap, 530, 1
\bibitem[Arnold et al.(2012)]{arn12} Arnold, T.J., Eisner, J.A., Monnier, J.D., \&\ Tuthill, P. 2012, \apj, 750, 119
\bibitem[Barsony et al.(2005)]{bar05} Barsony, M., Ressler, M.E., \&\ Marsh, K.A. 2005, \apj, 630, 381
\bibitem[Blandford \&\ Payne(1982)]{bla82} Blandford, R.D., \&\ Payne, D.G. 1982, \mnras, 199, 883
\bibitem[Bouvier et al.(2007)]{bou07} Bouvier, J., et al. 2007, \aap, 463, 1017
\bibitem[Brockwell \&\ Davis(2002)]{bro02} Brockwell, P.J., \&\ Davis, R.A., 2002, {\it Introduction to Time Series and Forecasting} (2nd ed.)
\bibitem[Calvet \&\ Gullbring(1998)]{cal98} Calvet, N., \&\ Gullbring, E. 1998, \apj, 509, 802
\bibitem[Chiang \&\ Goldreich(1997)]{chi97} Chiang, E.I. \&\ Goldreich, P. 1997, \apj, 490, 368
\bibitem[Cieza \&\ Baliber(2006)]{cie06} Cieza, L., \&\ Baliber, N. 2006, \apj, 649, 862
\bibitem[Carpenter et al.(2001)]{car01} Carpenter, J.M., Hillenbrand, L.A., \&\ Strutskie, M.F. 2001, \aj, 121, 3160
\bibitem[Cohen et al.(2004)]{coh04} Cohen, R.E., Herbst, W., Williams, E.C. 2004, \aj, 127, 1602
\bibitem[D'Alessio et al.(2006)]{dal06} D'Alessio, P., Calvet, N., Nartmann, L., Franco-Hern\'{a}ndez, R., \&\ Serv\'{i}n, H. 2006, \apj, 638, 314
\bibitem[Dullemond \&\ Monnier(2010)]{dul10} Dullemond, C.P. \&\ Monnier, J.D. 2010, \araa, 48, 205
\bibitem[Edelson \&\ Krolik(1988)]{ede88} Edelson, R.A., \&\ Krolik, J.H. 1988, \apj, 333, 646
\bibitem[Eisloffel et al.(2003)]{eis03} Eisloffel, J., Forebrich, D., Stanke, T., \&\ McCaughrean, M.J. 2003, \apj, 595, 259
\bibitem[Espaillat et al.(2010)]{esp10} Espaillat, C., et al. 2010, \apj, 717, 441
\bibitem[Espaillat et al.(2011)]{esp11} Espaillat, C., Furlan, E., D'Alessio, P., Sargent, B., Nagel, E., Calvet, N., Watson, D.M., \&\ Muzerolle, J. 2011, \apj, 728, 49
\bibitem[Espaillat et al.(2012)]{esp12} Espaillat, C., et al. 2012, \apj, 747, 103
\bibitem[Feigelson \&\ Montmerle(1999)]{fei99} Feigelson, E.D., \&\ Montmerle, T., 1999, \araa, 37, 363
\bibitem[Feigelson et al.(2007)]{fei07} Feigelson, E., Townsley, L., Gudel, M. \&\ Stassun, K. 2007, in Protostars and Planets V, ed. Reipurth, B., Jewitt, D., Keil, K. (Tucson, AZ: Univ. Arizona Press), 313
\bibitem[Flaccomio et al.(2010)]{flac10} Flaccomio, E., Micela, G., Favata, F. \&\ Alencar, S.P.H. 2010, \aap, 516, L8
\bibitem[Flaherty et al.(2007)]{fla07} Flaherty, K.M., Pipher, J.L., Megeath, S.T., Winston, E.M., Gutermuth, R.A., Muzerolle, J., Allen, L., \&\ Fazio, G.G. 2007, \apj, 663, 1069
\bibitem[Flaherty \& Muzerolle(2010)]{fla10} Flaherty, K.M., \& Muzerolle, J. 2010, \apj, 719, 1733
\bibitem[Flaherty et al.(2011)]{fla11} Flaherty, K.M., Muzerolle, J., Rieke, G., Gutermuth, R., Balog, Z., Herbst, W., Megeath, S.T., \&\ Kun, M. 2011, \apj, 732, 83
\bibitem[Flaherty et al.(2012)]{fla12} Flaherty, K.M., Muzerolle, J., Rieke, G., Gutermuth, R., Balog, Z., Herbst, W., Megeath, S.T., \&\ Kun, M. 2012, \apj, 748, 71
\bibitem[Forbich et al.(2008)]{for08} Forbich, J., Menten, K.M., \&\ Reid, M.J. 2008, \aap, 477, 267
\bibitem[Fragner \&\ Nelson(2010)]{fra10} Fragner, M.M., \&\ Nelson, R.P. 2010, \aap, 511, 77
\bibitem[Goodson \&\ Winglee(1999)]{goo99} Goodson, A.P., \&\ Winglee, R.M. 1999, \apj, 524, 159
\bibitem[Gutermuth et al.(2009)]{gut09} Gutermuth, R.A., Megeath, S.T., Myers, P.C., Allen, L.E., Pipher, J.L., \&\ Fazio, G.G. 2009, \apjs, 184, 18
\bibitem[Hartmann et al.(1998)]{har98} Hartmann, L., Calvet, N., Gullbring, E. \&\ D'Alessio, P. 1998, \apj, 495, 385
\bibitem[Herbst et al.(1994)]{her94} Herbst, W., Herbst, D.K., Grossman, E.J., \&\ Weinstein, D. 1994, \aj, 108, 1906
\bibitem[Herbst et al.(2002)]{her02} Herbst, W., Bailer-Jones, C.A.L., Mundt, R., Meisenheimer, K., \&\ Wackermann, R. 2002, \aap, 396, 513
\bibitem[Herbst(2008)]{her08} Herbst, W. 2008, in Handbook of Star Forming Regions, Volume I: The Northern Sky ed. B. Reipurth (San Francisco, California: ASP), 372
\bibitem[Hirose \&\ Turner(2011)]{hir11} Hirose, S. \&\ Turner, N.J. 2011, \apj, 732, 30L
\bibitem[Hosmer \&\ Lemeshow(2000)]{hos00} Hosmer, D.W., \&\ Lemeshow, S. 2000 {\it Applied Logistic Regression} (1st ed)
\bibitem[Hutchinson et al.(1994)]{hut94} Hutchinson, M.G., Albinson, J.S., Barrett, P., Davies, J.K., Evans, A., Goldsmith, M.J., \& Maddison, R.C. 1994, \aap, 285, 883
\bibitem[Jorgensen et al.(2006)]{jor06} Jorgensen, J.K., et al. 2006, \apj, 645, 1246
\bibitem[Jorgenson et al.(2008)]{jor08} Jorgenson, J.K., Johnstone, D., Kirk, H., Myers, P.C., Allen, L.E., Shirley, Y.L. 2008, \apj, 683, 822
\bibitem[Joy(1945)]{joy45} Joy, A.H. 1945, \apj, 102, 168
\bibitem[Juh\'{a}sz et al.(2007)]{juh07} Juh\'{a}sz, A., Prusti, T., \'{A}brah\'{a}m, P., Dullemond, C.P. 2007, \mnras, 374, 1252
\bibitem[Ke et al.(2012)]{ke12} Ke, T.T., Huang, H., \&\ Lin, D.N.C. 2012, \apj, 745, 60 
\bibitem[Kelly(2007)]{kel07} Kelly, B.C. 2007, \apj, 665, 1489
\bibitem[Kelly et al.(2009)]{kel09} Kelly, B.C., Bechtold, J., Siemiginowska, A. 2009, \apj, 698, 895
\bibitem[Kelly et al.(2011)]{kel11} Kelly, B.C., Sobolewska, M., \&\ Siemiginowska, A. 2011, \apj, 730, 52
\bibitem[Kenyon \&\ Hartmann(1995)]{ken95} Kenyon, S.J., \&\ Hartmann, L. 1995, \apjs, 101, 117
\bibitem[K\'{o}sp\'{a}l et al.(2012)]{kos12} K\'{o}sp\'{a}l, \'{A}, \'{A}brah\'{a}m, P., Acosta-Pulido, J.A., Dullemond, C.P., Henning, Th., Kun, M., Leinert, Ch., Mo\'{o}r, A., \&\ Turner, N. 2012, arXiv:/1204.3473
\bibitem[Kulkarni \&\ Romanova(2008)]{kul08} Kulkarni, A.K. \&\ Romanova M.M. 2008, \mnras, 386, 673
\bibitem[Lada et al.(2006)]{lad06} Lada, C.J., et al. 2006, \aj, 131, 1574
\bibitem[Luhman et al.(2003)]{luh03} Luhman, K.L., Stauffer, J.R., Muench, A.A., Rieke, G.H., Lada, E.A., Bouvier, J., Lada, C.J. 2003, \apj, 593, 1093
\bibitem[Luhman et al.(2005)]{luh05} Luhman, K.L., Lada, E.A., Muench, A.A., \&\ Elston, R.J. 2005, \apjl, 618, 810
\bibitem[Morales-Calder\'{o}n et al.(2009)]{mor09} Morales-Calder\'{o}n, M., et al. 2009, \apj, 702, 1507
\bibitem[Morales-Calder\'{o}n et al.(2011)]{mor11} Morales-Calder\'{o}n, M., et al. 2011, \apj, 733, 50
\bibitem[Morales-Calder\'{o}n et al.(2012)]{mor12} Morales-Calder\'{o}n, M., et al. 2012, \apj, 753, 149
\bibitem[Muench et al.(2007)]{mue07} Muench, A.A., Lada, C.J., Luhman, K.L., Muzerolle, J., \&\ Young, E. 2007, \aj, 134, 411
\bibitem[Muzerolle et al.(2003)]{muz03} Muzerolle, J., Calvet, N., Hartmann, L., \&\ D'Alessio, P. 2003, \apj, 597, L149
\bibitem[Muzerolle et al.(2006)]{muz06} Muzerolle, J., et al. 2006, \apj, 643, 1003
\bibitem[Muzerolle et al.(2009)]{muz09} Muzerolle, J., et al. 2009, \apj, 704, L15
\bibitem[Nagel et al.(2012)]{nag12} Nagel, E., Espaillat, C., D'Alessio, P., \&\ Calvet, N. 2012, \apj, 747, 139
\bibitem[Natta et al.(2004)]{nat04} Natta, A., Testi, L., Muzerolle, J., Randich, S., Comer\'{o}n, F., \&\ Persi, P. 2004, \aap, 424, 603
\bibitem[Nuth \&\ Ferguson(2006)]{nut06} Nuth, J.A. III, Ferguson, F.T. 2006, \apj, 649, 1178
\bibitem[Nordhagen et al.(2006)]{nor06} Nordhagen, S., Herbst, W., Rhode, K.L, \&\ Williams, E.C. 2006, \aj, 132, 1555
\bibitem[Preibisch \&\ Zinnecker(2002)]{pre02} Preibisch, T., \&\ Zinnecker, H. 2002, \aj, 123, 1613
\bibitem[Preibisch \&\ Zinnecker(2004)]{pre04} Preibisch, T., \&\ Zinnecker, H. 2004, \aap, 422, 1001
\bibitem[Rebull et al.(2006)]{reb06} Rebull, L.M., Stauffer, J.R., Megeath, S.T., Hora, J.L, \&\ Hartmann, L., \apj, 646, 297
\bibitem[Reegen(2007)]{ree07} Reegen, P. 2007, \aap, 467, 1353
\bibitem[Robitaille et al.(2007)]{rob07} Robitaille, T. P., Whitney, B.A., Indebetouw, R., \&\ Wood, K. 2007, \apjs, 169, 328
\bibitem[Shu et al.(1996)]{shu96} Shu, F., Shang, H., \&\ Lee, T. 1996, Science, 271, 1545
\bibitem[Sitko et al.(2008)]{sit08} Sitko, M.L., et al. 2008, \apj, 678, 1070
\bibitem[Skemer et al.(2010)]{ske10} Skemer, A.J., et al. 2010, \apj, 711, 1280
\bibitem[Stassun et al.(2007)]{sta07} Stassun, K.G., van den Berg, M. \&\ Feigelson, E. 2007, \apj, 660, 704
\bibitem[Stelzer et al.(2007)]{ste07} Stelzer, B., Flaccomio, E., Briggs, K., Micela, G., Scelsi, L., Audard, M., Pillitteri, I., \&\ Gudel, M., 2007, \aap, 468, 463
\bibitem[Stelzer et al.(2012)]{ste12} Stelzer, B., Preibisch, T., Alexander, F., Mucciarelli, P., Flaccomio, E., Micela, G., \&\ Sciortino, S. 2012, \aap, 537, 135
\bibitem[Stetson(1996)]{ste96} Stetson, P.B. 1996, \pasp, 108, 851
\bibitem[Tafalla et al.(2006)]{taf06} Tafalla, M., Kumar, M.S.N., \&\ Bachiller, R. 2006, \aap, 456, 179
\bibitem[Timmer et al.(1995)]{tim95} Timmer, J., \&\ K\"{o}nig 1995, \aap, 300, 707
\bibitem[Turner et al.(2010)]{tur10} Turner, N.J., Carballido, A., \&\ Sano, T. 2010, \apj, 708, 188
\bibitem[Walawander et al.(2006)]{wal06} Walawender, J., Bally, J., Kirk, H., Johnstone, D., Reipurth, B., \&\ Aspin, C. 2006, \aj, 132, 467
\bibitem[Waters \&\ Waelkens(1998)]{wat98} Waters, L.B.F.M. \&\ Waelkens, C. 1998, \araa, 36, 233
\bibitem[Wolk et al.(2005)]{wol05} Wolk, S.J., Harnden, F.R., Jr., Flaccomio, E., Micela, G., Favata, F., Shang, H., \&\ Feigelson, E.D. 2005, \apjs, 160, 423


\end{thebibliography}
\end{document}